\documentclass[a4paper,10pt,prd,aps,
longbibliography,
preprintnumbers,
nofootinbib,
floatfix,
superscriptaddress]{revtex4}
\usepackage{graphicx}
\usepackage{epsfig}
\usepackage{rotating}
\usepackage{amssymb}
\usepackage{dsfont}
\usepackage{psfrag}
\usepackage{amsmath,euscript,array,mathrsfs}
\usepackage{bbold}
\usepackage{bm}
\usepackage{multirow}
\usepackage{dcolumn}
\RequirePackage{doi}
\usepackage[skip=10pt plus1pt, indent=20pt]{parskip}
\usepackage{float}
\floatstyle{plaintop}
\usepackage{bbding}
\usepackage{xcolor}
\colorlet{RED}{red} 
\usepackage{orcidlink}
\usepackage{hyperref}
\usepackage{tikz}
\usepackage{bbm}
\usepackage{verbatim}
\usepackage{tikz}

\newcommand{\beq}{\begin{equation}}
\newcommand{\eeq}{\end{equation}}
\newcommand{\beqs}{\begin{eqnarray}}
\newcommand{\eeqs}{\end{eqnarray}}
\newcommand{\Tr}{{\rm Tr}}

\graphicspath{{./FIG/}}
\makeatletter
\def\input@path{{./TAB/}}
\makeatother

\newcommand{\orcidauthorBENNETT}{0000-0002-1678-6701}
\newcommand{\orcidauthorLUCINI}{0000-0001-8974-8266}
\newcommand{\orcidauthorPIAI}{0000-0002-2251-0111} 
\newcommand{\orcidauthorFORZANO}{0000-0003-0985-8858}
\newcommand{\orcidauthorVADACCHINO}{0000-0002-5783-5602}
\newcommand{\orcidauthorHILL}{0000-0003-2383-940X}
\newcommand{\orcidauthorHONG}{0000-0002-3923-4184}
\newcommand{\orcidauthorDELDEBBIO}{0000-0003-4246-3305}
\newcommand{\orcidauthorLUPO}{0000-0001-9661-7811}
\newcommand{\orcidauthorLIN}{0000-0003-3743-0840}
\newcommand{\orcidauthorLEE}{0000-0002-4616-2422}
\newcommand{\orcidauthorZIERLER}{0000-0002-8670-4054}
\newcommand{\orcidauthorHSIAO}{0000-0002-8522-5190}

\makeatletter
\def\@collaboration@present#1#2#3{%
 \par
 \begingroup
  \frontmatter@collaboration@above
  \@author@present{\ignorespaces#2\unskip}{#3}%
  \par
 \endgroup
 \set@listcomma@list#1%
}%
\makeatother

\begin{document}

\preprint{UTCCS-P168}
\preprint{UTHEP-807}
\preprint{PNUTP-25/A04}
\preprint{CTPU-PTC-25-24}

\title{Chimera baryons and mesons on the lattice: a spectral density analysis}

\author{Ed Bennett\,\orcidlink{\orcidauthorBENNETT}}
\email{E.J.Bennett@swansea.ac.uk}
\affiliation{Swansea Academy of Advanced Computing, Swansea University (Bay Campus), Fabian Way, Swansea SA1 8EN, United Kingdom}
\affiliation{ Centre for Quantum Fields and Gravity, Faculty  of Science and Engineering, Swansea University, Singleton Park, SA2 8PP, Swansea, United Kingdom}

\author{Luigi Del Debbio\,\orcidlink{\orcidauthorDELDEBBIO}}
\email{luigi.del.debbio@ed.ac.uk}
\affiliation{Higgs Centre for Theoretical Physics, School of Physics and Astronomy, 
The University of Edinburgh, Peter Guthrie Tait Road, Edinburgh EH9 3FD, United Kingdom}

\author{Niccolò Forzano\,\orcidlink{\orcidauthorFORZANO}}
\email{2227764@swansea.ac.uk}
\affiliation{Department of Physics, Faculty of Science and Engineering, Swansea University, Singleton Park, SA2 8PP, Swansea, United Kingdom}
\affiliation{ Centre for Quantum Fields and Gravity, Faculty  of Science and Engineering, Swansea University, Singleton Park, SA2 8PP, Swansea, United Kingdom}

\author{Ryan~C. Hill\,\orcidlink{\orcidauthorHILL}}
\email{ryan.hill@ed.ac.uk}
\affiliation{School of Physics and Astronomy, The University of Edinburgh, Edinburgh EH9 3FD, United Kingdom}

\author{Deog~Ki Hong\,\orcidlink{\orcidauthorHONG}}
\email{dkhong@pusan.ac.kr}
\affiliation{Department of Physics, Pusan National University, Busan 46241, Korea}
\affiliation{Extreme Physics Institute, Pusan National University, Busan 46241, Korea}

\author{Ho Hsiao\,\orcidlink{\orcidauthorHSIAO}}
 \email{hohsiao@ccs.tsukuba.ac.jp}
\affiliation{Center for Computational Sciences, University of Tsukuba, 1-1-1 Tennodai, Tsukuba, Ibaraki 305-8577, Japan}
\affiliation{Institute of Physics, National Yang Ming Chiao Tung University, 1001 Ta-Hsueh Road, Hsinchu 30010, Taiwan}

\author{Jong-Wan Lee\,\orcidlink{\orcidauthorLEE}}
\email{j.w.lee@ibs.re.kr}
\affiliation{ Particle Theory  and Cosmology Group, Center for Theoretical Physics of the Universe, Institute for Basic Science (IBS), Daejeon, 34126, Korea }

\author{C.-J. David Lin\,\orcidlink{\orcidauthorLIN}}
\email{dlin@nycu.edu.tw}
\affiliation{Institute of Physics, National Yang Ming Chiao Tung University, 1001 Ta-Hsueh Road, Hsinchu 30010, Taiwan}
\affiliation{Centre for High Energy Physics, Chung-Yuan Christian University, Chung-Li 32023, Taiwan}

\author{Biagio Lucini\,\orcidlink{\orcidauthorLUCINI}}
\email{b.lucini@qmul.ac.uk}
\affiliation{Swansea Academy of Advanced Computing, Swansea University (Bay Campus), Fabian Way, Swansea SA1 8EN, United Kingdom}
\affiliation{Department of Mathematics, Faculty of Science and Engineering, Swansea University (Bay Campus), Fabian Way, SA1 8EN Swansea, United Kingdom}
\affiliation{School of Mathematical Sciences, Queen Mary University of London, Mile End Road,
London, E1 4NS, United Kingdom}

\author{Alessandro Lupo\,\orcidlink{\orcidauthorLUPO}}
\email{alessandro.lupo@cpt.univ-mrs.fr}
\affiliation{Aix Marseille Univ, Université de Toulon, CNRS, CPT, Marseille, France}

\author{Maurizio Piai\,\orcidlink{\orcidauthorPIAI}}
\email{m.piai@swansea.ac.uk}
\affiliation{Department of Physics, Faculty of Science and Engineering, Swansea University, Singleton Park, SA2 8PP, Swansea, United Kingdom}
\affiliation{ Centre for Quantum Fields and Gravity, Faculty  of Science and Engineering, Swansea University, Singleton Park, SA2 8PP, Swansea, United Kingdom}

\author{Davide Vadacchino\,\orcidlink{\orcidauthorVADACCHINO}}
\email{davide.vadacchino@plymouth.ac.uk}
\affiliation{Centre for Mathematical Sciences, University of Plymouth, Plymouth, PL4 8AA, United Kingdom}

\author{Fabian Zierler\,\orcidlink{\orcidauthorZIERLER}}
\email{fabian.zierler@swansea.ac.uk}
\affiliation{Department of Physics, Faculty of Science and Engineering, Swansea University, Singleton Park, SA2 8PP, Swansea, United Kingdom}
\affiliation{ Centre for Quantum Fields and Gravity, Faculty  of Science and Engineering, Swansea University, Singleton Park, SA2 8PP, Swansea, United Kingdom}

\collaboration{(on behalf of the TELOS collaboration) \vspace{4pt}\\ 
\href{https://telos-collaboration.github.io}{ \includegraphics[height=1cm]{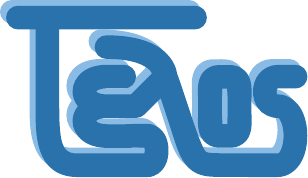} } }
\noaffiliation

\date{\today}

\begin{abstract}

We develop and test a spectral-density analysis method, based on the introduction of  smeared energy kernels,
 to extract physical information from
two-point correlation functions computed numerically in lattice field theory.
We apply it to a $Sp(4)$ gauge theory {\color{blue} with} fermion matter fields transforming in distinct representations, 
with $N_{\rm f}=2$ Dirac fermions in the fundamental and $N_{\rm as}=3$  in the 2-index antisymmetric representation.
The corresponding continuum theory provides the minimal candidate model for a composite Higgs boson with partial top compositeness. 
We consider a broad class of composite operators, that source flavored mesons and (chimera) baryons, for several finite 
choices of lattice bare parameters.
For the chimera baryons, which include candidate top-quark partners, we provide the first measurements, obtained with dynamical fermions, of the ground state and the lowest excited state masses, in all channels of spin, isospin, and parity. We also measure matrix elements and overlap factors, that are important to realize viable models of partial top compositeness, 
by implementing an innovative way of extracting this information from the spectral densities.
For the mesons, among which the pseudoscalars can be reinterpreted to provide an extension of the Higgs sector of the Standard Model of particle physics, our measurements of the renormalized matrix elements and decay constants are new results. We complement them with an update of existing measurements of the meson masses, obtained with higher statistics and improved analysis. The analysis software is made publicly available, and can be used in other lattice studies, including application to quantum chromodynamics (QCD).

\end{abstract}

\maketitle

\tableofcontents

\section{Introduction} 
\label{sec:introduction}

In the context of  strongly coupled lattice gauge theories with matter field content, 
such as quantum chromodynamics (QCD), 
the study of spectral densities provides a new tool for the analysis of numerical data
with numerous interesting applications~\cite{Hansen:2019idp,
Hansen:2017mnd,Bulava:2019kbi,Kades:2019wtd,Bailas:2020qmv,Gambino:2020crt,Bruno:2020kyl,Lupo:2021nzv,DelDebbio:2021whr,DelDebbio:2022qgu,Bulava:2021fre,Pawlowski:2022zhh,
Gambino:2022dvu,Lupo:2022nuj,Bulava:2023brj,DeSantis:2023rjl,Lupo:2023qna,Bergamaschi:2023xzx,Bonanno:2023thi,Frezzotti:2023nun,Frezzotti:2024kqk,Bruno:2024fqc,DelDebbio:2024lwm}.
It gives access to off-shell information encoded in the
correlation functions computed with lattices with finite spacings and sizes. It can be used to compute scattering amplitudes~\cite{Bulava:2019kbi,Patella:2024cto},
inclusive decay rates~\cite{Gambino:2020crt, Bulava:2021fre,ExtendedTwistedMassCollaborationETMC:2022sta,Blum:2024hyr}, in particular the
inclusive decays of the $\tau$ lepton~\cite{Evangelista:2023fmt,ExtendedTwistedMass:2024myu},
properties of glueballs~\cite{Panero:2023zdr} and of the 
quark-gluon plasma (see, e.g., the reviews in Refs.~\cite{Asakawa:2000tr,Meyer:2011gj, Ratti:2017qgq}, as well as Ref.~\cite{Aarts:2023vsf}), with special reference to electrical conductivity~\cite{Aarts:2007wj, Aarts:2014nba,Almirante:2024lqn}, 
and properties of heavy mesons travelling through the plasma~\cite{Bignell:2025bga,Smecca:2025hfw}.
The Backus-Gilbert algorithm for the reconstruction of spectral densities was originally proposed in Ref.~\cite{Backus:1968svk}, and for the purposes of this paper we adopt one of its promising refinement: the Hansen-Lupo-Tantalo (HLT) algorithm~\cite{Hansen:2019idp}.
Our first, general aim is to demonstrate the numerical implementation of the HLT algorithm 
to study fermionic bound states in strongly coupled field theories regulated on the lattice.
We present a set of analysis methods that are based on the reconstruction of 
spectral density information from lattice field theory. We  test them on a theory that represents a compelling candidate for new physics.
Our second aim is then to extract novel information about this new theory, useful for phenomenological studies, and to establish its viability.

Despite its astonishing successes and predictive power, 
the Standard Model of particle physics is not the complete theory of fundamental interactions at the microscopic level. 
Besides the fact that it does not include a theory of quantum gravity, there is also
conclusive theoretical evidence that many of the SM couplings are affected by the triviality problem.
On the other hand, because these couplings become strong at short length scales, they require the existence of a physical cut-off scale~\cite{Aizenman:2019yuo,Luscher:1987ay,Luscher:1987ek, Luscher:1988uq,Bulava:2012rb, Molgaard:2014mqa, Chu:2018ldw}, beyond which a new, more fundamental theory determines the dynamics.
Complementary, phenomenological evidence of new physics is provided by the fact that 
the Standard Model cannot explain the observed baryon asymmetry of the universe (the electroweak phase 
transition being too weak~\cite{Kajantie:1996mn, Laine:2012jy}), 
nor the origin of its dark matter (see the review~\cite{Cirelli:2024ssz}) and dark energy components~\cite{Planck:2013oqw}.  

An appealing framework for extensions of the Standard Model is obtained by postulating
 the existence of a new, strongly coupled, non-abelian and confining gauge theory, 
and identifying some of the bosons and fermions in the SM field content as
composite states of this new theory. This approach  is particularly suited to 
 the two heaviest among the SM particles: the Higgs boson and the top quark.
 In Composite Higgs Models (CHMs), in which the Higgs boson is described as a composite pseudo Nambu-Goldstone boson (PNGB)~\cite{Kaplan:1983fs,Georgi:1984af,Dugan:1984hq}, Top Partial Compositeness (TPC) also provides a mechanism for generating a large top quark mass. \footnote{See the reviews in Refs.~\cite{Panico:2015jxa,Witzel:2019jbe,Cacciapaglia:2020kgq,Bennett:2023wjw},
and the summary tables in Refs.~\cite{Ferretti:2013kya,Ferretti:2016upr,Cacciapaglia:2019bqz,Kaplan:1991dc,Grossman:1999ra,Gherghetta:2000qt,Chacko:2012sy}}.
The literature on these models is vast, see for instance Refs.~\cite{Katz:2005au,Barbieri:2007bh,
Lodone:2008yy,Gripaios:2009pe,Mrazek:2011iu,Marzocca:2012zn,Grojean:2013qca,Cacciapaglia:2014uja,
Ferretti:2014qta,Arbey:2015exa,Cacciapaglia:2015eqa,Vecchi:2015fma,Ma:2015gra,Feruglio:2016zvt,DeGrand:2016pgq,Fichet:2016xvs,
Galloway:2016fuo,Agugliaro:2016clv,Belyaev:2016ftv,Csaki:2017cep,Chala:2017sjk,Golterman:2017vdj,
Csaki:2017jby,Alanne:2017rrs,Alanne:2017ymh,Sannino:2017utc,Alanne:2018wtp,Bizot:2018tds,
Cai:2018tet,Agugliaro:2018vsu,Cacciapaglia:2018avr,BuarqueFranzosi:2018eaj,Gertov:2019yqo,Ayyar:2019exp,
Cacciapaglia:2019ixa,BuarqueFranzosi:2019eee,Cacciapaglia:2019dsq,Cacciapaglia:2020vyf,Appelquist:2020bqj,
Dong:2020eqy,Cacciapaglia:2021uqh,Banerjee:2022izw,Ferretti:2022mpy, Cai:2022zqu,Appelquist:2022qgl, Cacciapaglia:2024wdn,Banerjee:2024zvg}, the bottom-up
 holographic models in
Refs.~\cite{Contino:2003ve,Agashe:2004rs,Agashe:2005dk,Agashe:2006at,
Contino:2006qr,Falkowski:2008fz,Contino:2010rs,Contino:2011np,Erdmenger:2020lvq,Erdmenger:2020flu,Elander:2020nyd,Elander:2021bmt,Elander:2023aow,Erdmenger:2023hkl,Elander:2024lir,Erdmenger:2024dxf}, and the top-down holographic theories in 
 Refs.~\cite{Imoto:2009bf,Elander:2021kxk}. 
In all of these models,
the new strong dynamics is coupled with SM fields and interactions in such a way as to
trigger electroweak symmetry breaking 
via a mechanism referred to as
vacuum misalignment, in juxtaposition  to the classical work on vacuum alignment~\cite{Das:1967it,Peskin:1980gc,Preskill:1980mz}.

The signature of CHMs and TPC is the appearance of new particles, belonging to towers of  bound states, with masses at the electroweak scale and above. Lattice field theory is the natural instrument to study the properties of such states, emerging in strongly coupled theories, and gain predictive power to guide experimental searches.
A number of investigations of candidate completions of CHMs
exist, for gauge theories with $SU(2)$ group and  matter fields transforming in the fundamental representation~\cite{Hietanen:2014xca,Detmold:2014kba,
Arthur:2016dir,Arthur:2016ozw,Pica:2016zst,Lee:2017uvl,Drach:2017btk,Drach:2020wux,
Drach:2021uhl,Bowes:2023ihh}, with $SU(4)$ group and  fermion content in an admixture of fundamental and 2-index antisymmetric representations~\cite{DeGrand:2015lna,DeGrand:2016htl,Ayyar:2017qdf,Ayyar:2018zuk,Ayyar:2018ppa,Ayyar:2018glg, Cossu:2019hse, Lupo:2021nzv,DelDebbio:2022qgu,Hasenfratz:2023sqa}, with $SU(2)$ 
group and fermions transforming as an admixture of fundamental and adjoint representations~\cite{Bergner:2020mwl, Bergner:2021ivi}.
The TELOS collaboration has developed an extensive program of lattice studies of the $Sp(2N)$ gauge theories with 
fermion matter fields transforming in an admixture of fundamental and 2-index antisymmetric representations~\cite{Bennett:2017kga,
  Bennett:2019jzz, Bennett:2019cxd, Bennett:2020hqd, Bennett:2020qtj,
  Bennett:2022yfa, Bennett:2022gdz, Bennett:2022ftz, Bennett:2023wjw,
  Bennett:2023gbe, Bennett:2023mhh, Bennett:2023qwx, Bennett:2024cqv,
  Bennett:2024wda, Bennett:2024tex},  measuring  masses and decay constants of bound states, topological observables, and spectral densities---see also Refs.~\cite{Hong:2017suj,Kulkarni:2022bvh,Bennett:2023rsl, Dengler:2024maq,Bennett:2024bhy}.

The theory of interest in this paper has $Sp(4)$ gauge group, $N_{\rm f}=2$  Dirac fermions (hyperquarks) transforming in the fundamental representation, and $N_{\rm as}=3$ in the 2-index antisymmetric representation. 
We occasionally refer to the two species of fermions as type-$({\rm f})$ and type-$({\rm as})$, respectively.
These choices yield the minimal theory 
that is amenable to lattice calculations and
can be used to combine the CHM and TPC paradigms~\cite{Barnard:2013zea,Ferretti:2013kya}. At low energies, the global symmetry-breaking pattern in the type-$({\rm f})$ fermion sector is described by the $SU(4)/Sp(4)$ coset, and the SM electroweak gauge symmetries can be embedded so that the five PNGBs are reinterpreted in terms of the Higgs doublet supplemented by a real singlet.
Among the bound states are fermions, made of two type-$({\rm f})$ and one type-$({\rm as})$ hyperquarks, called chimera baryons.
As the coset describing symmetry breaking in the  type-$({\rm as})$ fermion sector is $SU(6)/SO(6)$, it is possible to embed the $SU(3)$ gauge symmetry of the Standard Model in the unbroken $SO(6)$, so that  some of the chimera baryons have the same quantum numbers as the top quark, and hence they can act as top partners in the TPC mechanism. Of the two additional $U(1)$ factors acting on the two types of fermions, one linear combination is anomalous, while the complementary one is broken explicitly only by the hyperquark masses---the phenomenology of the associated, unflavored PNGBs is of interest in itself~\cite{Cacciapaglia:2019bqz} (see also Refs.~\cite{DeGrand:2016pgq,Belyaev:2016ftv,Cai:2015bss,Belyaev:2015hgo,Cacciapaglia:2017iws} and~\cite{Bennett:2023rsl,Bennett:2024wda}), but will be further discussed elsewhere~\cite{Bennett:2025singlets}, while we focus only on flavored states in this paper.

Extensive lattice studies of this theory have been performed both in the quenched approximation~\cite{Bennett:2017kga,Bennett:2019cxd, Bennett:2023mhh, Bennett:2023qwx} or with dynamical treatment of  one of the fermion species~\cite{Bennett:2019jzz, Bennett:2024tex}. Work on the theory with dynamical fermion in the case of $N_{\rm f}=2$ and $N_{\rm as}=3$ is more recent~\cite{Bennett:2022yfa,Bennett:2024cqv,
  Bennett:2024wda}. In particular, Ref.~\cite{Bennett:2024cqv} considers a selection of lattice ensembles, with several fermion masses and finite  lattice coupling, and presents the spectrum of flavored mesons,  made of either type-$({\rm f})$ and  type-$({\rm as})$ fermions.
In this paper, we improve the statistics by enlarging the ensembles, and update the mass measurements for the flavored mesons.
As in Ref.~\cite{Bennett:2024cqv}, we work with Wilson fermions, and  generate the relevant ensembles with an admixture of Hybrid Monte Carlo (HMC)~\cite{Duane:1987de} and Rational Hybrid Monte Carlo (RHMC) algorithms~\cite{Clark:2006fx}, performed within the 
Grid software suite~\cite{Boyle:2015tjk,Boyle:2016lbp,Yamaguchi:2022feu}, with the adaptations needed to implement symplectic groups~\cite{Bennett:2023gbe}.
Observables are computed from gauge configurations using the HiRep code~\cite{DelDebbio:2008zf,HiRepSUN,HiRepSpN}.
The (flavored meson) correlation functions of interest are computed by applying Wuppertal smearing ~\cite{Gusken:1989qx,Roberts:2012tp,Alexandrou:1990dq} supported by APE smeared~\cite{APE:1987ehd,Falcioni:1984ei} gauge links.
The spectroscopic analysis uses both a variational analysis based on the generalized eigenvalue problem (GEVP) algorithm~\cite{Blossier:2009kd}, and an implementation of 
the HLT spectral density algorithm that uses Gaussian as well as Cauchy kernels. We compare these approaches  for consistency,
and to estimate the methodology systematics.

There are two major elements of novelty 
to the analysis presented in this paper, besides the aforementioned improved statistics.
First, we apply a combination of variational analysis and HLT algorithms to the spectroscopy of chimera baryons. We obtain the first measurements 
of the masses of such composite fermions in both spin-$1/2$ and spin-$3/2$ channels, for both even- and odd-parity projection, and for 
composite states transforming as the $5$ and $10$ representations of the global, unbroken $Sp(4)$ symmetry group.
Second, we make the first non-trivial steps towards implementing 
a genuine off-shell treatment of the correlation functions, by computing vacuum matrix elements for meson as well as chimera baryon operators, extracting decay constants and overlap factors,  by combining the use of smeared and
 point-like sources.\footnote{For this work, we do not use the alternative approach, based on local, unsmeared operators and 
stochastic wall sources~\cite{Boyle:2008rh}, but see Appendix~\ref{sec:Wall}.} We renormalize the matrix elements and overlap factors by matching them at the 1-loop order~\cite{Martinelli:1982mw,Lepage:1992xa}. We critically discuss our results and compare them to the 
literature on other theories, to assess both the validity of our approach, and the viability of this specific theory as a new physics candidate.
Our results are currently at a single lattice spacing, but the continuum limit---deferred to future work---will enable more definitive statements about the viability of the theory with a  view to partial compositeness.

Spectral densities can also be used to test non-trivial properties of interacting field theories.
For example, the off-shell information they encode could be used in relation to the
Weinberg sum rules, and  the properties of spectral functions~\cite{Weinberg:1967kj,Dash:1967fq,Bernard:1975cd}.
While motivated by a specific candidate for new physics beyond the Standard Model, our methods—particularly the reconstruction of spectral densities (Sect.~\ref{sec:fitting_procedure})—are broadly applicable to strongly coupled gauge theories, including but not limited to QCD. These techniques offer valuable tools for model building and phenomenology across a wide class of strongly interacting theories,
and our results demonstrate the effectiveness of the methodology. 

The paper is organized as follows.
In Sect.~\ref{sec:lattice} we introduce the field theory of interest and its lattice discretization, discuss our choices and conventions, 
and describe in detail our ensemble generation process, including the checks we performed on thermalization,  topology,
and autocorrelation.  We present the procedures we adopt for the extraction of masses,  matrix elements, decay constants, and overlap factors in Sect.~\ref{sec:spect_decay_const}. The HLT algorithm used in our spectral density analysis  is presented in Sect.~\ref{sec:sp_dens},
together with some details about its numerical implementation.
The main numerical results are summarized in Sect.~\ref{sec:numerical_results}, both for mesons and baryons, by comparing spectra, matrix elements and overlap factors across all available channels.
We conclude with an outlook, in Sect.~\ref{sec:outlook}, outlining future research.
The paper is completed by four Appendices, which provide technical details on the smearing techniques (\ref{sec:Wuppertal_APE}),
a description of the renormalization of chimera-baryon overlap factors (\ref{sec:renormalisation_CB}), 
 a comprehensive tabulation  of  numerical results (\ref{sec:tables}), and
a comparison with matrix elements computed with stochastic wall sources (\ref{sec:Wall}).

\section{Lattice field theory} \label{sec:lattice}
In this section, we introduce the theory of interest, by defining its continuum and discretized actions. We provide a characterization of the ensembles used in the analysis, which have higher statistics with respect to those used for Ref.~\cite{Bennett:2024cqv}. We report the number of configurations as well as an updated and detailed discussion of the topological charge and autocorrelation times. 

\subsection{Continuum action and global symmetries}
\label{sec:continuum}

The $Sp(4)$ gauge theory is coupled to $N_{\rm f}=2$ Dirac fermions, $Q^I$,  transforming in the fundamental representation (with $I=1,\,2$), together with $N_{\rm as}=3$ Dirac fermions, $\Psi^k$, in the two-index antisymmetric representation (with $k=1,\,2,\,3$). The Lagrangian density, in the continuum theory and with flat Minkowski geometry, is given by
\begin{align}
    \label{eq:Lagrangian} 
    \mathcal{L} = -\frac{1}{2} \Tr\, G_{\mu\nu} G^{\mu\nu} + \sum_{I=1}^{N_{\rm f}} \bar Q^I \left( i\gamma^\mu {\cal D}^{({\rm f})}_\mu - m_{I}^{\rm f} \right) Q^I + \sum_{k=1}^{N_{\rm as}} \bar \Psi^k \left( i\gamma^\mu {\cal D}^{({\rm as})}_\mu - m_{k}^{\rm as} \right) \Psi^k,
\end{align} 
where  $G_{\mu\nu}$ is the field-strength tensor of $Sp(4)$: the trace is over gauge indices, which we leave implicit: ${\cal D}^{({\rm f})}_{\mu}$ and ${\cal D}^{({\rm as})}_{\mu}$ denote the covariant derivatives in the two fermion representations, respectively:  $\gamma^{\mu}$ are Dirac gamma matrices, while $I$ and $k$ are flavor indices. We assume mass-degenerate fermions, i.e. $m_{I}^{\rm f} = m^{\rm f}$ and $m_{k}^{\rm as} = m^{\rm as}$.

The fundamental representation is pseudo-real, while the antisymmetric representation is real. Thus, the Lagrangian has a global $SU(2N_{\rm f}) \times SU(2N_{\rm as}) \times U(1)_{\rm f} \times U(1)_{\rm as}$ symmetry.
The bilinear condensates of the fermions break $SU(2N_{\rm f}) \to Sp(2N_{\rm f})$ and $SU(2N_{\rm as}) \to SO(2N_{\rm as})$~\cite{Kosower:1984aw,Peskin:1980gc}. Concurrently, the mass terms explicitly break the symmetry, giving masses to the associated PNGBs. For $N_{\rm f}=2$ and $N_{\rm as}=3$, there are  $5 + 20$ PNGBs associated with the $(SU(4)\times SU(6))/(Sp(4)\times SO(6))$ coset.
One linear combination of the  $U(1)$ factors is broken by the axial anomaly, whereas the orthogonal $U(1)$ is expected to break spontaneously, producing an additional PNGB, the mass and composition of which is controlled by the mass terms in Eq.~(\ref{eq:Lagrangian}) and the non-perturbative dynamics. The non-anomalous combination has potentially interesting phenomenological implications~\cite{Belyaev:2015hgo,Cai:2015bss,Belyaev:2016ftv,Cacciapaglia:2017iws,Cacciapaglia:2019bqz}, but will not play a role in this paper, as its analysis requires implementing dedicated technology, which we will pursue elsewhere~\cite{Bennett:2025singlets}.

\subsection{Lattice action}
\label{sec:lattice_action}

We discretise the Wick-rotated, Euclidean action, on hyper-cubic lattices with $N_s$ sites in the spatial and $N_t > N_s$ sites in the temporal direction. We denote the lattice spacing as $a$, so that the total lattice volume is $V_4 = L^3 \times T = a^4 N_s^3 N_t$. We impose periodic boundary conditions for the gauge fields and the spatial boundaries of the fermion fields. We use anti-periodic boundary conditions for the temporal boundaries of the fermion fields.  
We adopt the standard Wilson plaquette action for the $Sp(2N)$ gauge fields, which we write~\cite{Bennett:2022yfa,Bennett:2023wjw,Bennett:2023gbe}, as follows:
\begin{equation} 
S_g = \beta \sum_x \sum_{\mu<\nu} \left(1 - \frac{1}{2N} {\rm Re}~ \Tr \left( U_\mu(x) U_\nu(x+\hat{\mu}) U_\mu^\dagger(x+\hat{\nu}) U_\nu^\dagger(x) \right) \right)\,.
\end{equation} 
The lattice coupling, $\beta$, is  related to the bare gauge coupling, $g$,  $\beta\equiv 4N/g^2$.
The gauge links, $U_\mu(x)$, are labelled by the directions on the lattice, $\mu,\, \nu$, with $\hat \mu, \,\hat \nu$ denoting unit vectors. 
Fermions are described by the  standard Wilson action (without a clover term)~\cite{Wilson:1974sk}:
\begin{align} 
\label{eq:lattice_fermion_action}
    S_f = a^4 \sum_{I=1}^{N_{\rm f}} \sum_{x,y} \overline{Q}^I(x) D^{\mathrm{(f)}}(x,y) Q^I(y) + a^4 \sum_{k=1}^{N_{\rm as}} \sum_{x,y} \overline{\Psi}^k(x) D^{\mathrm{(as)}}(x,y) \Psi^k(y)\,.
\end{align} 
Here,  $D^{\rm (f)}(x, y)$ and $D^{\rm (as)}(x, y)$ denote the lattice Dirac operators in the associated representations. Following the notation in Ref.~\cite{Bennett:2023gbe}, the Dirac operator in a given representation $\rm R$ is defined as 
\begin{equation} 
D^{({\rm R})}(x, y) = \left( \frac{4}{a} + m_0^{\rm R} \right) \delta(x, y) \nonumber \ - \frac{1}{2a} \sum_{\mu=1}^{4} 
\left\{ (1 - \gamma_\mu) U^{({\rm R})}_\mu(x) \delta(x+\hat{\mu}, y) + (1 + \gamma_\mu) U^{({\rm R})\dagger}_\mu(x) \delta(x-\hat{\mu}, y) \right\}\,,
\end{equation} 
with $m_0^{\rm R=f,\,as}$  the bare masses, and $U^{({\rm R=f,\,as})}_\mu(x)$ the link variables for each representation.

The effects of dynamical fermions are included in the ensemble generation via the Hybrid Monte Carlo (HMC) algorithm~\cite{Duane:1987de}   for the ${\rm (f)}$-type fermions and the Rational HMC (RHMC) algorithm~\cite{Clark:2006fx}  for the ${\rm (as)}$-type ones. Despite the odd number of ${\rm (as)}$ fermions, the determinant of  the Dirac operator is positive and real~\cite{Hands:2000ei} (see also Ref.~\cite{Bennett:2022yfa}). In summary, the action has three parameters: the inverse coupling, $\beta = 8/g^2$, and the fermion masses, $a m_0^{\rm f}$ and $a m_0^{\rm as}$. All the ensembles used in this work have the same  values for  $\beta = 6.5$ and  $a m_0^{\rm as} = -1.01$, while $a m_0^{\rm f}$ is varied between different ensembles. The choice of the coupling $\beta$ is at sufficiently weak coupling regime and thus free from the unwanted systematics associated with the bulk phase transition at strong coupling of the lattice theory. We refer to Ref.~\cite{Bennett:2022yfa} for further details. In addition, we vary the size of space and time directions, $N_s$ and $N_t$.

\subsection{Scale setting and topology}
\label{sec:technicalities}

We set the scale using the gradient flow~\cite{Luscher:2011bx,BMW:2012hcm,Luscher:2013vga}, and its lattice counterpart, the Wilson flow~\cite{Luscher:2010iy}. For this purpose, we define a new observable, $\mathcal W(t)$, as a function of gradient-flow time, $t$, as follows:
\beqs
  \label{eq:gradient_flow_definitions}
  \mathcal W(t) &\equiv  &\frac{\rm d }{\rm d \ln t}\left\{ t^2 \langle E(t) \rangle\right\}\,, 
  \eeqs
  where $\langle E(t)\rangle$ is the space and ensemble average of the quantity
  \beqs
  \label{eq:gradient_flow_energy_density}
  E(t,x) &\equiv &- \frac{1}{2} ~\Tr \, G_{\mu\nu}(t,x) G_{\mu\nu}(t,x)\,.
\eeqs
In this expression, the field-strength tensor in Euclidean space-time, $G_{\mu\nu}(t,x)$, is
 defined starting from a new gauge field, $A_\mu(t,x)$, living in five dimensions, which themselves 
are determined by identifying the four-dimensional
  gauge field as the five-dimensional one evaluated at vanishing flow time, $t=0$, and then evolving it by solving  the differential flow equation
\beqs \label{eq:gradient_flow_differential_equation}
    \frac{{\rm d} A_\mu(t, x)}{{\rm d} t}& =& D_\nu G_{\nu \mu} (t,x)\,,~~{\rm with}~~ A_\mu(t=0,x) = A_\mu(x)\,.
\eeqs

\begin{table}[t]
    \caption{Ensembles studied in this paper. For each ensemble, we list the lattice parameters, $\beta$, $am_0^{\rm f}$, $am_0^{\rm as}$, $N_t$, and $N_s$, as well as the number of thermalization trajectories we discarded, $N_{\rm therm}$, the number of trajectories between configurations we retained, $n_{\rm skip}$, and the number of remaining configurations, $N_{\rm conf}$,  while skipping every $n_{\rm skip}$ trajectories. The length for each trajectory is set to unit, $\tau = 1$. We also report the average plaquette, $\langle P \rangle$, and the Wilson flow scale, $w_0 / a$.  The integrated autocorrelation time is estimated in four ways, using four different reference observables: the average plaquette, $\tau^{P}_{\rm int}$,  the 2-point correlation function of the 2-index antisymmetric pseudoscalar ($\rm ps$) meson,  $\tau^{\rm ps}_{\rm int}$,  the energy density at  flow time $t = (w_0 / a)^2$,  $\tau^{w_0}_{\rm int}$, and the topological charge, $\tau^{Q}_{\rm int}$.  All results for the autocorrelation time are expressed in units of $n_{\rm skip}$.  The last column shows the average topological charge, $\bar{Q}$, obtained with a Gaussian fit of the distribution of topological charges of the configurations.
    \label{tab:ensembles}}
    \begin{tabular}{|c|c|c|c|c|c|c|c|c|c|c|c|c|c|c|c|}
\hline \hline
Label & $\beta$ & $am_0^{\rm f}$ & $am_0^{\rm as}$ & $N_t$ & $N_s$ & $N_{\rm therm}$ & $n_{\rm skip}$ & $N_{\rm conf}$ & $\langle P \rangle$ & $w_0 / a$ & $\tau_{\rm int}^{\langle P \rangle}$ & $\tau_{\rm int}^Q$ & $\tau_{\rm int}^{w_0}$ &  $\tau_{\rm int}^{\rm ps}$ & $\bar{Q}$ \\ 
\hline
M1 & 6.5 & -0.71 & -1.01 & 48 & 20 & 3006 & 14 & 873 & 0.5851679(88) & 2.5222(46) & 3.3(1.8) & 12.7(7.0) & 14.8(8.2) & 1.93(67) & 0.052(73) \\
M2 & 6.5 & -0.71 & -1.01 & 64 & 20 & 1000 & 28 & 737 & 0.5851719(84) & 2.5297(42) & 2.7(1.4) & 7.6(4.5) & 10.4(5.8) & 1.8(1.0) & 0.490(98) \\
M3 & 6.5 & -0.71 & -1.01 & 96 & 20 & 4000 & 26 & 1471 & 0.5851565(48) & 2.5245(25) & 2.12(65) & 6.9(1.6) & 6.8(2.0) & 1.55(45) & -0.541(80) \\
M4 & 6.5 & -0.7 & -1.01 & 64 & 20 & 1000 & 20 & 888 & 0.5842346(75) & 2.3672(35) & 2.32(89) & 10.2(4.3) & 10.3(4.2) & 1.97(57) & -0.05(11) \\
M5 & 6.5 & -0.72 & -1.01 & 64 & 32 & 3020 & 20 & 295 & 0.5860810(66) & 2.6936(31) & 1.90(87) & 11.6(7.3) & 7.0(3.3) & 2.1(1.2) & 0.80(24) \\
\hline \hline
\end{tabular}
\end{table}

The gradient flow scale, $w_0$, is defined by setting a convenient reference value, ${\mathcal W}_0$, and measuring the flow time it takes for the flow to reach it~\cite{Fodor:2012td}. We adopt the conventional choice  $\mathcal W(t)\vert_{t = w_0^2} = {\mathcal W}_0 \equiv 0.2815$~\cite{Bennett:2022ftz}. All dimensional quantities are expressed in units of $w_0$. Lattice measurements of $w_0/a$, obtained by discretizing Eq.~\eqref{eq:gradient_flow_differential_equation} using the clover operator, are shown in Tab.~\ref{tab:ensembles}, which also displays a complete characterization of the ensembles.

The topological charge, $Q$, of the continuum theory is defined as:
\beqs
    Q(t) &\equiv & \frac{1}{32\pi^2} \int {\rm d}^4x\, \epsilon^{\mu \nu \rho \sigma} \, \Tr \,G_{\mu\nu} (t,x)G_{\rho\sigma} (t,x)\,.
\eeqs
On the lattice, we follow the procedure described in details in Ref.~\cite{Bennett:2022ftz}, and compute $Q$ on gauge configurations smoothened  via the gradient flow, in order to suppress short-distance fluctuations. By inspection, we find that the distribution of $Q$ appears Gaussian in all the ensembles. We report in Tab.~\ref{tab:ensembles} the average value of the topological charge, $\bar{Q}$, obtained by a Gaussian fit over the ensemble distribution of $Q$.

\subsection{Ensemble updates and autocorrelation}
\label{sec:ensembles}

Gauge configurations are generated using the Grid software~\cite{Boyle:2015tjk, Boyle:2016lbp, Yamaguchi:2022feu}, supplemented by the $Sp(2N)$ dedicated package~\cite{Bennett:2023gbe}. For this work, we expand the statistics of the ensembles M1-M5 considered in Ref.~\cite{Bennett:2024cqv}. 
We adopt the (R)HMC algorithms to include the effect of the fermions. In each Monte-Carlo update (or trajectory) for the gauge configurations, the numerical integration of molecular dynamics equations (which include contributions from both the gauge bosons and the two types of fermions) is followed by a  Metropolis test to accept/reject the trajectories. 
For each ensemble, we  measure and report in Tab.~\ref{tab:ensembles} all the lattice parameters, as well as the average plaquette, $\langle P \rangle$, defined as
\begin{equation}
\langle P \rangle \equiv\frac{1}{6 N_t N_s^3}\sum_{x}\sum_{\mu>\nu}
{\rm Re}\, {\Tr}\, \frac{1}{2N} \left[\frac{}{} U_\mu (x) U_\nu (x+\hat{\mu}) U_\mu^\dagger(x+\hat{\nu}) U_\nu^\dagger(x) \frac{}{}\right]\,.    
\end{equation}

In order to assess  the degree of residual autocorrelation in our ensembles, we proceed to study the integrated autocorrelation time, $\tau_{\rm int}$, in each ensemble. We do so by exploiting four different observables, hence obtaining four estimates of $\tau_{\rm int}$. 
For each observable $X$, the integrated autocorrelation time, $\tau_{\rm int}^X$, is defined as~\cite{Madras:1988ei,Wolff:2003sm,Luscher:2004pav}:
\begin{align}
    \tau^{X}_{\rm int} &= \frac{1}{2} + \sum_{\tau=1}^{\tau_{\rm max}} \Gamma^{X}(\tau)\,, 
\end{align}
where $\Gamma^{X}(\tau)$ is the autocorrelation function of the observable $X$ 
    \beqs
    \Gamma^X(\tau) &=& \sum_{i=1}^{N-\tau} \frac{\left( X_{i} - \bar X \right)\left(  X_{i+\tau} - \bar X  \right)}{N-\tau}\,.
\eeqs
Here, $\tau=1,\,\dots,\,N$ is the Monte-Carlo time,  $X_i$
denote measurements of the observable in consideration, 
 and $\bar{X}$ their arithmetic mean.
 We skip $n_{\rm skip}$ Monte Carlo trajectories and retain $N_{\rm conf}$ configurations in each ensemble.
 We report in Tab.~\ref{tab:ensembles} the resulting four measurements of the autocorrelation, for all the ensembles, as well as our choices of $n_{\rm skip}$ and $N_{\rm conf}$. Although the four estimates of the autocorrelation $\tau_{\rm int}$ are computed on the remaining $N_{\textrm{conf}}$ configurations only, our results still present $\tau_{\rm int}>1$. Therefore, we conclude that the configurations used for this paper are still affected by a moderate amount of residual autocorrelation.

In particular, by inspecting the  autocorrelation time in the topological charge, $\tau_{\rm int}^Q$, and Wilson scale, $\tau_{\rm int}^{w_0}$,  shown in Tab.~\ref{tab:ensembles}, we observe the presence of some residual autocorrelation in these observables, a sign of mild topological freezing. For further discussion, we refer to Ref.~\cite{Bennett:2024cqv}.  We also discarded the first $N_{\rm therm}$ updates, to ensure thermalization, and checked that no residual thermalization is present, by partitioning the ensembles and repeating the measurement of observables on different portions of the ensembles, to find consistency among different partitions.

\begin{table}
    \centering
    \caption{Interpolating operators used for flavored mesons, classified by spin, $J$, parity, $P$, and irreducible representations under the action of the unbroken global $Sp(4)$ and $SO(6)$ symmetries, acting on $({\rm f})$-type and $({\rm as})$-type fermions, respectively. We follow the naming conventions of Ref.~\cite{Bennett:2024cqv}, except that we denote as $20^{\prime}$ the traceless symmetric  self-conjugate 2-index representation of $SO(6)\sim SU(4)$, to distinguish it from the $20$ and $20^{\prime\prime}$ of $SU(4)$. \label{tab:operators}
    \\}
    \begin{tabular}{ |c|c|c|c|c| }
        \hline\hline
        Label & Interpolating operator $\mathcal{O}$ & $J^P$ & $Sp(4)$ & $SO(6)$ \\
        \hline
        PS  & $\bar{Q}_1 \gamma_5 Q_2$ & $0^{-}$ & $5$ & $1$ \\
        V  & $\bar{Q}_1 \gamma_i Q_2$ & $1^{-}$ & $10$ & $1$  \\
        T  & $\bar{Q}_1 \gamma_0 \gamma_i Q_2$ & $1^{-}$  & $10$ & $1$\\ 
        AV  & $\bar{Q}_1 \gamma_5 \gamma_i Q_2$ & $1^{+}$ & $5$ & $1$  \\ 
        AT  & $\bar{Q}_1 \gamma_0 \gamma_5 \gamma_i Q_2$ & $1^{+}$ & $10$ & $1$  \\ 
        S  & $\bar{Q}_1 Q_2$ & $0^{+}$ & $5$ & $1$  \\
        \hline
        ps & $\bar{\Psi}^k \gamma_5 \Psi^{\ell}$ & $0^{-}$  & $1$ & $20^{\prime}$  \\
        v & $\bar{\Psi}^k \gamma_i \Psi^{\ell}$ & $1^{-}$& $1$ & $15$   \\
        t& $\bar{\Psi}^k \gamma_0 \gamma_i \Psi^{\ell}$ & $1^{-}$ & $1$ & $15$  \\ 
        av& $\bar{\Psi}^k \gamma_5 \gamma_i \Psi^{\ell}$ & $1^{+}$  & $1$ & $20^{\prime}$ \\
        at& $\bar{\Psi}^k \gamma_0 \gamma_5 \gamma_i \Psi^{\ell}$ & $1^{+}$ & $1$ & $15$  \\ 
        s & $\bar{\Psi}^k \Psi^{\ell}$ & $0^{+}$  & $1$ & $20^{\prime}$  \\
        \hline  \hline
    \end{tabular}
    \label{tab:operators}
\end{table}

\section{Mass spectra, matrix elements and overlap factors from correlation functions}
\label{sec:spect_decay_const}

In this section, we describe the interpolating operators used to construct the correlation functions from which we extract the spectrum and the matrix elements of mesons and (chimera) baryons, and the analysis we employ.
The meson and chimera baryon correlation functions were measured using the HiRep code~\cite{DelDebbio:2008zf,HiRepSUN}, extended to symplectic gauge groups~\cite{HiRepSpN}, after converting configurations, generated with Grid, using the GLU library~\cite{GLU}.

\subsection{Interpolating operators and correlation functions for mesons}
\label{sec:operators}

The interpolating operators of flavored mesons are listed in Tab.~\ref{tab:operators}, following the conventions of Ref.~\cite{Bennett:2024cqv}. They take the general form 
\begin{align} 
    \mathcal{O}_{\rm f} (x)  &= \bar Q_1(x) \Gamma Q_2(x)\,, \\
    \mathcal{O}_{\rm as} (x) &= \bar \Psi_1(x) \Gamma \Psi_2(x)\,,
    \label{eq:meson_ops}
\end{align}
where $\Gamma$ is a product of Dirac gamma matrices that selects a channel with given spin and parity, $J^P$. When using $\gamma_i$, the index takes values $i=1,2,3$. Starting from these localized operators, we construct spatially smeared ones, by applying Wuppertal smearing to the fermion fields. We also smoothen the gauge fields using APE smearing. We follow the smearing procedure outlined in Appendix~\ref{sec:Wuppertal_APE}, and we use the same choice of parameters as in Ref.~\cite{Bennett:2024cqv}: the APE smearing parameters are $\alpha_{\rm APE} = 0.4$ and $N_{\rm APE} = 50$, the Wuppertal smearing parameters are $\varepsilon_{\rm f} = 0.20$ and $\varepsilon_{\rm as} = 0.12$ for ensembles M1--M4, and $\varepsilon_{\rm f} = 0.24$ and $\varepsilon_{\rm as} = 0.12$ for ensembles M5. Different smeared operators, obtained with different number of smearing steps at sink and source, are retained in the variational analysis used in the measurements.

For general momentum, the correlation functions of interest are defined as:
\begin{equation}
    C_{AB}(t, \vec{p}) \equiv \frac{1}{N_s^3}\sum_{\vec x} e^{-i\vec{p} \cdot \vec{x}} \langle  \mathcal{O}_A(t,\vec x) \bar{\mathcal{O}}_B(0,0) \rangle \,, 
\end{equation}
and setting the spatial momentum, $\vec{p}$, to zero, we find the 
 two-point correlation functions of the interpolating operators
\begin{align}\label{eq:two-point-correlator}
    C_{AB}(t) \equiv \frac{1}{N_s^3}\sum_{\vec x} \langle  \mathcal{O}_A(t,\vec x) \bar{\mathcal{O}}_B(0,0) 
    \rangle\,.
\end{align}
From these, we can measure masses and matrix elements as follows.
Taking the lattice periodicity into account, at large Euclidean times the correlation function, for mesons with $\mathcal O_A = \mathcal O_B$, behaves  as follows,
\begin{align}\label{eq:large_t_meson}
    C_{\rm meson}(t) \approx A \left( e^{-mt} + e^{-m(T-t)}  \right)\,,
\end{align}
where $m$ is the energy of the ground state, and the normalization coefficient, $A$, contains the information of the matrix element of the interpolating operator between the meson ground state and the vacuum. Once the two-point correlation function is measured, we can then extract $m$ and $A$ by fitting the above functional form.

\subsection{Interpolating operators and correlation functions for chimera baryons: spin and parity projection}
\label{sec:chimera_projections}

\begin{table}
    \centering
    \caption{Quantum numbers of the chimera baryons studied for this paper. For each chimera baryon, we list the matrix appearing in the interpolating operator, $\Gamma_1$, appearing in Eq.~(\ref{eq:baryon_ops}), the spin, $J$, and the irreducible representations of the unbroken global $Sp(4)$ and $SO(6)$ symmetries, acting on $({\rm f})$-type and $({\rm as})$-type fermions, respectively. 
    \label{tab:operators_baryons} \\    }
    \begin{tabular}{ |c|c|c|c|c| }
        \hline\hline
        $~~~$baryon$~~~$ & $~~~\Gamma_1~~~$ & $~~~J~~~$ & $~~~Sp(4)~~~$ & $~~~SO(6)~~~$ \\
        \hline
        $\Lambda_{\rm CB}$     & $\gamma_5$ & $1/2$ & $5$  & $6$   \\
        $\Sigma_{\rm CB}$      & $\gamma_i$ & $1/2$ & $10$ & $6$  \\
        $\Sigma_{\rm CB}^\ast$ & $\gamma_i$ & $3/2$ & $10$ & $6$ \\ 
        \hline  \hline
    \end{tabular}
\end{table}

Following the notations and conventions of Ref.~\cite{Bennett:2022yfa},  we write the  chimera baryon operators in the general form:
\begin{align} \label{eq:baryon_ops}
    \mathcal O_{\rm CB, \alpha}^{\Gamma_1 \Gamma_2}(x) = \left( Q_1^{a,\beta_1}(x) \left(  \mathcal{C} \, \Gamma_1\right)_{\beta_1 \beta_2} Q_2^{b,\beta_2}(x) \right) \Omega^{ad} \Omega^{bc} \Gamma_{2, \alpha \beta_3} \Psi^{cd,\beta_3}(x)\,,
\end{align}
where $\Omega$ is the symplectic matrix and $\mathcal{C}$ is the charge-conjugation matrix. Latin letters are used for color indices, and Greek letters for spinor indices. While $\Gamma_{1,2}$ could, in principle, denote any product of gamma matrices, in this paper
we set $\Gamma_2 = \mathbbm{1}$, while  $\Gamma_1 = \gamma_5$ or $\Gamma_1 = \gamma_i$. Hence, throughout this paper,  we denote the baryon operators as $\mathcal O_{\rm CB, \alpha}^{ \Gamma}$, with $\Gamma=\Gamma_1$---we occasionally omit an index, to indicate that the relations we write apply to all operators that differ only by that index.

We expect the chimera baryon states to have definite parity; hence we project the correlator,
$C_{\rm {CB}, \alpha \beta}(t) \equiv \langle \mathcal{O}_{\rm CB, \alpha} (t) \mathcal{O}_{\rm CB, \beta} (0)  \rangle$,  defined in analogy with Eq.~(\ref{eq:two-point-correlator}),
 to its parity-even and parity-odd components using the projectors $P_\pm = (1 \pm \gamma_0)/2$:
\begin{equation}
\label{eq:parity_projection}
    C_{\rm CB, \alpha \beta}^\pm(t) \equiv P_{\pm , \alpha \alpha_1} C_{\rm CB , \alpha_1 \beta}(t)\,.
\end{equation}
At large Euclidean times, on a lattice with anti-periodic boundary conditions for the fermions in the time direction, the projected correlation functions for chimera baryons behave  as
\begin{align}\label{eq:large_t_baryon}
    C_{\rm CB}^\pm(t) \approx (A_\pm e^{-m^\pm t} - A_\mp e^{-m^\mp(T-t)}) P_{\pm}\,,
\end{align}
where we understand spinorial indices, $A_\pm$ are  proportional to the vacuum-to-hadron matrix elements, for the parity-even and parity-odd state, respectively, and $m_\pm$ are the ground state masses of even and odd states.

Since the baryon operators defined with $\Gamma_1 = \gamma_i$ source both spin-$1/2$ and spin-$3/2$ states, we decompose the correlation functions involving such baryon operators by projecting them onto components with definite spin. Starting from  the zero-momentum correlation function
\begin{align}
    C^{ij}_{\rm CB , \alpha \beta}(t) \equiv& \frac{1}{N_s^3}\sum_{\vec x} \langle  \mathcal{O}_{\rm CB , \alpha}^{\gamma_i}(t,\vec x) \bar{\mathcal{O}}_{\rm CB ,  \beta}^{\gamma_j}(0,0) \rangle\,,
\end{align}
we project to states with a definite spin quantum number with the following definitions:
\begin{align}
    C_{\rm CB, \alpha \beta}^{(1/2)}(t) \equiv& P_{ij}^{(1/2)} C^{ij}_{\rm CB, \alpha \beta}(t) = \frac{1}{3} \gamma_i \gamma_j C^{ij}_{\rm CB , \alpha \beta}(t)\,, \\ 
    C_{\rm CB , \alpha \beta}^{(3/2)}(t) \equiv& P_{ij}^{(3/2)} C^{ij}_{\rm CB, \alpha \beta}(t) = \left(\delta_{ij} - \frac{1}{3} \gamma_i \gamma_j \right) C^{ij}_{\rm CB , \alpha \beta}(t)\,.
\end{align}
We report  the quantum numbers of the three operators of interest  in Tab.~\ref{tab:operators_baryons}. By analogy with hadrons in QCD, we denote the state sourced with $\Gamma_1=\gamma_5$ as $\Lambda_{\rm CB}$, the spin-$1/2$ projection of the state interpolated by $\Gamma_1=\gamma_i$ as $\Sigma_{\rm CB}$, and its spin-$3/2$ partner as $\Sigma_{\rm CB}^\ast$.

\subsection{Spectroscopy: Generalized Eigenvalue Problem}
\label{sec:GEVP_corr}

For the numerical analysis, after applying APE smearing to the configurations, we build a basis of interpolating operators by (Wuppertal) smearing those given in Eqs.~\eqref{eq:meson_ops}-\eqref{eq:baryon_ops}, and  perform a variational analysis of all channels under investigation, aimed at
optimizing the signal of the lowest-lying states, while also  gaining access to excited states.
Following the procedure applied for mesonic operators in Ref.~\cite{Bennett:2024cqv}, we construct three different smeared operators, $\{\mathcal O_i\}$,  for each meson and chimera baryon operator, by choosing $N_{\rm source}, \, N_{\rm sink} = 0, \, 40,\, 80$. This process generates a variational basis of correlation functions, so that the correlation matrix $\mathcal{C}(t)$ has nine elements, which we denote as $C_{N_{\rm source}, \, N_{\rm sink}}(t)$, with $N_{\rm source}, \, N_{\rm sink} \in \{0,\,40,\,80\}$. In the special case of the $J^P = 1^-$ mesons, the same tower of states is sourced by two of the meson operators listed in 
Tab.~\ref{tab:operators}, hence we further use the cross-channels interpolating operators $\rm V$ and $\rm T$, for $({\rm f})$-type fermions, and $\rm v$ and $\rm t$, for $({\rm as})$-type ones, widening the correlation matrix to $36$ elements.

For any given  correlation matrix of interest, generically denoted as  $ C(t)$, we extract the energy levels from the eigenvalues, $\lambda_n(t,t_0)$, which solve the Generalized Eigenvalue Problem (GEVP) defined as:
\begin{align}
     C(t) v_n(t,t_0) = \lambda_n(t,t_0) C(t_0) v_n(t,t_0)\,,
\end{align}
where $v_n$ are the GEVP eigenvectors. A discussion of how systematic effects depend on the energy gap, $\Delta E_n = \min_{m \neq n} |aE_m - aE_n|$,  proportionally to terms $\mathcal O\left(e^{-\Delta E_n t_0}\right)$ 
can be found in Ref.~\cite{Blossier:2009kd}. Our choice is $t_0 = 1$.

We can then extract the energy levels by performing a fit to the eigenvalues at large Euclidean time. The fit functions for the mesons and (parity-projected) baryons are given by Eqs.~\eqref{eq:large_t_meson} and~\eqref{eq:large_t_baryon}, respectively. In order to determine the fit interval for the chimera baryons, it is convenient to introduce the difference of eigenvalues
\begin{align}
    \tilde  \lambda_{\rm CB}^\pm(t) =  \frac{1}{2} \left[\lambda_{\rm CB}^\pm(t) -  \lambda_{\rm CB}^\mp(T-t) \right]\,,
\end{align}
which selects only the parity-even or parity-odd contributions in Eq.~\eqref{eq:large_t_baryon}. We then examine the effective mass, $m_{\rm eff}(t)$, defined implicitly by the relation 
\begin{align}
    \frac{\lambda(t-1)}{\lambda(t)} = \frac{e^{- m_{\rm eff}(t) \cdot (T-t+1)} \pm e^{- m_{\rm eff}(t) \cdot (t-1)}}{e^{- m_{\rm eff}(t) \cdot (T-t)} \pm e^{- m_{\rm eff}(t) \cdot t}}\,,
\end{align}
where the relative sign of the exponential terms is chosen to take the lattice periodicity of the eigenvalues into account. We identify the fitting range of the correlation function by the range of $t$ for which the effective mass displays an approximately  constant behavior (plateau).

\subsection{Matrix elements and overlap factors}
\label{sec:f_p_corr}

For the mesons and chimera baryon states that are accessible to our spectroscopic measurements, it is also possible to 
provide estimates of the corresponding hadron-to-vacuum matrix elements. They are extracted from
 the coefficients in front of the exponential terms in the correlation functions, in Eqs.~\eqref{eq:large_t_meson} and~\eqref{eq:large_t_baryon}. For mesons, we  write\footnote{We denote by $\left| e_{M,n}\right\rangle$ a complete set of energy eigenstates  associated with the two-point correlation function of interest, with given meson channels labelled by $M$. It may be helpful  to highlight that  $\left| e_{M,0} \right\rangle \neq |0\rangle$, as the 
 former is the lowest-lying state, and the latter is the vacuum.}
\begin{align}\label{eq:correlation_matrix}
    C^{{M}}_{N_{\rm source}, \, N_{\rm sink}}(t)  = \sum_n \frac{1}{2E_n} \langle 0 \vert \mathcal{O}_{N_{\rm source},} \vert e_{M,n} \rangle \langle e_{M,n} \vert  \bar{\mathcal{O}}_{N_{\rm sink}} \vert 0 \rangle \left[e^{-E_n t} + e^{-E_n (T-t)}\right].
\end{align}
For all of the meson operators listed in Tab.~\ref{tab:operators},  one can define the decay constants of the particles they source in terms of the matrix elements of local operators. We restrict our attention to pseudoscalar, vector and axial-vector channels, for which the numerical signal is under good control, and use  the following definitions:\footnote{The normalizations are chosen so that, when applied to 2-flavor QCD, one finds $f_{\rm PS}=f_{\pi} \simeq 93$ MeV. In the definition of $f_{\rm PS}$, in practice, we set $\mu = 0$ and $p_{\mu = 0} = m_{\rm PS}$.} 
\begin{align}
    \label{eq:fpi} 
    \langle 0 \vert \bar{Q}_1 \gamma_5 \gamma_\mu Q_2 \vert {\rm PS} \rangle &= \sqrt{2} f_{\rm PS} p_\mu 
    & \langle 0 \vert \bar{\Psi}_1 \gamma_5 \gamma_\mu \Psi_2 \vert {\rm ps} \rangle &= \sqrt{2} f_{\rm ps} p_\mu \\
    \label{eq:fV} 
    \langle 0 \vert \bar{Q}_1 \gamma_\mu Q_2 \vert {\rm V} \rangle &= \sqrt{2} f_{\rm V} m_{\rm V} \epsilon_\mu,
    & \langle 0 \vert \bar{\Psi}_1 \gamma_\mu \Psi_2 \vert {\rm v} \rangle &= \sqrt{2} f_{\rm v} m_{\rm v} \epsilon_\mu, \\
    \label{eq:fAV} 
    \langle 0 \vert \bar{Q}_1 \gamma_5 \gamma_\mu Q_2 \vert {\rm AV} \rangle &= \sqrt{2} f_{\rm AV} m_{\rm AV} \epsilon_\mu,
    & \langle 0 \vert \bar{\Psi}_1 \gamma_5 \gamma_\mu \Psi_2 \vert {\rm av} \rangle &= \sqrt{2} f_{\rm av} m_{\rm av} \epsilon_\mu.
\end{align}
In these expressions, $\epsilon_\mu$ is the polarization vector, transverse to the momentum, $p_\mu$, so that  $p_\mu \epsilon^\mu =0$ and normalized so that  $\epsilon^\ast_\mu \epsilon^\mu =1$. In contrast to what is done for the mass determination,  in the process of measuring matrix elements we do not carry out a variational analysis with multiple smearing levels. Rather, we perform a simultaneous fit based on Eq.~\eqref{eq:correlation_matrix}, applied to the following selected subset of matrix elements:
\begin{equation} \label{eq:decay_const_corr}
    C_{80, \, 80} (t) \equiv \langle \mathcal{O}_{80}(t) \, \bar{\mathcal{O}}_{80}(t) \rangle\,, \quad \hbox{and} \quad C_{80, \, 0} (t) \equiv \langle \mathcal{O}_{80}(t) \,\bar{\mathcal{O}}_0(t) \rangle\,,
\end{equation}
which contain both smeared and local operators. From the result, we
 extract the local matrix elements in Eqs.~\eqref{eq:fpi}-\eqref{eq:fAV}.

The decay constants renormalize multiplicatively, and hence, following Refs.~\cite{Martinelli:1982mw, Lepage:1992xa}, we define
\begin{align}
\label{eq:ren_const_meson}
    f_{\rm PS}^{\rm ren} =&  Z^{\rm f}_A f_{\rm PS}\,, & f_{\rm ps}^{\rm ren} =&  Z^{\rm as}_A f_{\rm ps}\,, \\
    f_{\rm V}^{\rm ren}  =&  Z^{\rm f}_V f_{\rm V}\,,  & f_{\rm v}^{\rm ren} =&  Z^{\rm as}_V f_{\rm v}\,, \\
    f_{\rm AV}^{\rm ren} =&  Z^{\rm f}_A f_{\rm AV}\,, & f_{\rm av}^{\rm ren} =&  Z^{\rm as}_A f_{\rm av}\,.
\end{align}
The  renormalization coefficients, $Z_M^{\rm R}$, depend on the fermion representation, $\rm R$. We determine them by a 1-loop calculation in lattice perturbation theory, and matching it with the calculation in the $\overline{\textrm{MS}}$ scheme from the vertex renormalization, consistently with earlier work reported in Ref.~\cite{Bennett:2019jzz,Bennett:2024tex}. The resulting coefficients can be written as follows~\cite{Martinelli:1982mw}: 
\begin{align}
    &Z_{\rm A}^{\rm R} = 1 + \frac{C^R g^2}{16 \pi^2 \langle P \rangle} \left( \Delta_{\Sigma_1} +  \Delta_{\rm AV} \right)\,,
    &Z_{\rm V}^{\rm R} = 1 + \frac{C^R g^2}{16 \pi^2 \langle P \rangle} \left( \Delta_{\Sigma_1} +  \Delta_{\rm V} \right)\,,
\end{align}
where $\langle P \rangle$ is the average plaquette, that appears in these expressions to implement tadpole improvement of the gauge coupling~\cite{Lepage:1992xa}, $C^R$ is the quadratic Casimir, with $C^{\rm f}=5/4$ and $C^{\rm as}=2$, and the numerical coefficients are $\Delta_{\Sigma_1}=-12.82$, $\Delta_{\rm AV}=-3.0$, and $\Delta_{\rm V}=-7.75$.

The matrix elements of the (chimera) baryon operators are defined in analogy to the baryons in QCD---see, e.g., Ref.~\cite{Erben:2022tdu} and references therein. We write the operators in the form of Eq.~\eqref{eq:baryon_ops}, and define the overlap factor, $K_B$, for each chimera baryon, $B$,  by using local (unsmeared) interpolating operators. In particular, we are interested in the following matrix elements: 
\begin{align}
    \label{eq:overlap_factors} 
    \langle 0 \vert \mathcal{O}^{\gamma_5, \mathbb{1}}_{\rm CB, \alpha} \vert {\Lambda} (\vec{p},s) \rangle &\equiv K_{\Lambda}\, u_{s, \alpha} (\vec{p})\,,
    & \langle 0 \vert \mathcal{O}^{\gamma_i, \mathbb{1}}_{\rm CB, \alpha} \vert {\Sigma} (\vec{p},s) \rangle &\equiv K_{\Sigma}\, u_{s, \alpha} (\vec{p})\,,
\end{align}
where \( u_s (\vec{p})\) is an on-shell Dirac spinor with momentum $\vec{p}$ and spin projection $s$ associated to the chimera baryon states with defined parities, \( |{\Lambda}(\vec{p},s) \rangle \) and \( |{\Sigma} (\vec{p},s) \rangle \), respectively. The overlap factors quantifies the coupling strength between interpolating operators and physical  states.

We write the baryon correlation function at finite momentum as follows~\cite{Erben:2022tdu}:
\begin{align}\label{eq:correlation_matrix_baryonic}
    C_{\rm{CB}, \alpha \beta}(t, \vec{p}) = \dfrac{1}{N^3_s} \sum_{\vec{x}} e^{-i \vec{p} \cdot \vec{x} } \, \langle  \mathcal{O}_{\rm{CB}, \alpha}(t, \vec{x}) \bar{\mathcal{O}}_{\rm{CB}, \beta}(0, 0) \rangle = \sum_n \sum_s \dfrac{|K_{B,n}|^2 u_{s, \alpha} \bar{u}_{s, \beta}}{2E_n(\vec{p})}  e^{-E_n (\vec{p}) \, t}\,,
\end{align}
where we show explicitly the dependence on the momentum and the 
 summations over spin projection labels, $s$, and energy levels, $n$. In the  zero-momentum limit and for single particle states, $E_n\rightarrow m_n$ is the mass, 
 and the sum over the spinors is $\sum_s u_s\bar{u}_s= \pm 2m_n P_{\pm}$.
The overlap factors, associated with the spin-projected operators and correlation functions described in Sect.~\ref{sec:chimera_projections}, can be found by using the fitting form:
\begin{align}\label{eq:correlation_matrix_baryonic}
    C^{\pm}_{\rm{CB}}(t) = \pm \sum_n |K_{B,n}|^2  e^{-E_n t} P_{\pm}\,.
\end{align}
The numerical results for the overlap factors are extracted by simultaneous fits of correlators containing smeared and local operators, following the meson case described in Eq.~\eqref{eq:decay_const_corr}.

 In analogy to the mesons, to account for multiplicative renormalization of baryon operators in Tab.~\ref{tab:operators_baryons}, we define the renormalized overlap factors via the following relations:
\begin{align}
    \label{eq:ren_const_baryons}
    K_{\Lambda}^{\rm ren} &= Z_{\rm{CB}, \gamma_5}\, |K_{\Lambda}|\,,
    & K_{\Sigma}^{\rm ren} &= Z_{\rm{CB}, \gamma_i} \, |K_{\Sigma}|\,,
\end{align}
in which we identify the overlap factors of interest with the ground state elements,  $K_{\Lambda}=K_{\Lambda,0}$ and $K_{\Sigma}=K_{\Sigma,0}$.\footnote{The examination of the overlap factors and matrix elements for the excited states is also an interesting development, but it requires high statistics and therefore it is deferred to future investigations.} 
The detailed derivation of the renormalization factors, \( Z_{\rm{CB}, \gamma_5} \) and \( Z_{\rm{CB}, \gamma_i} \), is provided in Appendix~\ref{sec:renormalisation_CB}. They are given by:
\begin{align}
\label{eq:renorm_const_baryons}
    Z_{\rm{CB}, \gamma_5} &= 1 + \frac{g^2}{16 \pi^2 \langle P \rangle} \left[\left(C^{\rm f} + \frac{1}{2} C^{\rm as}\right) \Delta_{\Sigma_1} + \Delta_{{\rm CB}}[\gamma_5]\right], \\
    Z_{\rm{CB}, \gamma_i} &= 1 + \frac{g^2}{16 \pi^2 \langle P \rangle} \left[\left(C^{\rm f} + \frac{1}{2} C^{\rm as}\right) \Delta_{\Sigma_1} + \Delta_{{\rm CB}}[\gamma_i]\right],
\end{align}
where $\langle P \rangle$, $C^{\rm f}$,  $C^{\rm as}$, and $\Delta_{\Sigma_1}$ are the same constants appearing in the renormalization of the  mesons, while the numerical values of the vertex corrections are determined for the first time in this work and at the renormalization scale $\mu = 1/a$ in the $\overline{\rm{MS}}$ scheme are the following:
\begin{equation}
\Delta_{\rm CB}[ \gamma_5] = -26.67, \quad\quad \Delta_{\rm CB}[\gamma_i] = 18.12\,.
\end{equation}

\section{Spectral densities}
\label{sec:sp_dens}
In this section, we discuss the spectral density reconstruction algorithm, and apply it to the extraction of masses and matrix elements of composite states. This work expands on the study presented in Ref.~\cite{Bennett:2024cqv}, which focused on the mesons and their masses, by including the (chimera) baryon spectrum and matrix elements. As we exemplify the algorithm by performing measurements on the same theory as in Ref.~\cite{Bennett:2024cqv}, but with enhanced statistics, our results represent also an update for the masses of the mesons. We have implemented our method in the Python package \texttt{LSDensities}, publicly available in Ref.~\cite{Forzano:2024}. We devote special attention to the presentation  of our estimates of the systematic uncertainties in the reconstruction process and to the discussion of the parameter choices.

\subsection{The Hansen-Lupo-Tantalo method} \label{sec:HLT_method}

The spectral density, $\rho(E)$, of a generic two-point correlation function, $C(t)$, is defined by: 
\begin{equation} \label{eq:inverse_laplace}
C(t) \equiv \int_{E_{\rm min}}^{\infty} dE \, \rho(E) \, b(t,E)\,,
\end{equation}
where 
\begin{equation}
\label{eq:basis_functions}
b(t, E) \equiv b_+(E) e^{-t \,E} + b_-(E) e^{-(T - t)E}\,.
\end{equation}
Eq.~\eqref{eq:basis_functions} is a generalization of the Laplace transform to a finite temporal extent, while
 $E_{\rm min}$ can be chosen between zero and the ground state energy for the channel under consideration.

Lattice data is provided as measurements, $C_k(t)$, labeled by $k = 1,\, \dots,\, N_m$ (with $N_m \leq N_{\textrm{conf}}$ bounded by the number of configurations in the ensembles listed in Tab.~\ref{tab:ensembles}), where $t$ is discrete and $t \leq t_{\rm max}\leq a N_t/2$.
Reconstructing spectral densities from the average of a finite number of noisy measurements is an ill-posed problem. Yet, several methods have been proposed to reliably compute spectral densities, among which we adopt the HLT method, which  is based on a modification of the Backus-Gilbert algorithm, tailored for lattice simulations~\cite{Backus:1968svk, Hansen:2019idp}.

The HLT method requires to introduce the smeared spectral density, $\rho_\sigma (\omega)$, defined  as: 
\begin{equation}
\label{eq:smeared_sp_dens}
\rho_\sigma (\omega) \equiv \int_{E_{\rm min}}^{\infty} \, dE \, \Delta_\sigma (E - \omega) \, \rho(E)\,,
\end{equation}
where $\Delta_\sigma (E - \omega)$ is a smearing kernel with finite smearing radius, $\sigma$. In a finite volume, $\rho(E)$ becomes a sum of Dirac $\delta$ functions, peaked across the discrete eigenvalues of the Hamiltonian of the physical system considered; the chosen kernels, $\Delta_\sigma$, are designed to smoothen the $\delta$ functions, but  approach $\delta$ functions in the limit $\sigma\rightarrow 0$. Hence, using  Eq.~\eqref{eq:smeared_sp_dens}, the physical spectral density can be obtained as a limiting case, $\sigma \to 0$.

Given a choice of smearing kernel, one approximates it as a sum of $t_{\rm max}$ basis functions, hence introducing the reconstructed kernel, $\bar{\Delta}_\sigma$, defined as
\begin{equation}
\label{eq:basis_expansion}
    \bar{\Delta}_\sigma(E-\omega) \equiv \sum_{t=a}^{t_{\rm max}} g_t(\omega, \sigma) \,  b(t,E)\,.
\end{equation}
The coefficients are determined by minimizing the following functional: 
\begin{equation}
\label{eq:A_functional}
A[\vec{g}] \equiv \int_{E_{\rm min}}^{\infty} dE \,  e^{\alpha E} \, \left| \bar{\Delta}_\sigma (E - \omega) - \Delta_{\sigma} (E - \omega) \right|^2\,,
\end{equation}
which measures the difference between the reconstructed kernel, $\bar{\Delta}_\sigma (E - \omega)$, and the target one, $\Delta_\sigma (E - \omega)$. The strength of this representation is that it is exact for $t_{\rm max} \to \infty$, where the minimum of $A[\vec{g}]$ is zero.
The unphysical parameter, $\alpha$, characterizes different choices of norm in the functional space~\cite{DelDebbio:2022qgu}: changing its value and ensuring that the results are compatible provides a useful cross-check. 
\begin{center}
   \begin{figure}
    \includegraphics[width=0.5\linewidth]{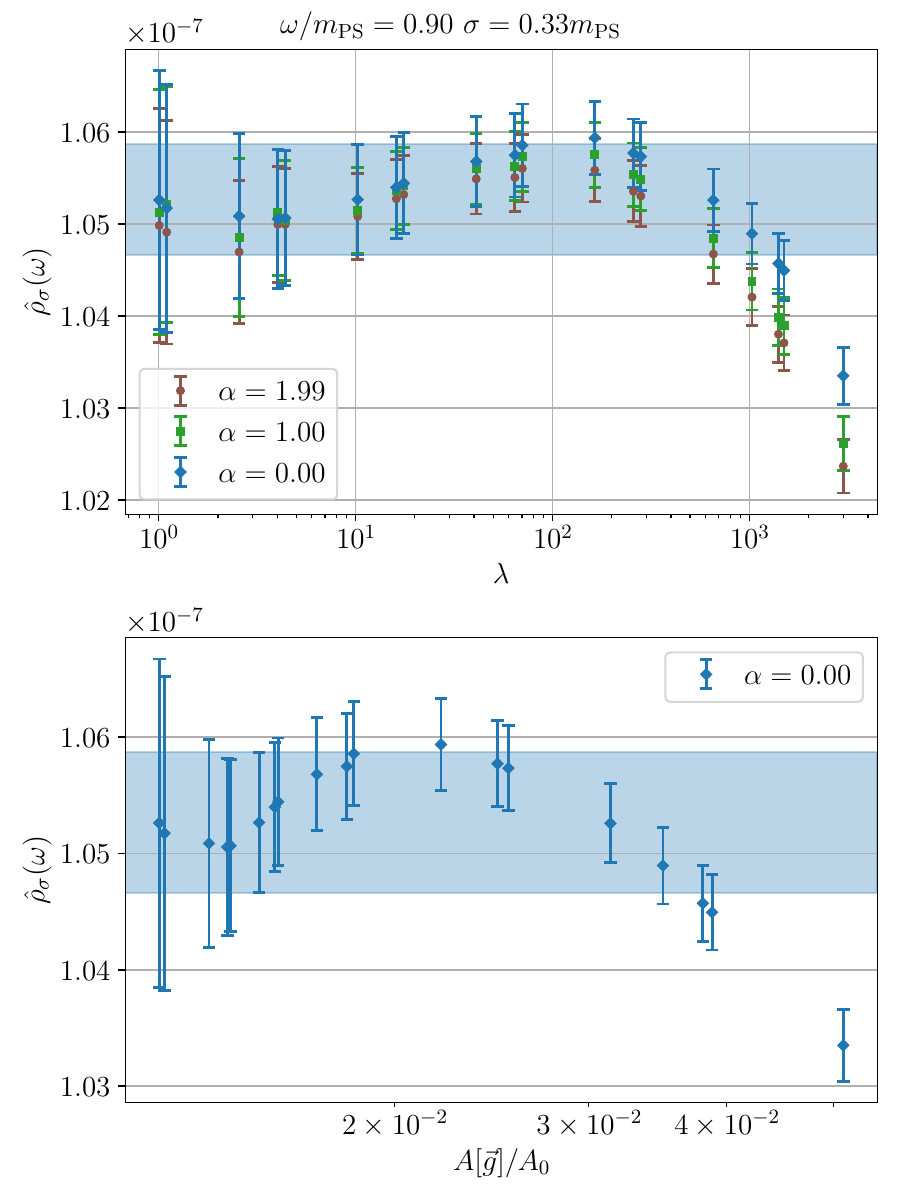}
    \caption{Illustrative examples of the plateaux in the spectral density reconstructed with the HLT method, for a fixed value of the energy, $\omega$. The underlying data correspond to the correlation function of pseudoscalar mesons made of $({\rm f})$-type fermions,  
    in ensemble M1 in Tab.~\ref{tab:ensembles}, using $t_{\rm max} =a  N_t / 2$,  $\sigma = 0.33 a m_{\mathrm{PS}}$. 
In the top panel, the reconstructed spectral density, $\hat{\rho}_{\sigma}(\omega)$, is shown as a function of the trade-off parameter, $\lambda$, for three choices of  $\alpha = 0, \, 1.00, \, 1.99$. The horizontal band is our best estimate, its width representing the statistical error. All estimates obtained for asymptotically small choices of $\lambda$ are compatible within statistical errors, but affected by larger uncertainties, which decrease with larger $\lambda$, until the discrepancy exceeds the statistical uncertainty, for large enough $\lambda$.
The bottom panel shows the same results, but restricted to one value of $\alpha$, and plotted as a function of 
$A[\vec{g}] /(A_0 = A[\vec{g} = \vec{0}])$, obtained by minimizing the functional   $W[\vec{g}]$, for the same 
selection of values of $\lambda$ as in the top panel. Again, the plot shows that for small values of 
$\lambda$ the reconstructed density is independent of the correspondingly small value of $A[\vec{g}] /(A_0)$, 
but discrepancies larger than the statistical uncertainty are visible for large $\lambda$, in which 
case also $A[\vec{g}] /A_0$ is large.
    \label{fig:plateauscan}}
     \end{figure}
\end{center}

If the two-point correlation functions were known with infinite precision, minimising Eq.~\eqref{eq:A_functional} would be sufficient to solve the inverse problem exactly.
 However, solving the inverse problem requires dealing with a highly ill-conditioned system. For this reason, even minor uncertainties or errors in the input correlation functions will severely destabilize the numerical solution, posing a significant obstacle to the calculation. A solution to this problem was proposed in Ref.~\cite{Backus:1968svk}, and a regularization of this process can be built by introducing an additional functional: 
\begin{equation} \label{eq:B_functional}
    B[\vec{g}] \equiv \sum_{t, \, t^{\prime} = a}^{t_{\rm max}} \, g_{t} \, \hbox{Cov}_{t t^{\prime}} [C] \, g_{t^{\prime}}\,,
\end{equation}
derived from the covariance matrix associated with the correlation functions, $C(t)$, defined as
\begin{equation}
\label{Eq:cov}
\hbox{Cov}_{t t{\prime}} [C] \equiv \dfrac{1}{N_m} \sum_{k = 1}^{N_m } \, \Big( C_k (t) - \langle C(t) \rangle \Big) \Big( C_k (t^{\prime}) - \langle C(t^{\prime}) \rangle \Big)\,,
\end{equation}
where $\langle C(t) \rangle$ is the average over the measurements at given $t$. We then minimize the following combination of functionals:
\begin{equation} \label{eq:W_functional} 
    W[\vec{g}] = \frac{A[\vec{g}]}{A[0]} + \lambda \frac{B[\vec{g}]}{B_{\rm norm}(\omega)}\,,
\end{equation} 
where $B_{\rm norm}(\omega) \equiv C^2(t=a) / \omega^2$ is introduced a dimensionless term. Having a non-zero value for $\lambda$ introduces a systematic error in our reconstruction. This effect has to be carefully taken into account in order to provide a reconstruction with reliable uncertainties. For each energy $\omega$, minimizing $W[\vec{g}]$ yields a set of coefficients that are used to construct the following estimator for the smeared spectral density
\begin{equation}
\label{Eq:hat}
\hat{\rho}_{\sigma}(\omega) = \sum_{t = a}^{t_{\rm max}} g_{t} (\omega, \sigma) \, C(t)\,.
\end{equation}

In order to remove the effect of working with a non-zero value of lambda, we adopt the procedure described in Refs.~\cite{Bulava:2021fre, DelDebbio:2024lwm}, where the value of $\lambda$ is chosen in such a way that its effect is absorbed within the statistical fluctuations. This procedure has been shown~\cite{Bulava:2021fre, DelDebbio:2024lwm} to be able to successfully account for systematics effect due to the introduction of the $B[\vec{g}]$ functional. We illustrate this procedure in Fig.~\ref{fig:plateauscan}, which shows $\hat{\rho}_\sigma(\omega)$ for a representative value of $\omega$ as a function of 
$\lambda$, at three values of $\alpha$, obtained by minimizing $W[\vec{g}]$. The figure also displays the relation between $\hat{\rho}_\sigma(\omega)$ and
$A[\vec{g}] / A_0$, both evaluated at the minimum of $W$, for each choice of  $\lambda$ and $\alpha$. At larger values of $\lambda$, we find that the systematic errors exceed the statistical fluctuations\footnote{An indicative measure of the size of systematic effects in the reconstruction is given by the ratio $\sqrt{A[\vec{g}] / A[0]}$, which captures the deviation between the target, $\Delta_\sigma (E - \omega)$, and the reconstructed approximation, $\bar{\Delta}_{\sigma} (E - \omega)$,
in the presence of a finite truncation of the basis, with $t_{\rm max}/a$ elements,  and  finite statistics encoded in  $N_{\rm conf}$. On the other hand, the size of  statistical errors is related to $B[\vec{g}] / B_{\rm norm}(\omega)$, which depends on the covariance matrix of the input lattice data. }: this is signalled by the fact that values for the reconstructions at different values of $\lambda$ are not compatible with each other.

On the other hand, reducing $\lambda$ decreases $A[\vec{g}] / A_0$ and the systematic effects related to the reconstruction algorithm, yet, by doing so one increases the statistical uncertainties. This is expected, as a smaller value of 
$\lambda$ prioritizes the minimization of $A[\vec{g}] / A_0$, reducing systematic errors at the expense of constraining $B[\vec{g}] / B_{\rm norm}$ less effectively.

When the quality of the data permits, one identifies a region in $\hat{\rho}_\sigma(\omega)$---a plateau---emerging at low vales of $\lambda$, indicating a window in which systematic effects are controlled, and  the signal remains meaningful, before eventually deteriorating into noise at extremely small $\lambda$. In the plateau region, the size of statistical and systematic effects are comparable. Figure~\ref{fig:plateauscan} illustrates this plateau-seeking analysis, showing that values exist for which the dependence on unphysical parameters is minimized, while reconstruction remains robust against statistical noise.

While the plateau analysis should ensure that the error is purely statistical, as an additional precaution we evaluate the following term: 
\begin{equation} 
    \sigma_{1,\rm sys}(E) \equiv |\rho_{\lambda^\star}(E) - \rho_{\lambda^\star/10}(E)|\,,
\end{equation} 
where $\lambda^\star$ is the value of the trade-off parameter $\lambda$ determined according to the plateau analysis. This term is introduced to check for a residual dependence on $\lambda$, and take it as an estimate for a possible residual systematic error. Moreover, as shown in Fig.~\ref{fig:plateauscan}, we repeat the calculation for different values of $\alpha$, and again we interpret variations in the output, captured by the term:
\begin{equation} 
    \sigma_{2,\rm sys}(E) \equiv |\rho_{\lambda^\star, \alpha}(E) - \rho_{\lambda^\star, \alpha'}(E)| \,,
\end{equation} 
as a residual systematic error. We denote as $\alpha$ and $\alpha^{\prime}$ any representative values of the range of $\alpha$ used in the analysis. The two aforementioned terms, however, are negligible in our analysis, since our estimates are statistically dominated: this is a healthy sign that the plateau analysis is working as intended.
The total error is given by the sum in quadrature of these two errors and the statistical one, which is found by bootstrapping.

 The choice of the general form of the smearing kernel entering into 
Eq.~\eqref{eq:smeared_sp_dens} is dictated by  convergence properties (and convenience of use). We adopt two alternative functional forms as smearing kernels:  a Gaussian kernel, parameterized as 
\begin{equation} \label{eq:Gaussian_kernel}
    \Delta^{(1)}_\sigma(E - \omega) \equiv \frac{1}{Z(\omega)} \exp\left[-\frac{(E-\omega)^2}{2\sigma^2}\right] \,,
\end{equation} 
where $Z(\omega) \equiv \int_0^\infty dE \,  \exp\left[-(E-\omega)^2/(2\sigma^2)\right]$, and a Cauchy-type kernel, that we write as
\begin{equation} \label{eq:Cauchy_kernel}
    \Delta^{(2)}_{\sigma} (E - \omega) \equiv \frac{\sigma}{ (E - \omega)^2 + \sigma^2}\,.
\end{equation} 
In the numerical analysis, we measure the smearing radius,  $\sigma$, in units of the mass of the ground state meson or chimera baryon, $m_0$, choosing it to lie in the range $0.18m_0 \leq \sigma \leq 0.35m_0$. The  smaller values are used  when attempting to resolve closely spaced energy levels. We inspect our intermediate results to ensure that at the minima of $W[\vec{g}]$ the quantity $A[\vec{g}]/A_0 < 0.1$, so that we can be reasonably confident about our estimates of the systematic uncertainties.

\subsection{Fitting procedure and systematic uncertainties}
\label{sec:fitting_procedure}
In this subsection, we explain how smeared spectral densities can be used to extract physical quantities, such as the meson and (chimera) baryon masses, matrix elements and overlap factors. The numerical results are presented in Sec.~\ref{sec:numerical_results}.
The starting point is the definition of the correlated $\chi^2$ functional~\cite{DelDebbio:2022qgu}:
\begin{equation}
\label{eq:correlated_chisquare}
    \chi^2 \equiv \sum_{E, \, E^{\prime}} \left( f^{(k)}_\sigma (E) - \hat{\rho}_\sigma (E) \right) \text{Cov}^{-1}_{E\,E^{\prime}} [\rho_\sigma] \left( f^{(k)}_\sigma (E^{\prime}) - \hat{\rho}_\sigma (E^{\prime}) \right)\,,
\end{equation}
where $\text{Cov}_{E,E^{\prime}}$ is the covariance matrix in energy space for the smeared spectral densities, and the fitting function $f^{(k)}_\sigma(E)$ is a weighted sum of (either Gaussian or Cauchy) kernels:
\begin{equation}
\label{eq:sum_of_kernels}
    f^{(k)}_\sigma (E) = \sum_{n = 0}^{k-1} \mathcal{A}_n \, \, \Delta^{(i)}_\sigma (E - E_n)\,.
\end{equation}

The fit parameters, $E_n$, are identified with the eigenvalues of the finite-volume Hamiltonian, linking the position of the resolved fitted peaks in the spectral densities to the mass spectroscopy of interest. The number of states, $k$, is determined \emph{a posteriori}; given the finite smearing radius and the deterioration of reconstruction at high energies, only a limited number of states in the spectrum are accessible. In addition, fit results may also be contaminated by additional states not included in the analysis. To address both considerations, we iteratively repeat the fit for different choices of $k$, by adding one extra state, and attempt the reconstruction of $k+1$ peaks in Eq.~(\ref{eq:sum_of_kernels}). As long as the target states remain stable and the $\chi^2$ per degree of freedom does not deteriorate significantly, we consider the results reliable. Otherwise, we stop the process having determined the first $k$ peaks.

The use of both Gaussian and Cauchy smearing kernels adds an additional consistency check, and an estimate of the systematic effects related to this choice for the determination of the $n^{th}$ energy level can be given:
\begin{equation} \label{eq:fitting_systematics} 
    \sigma_{\rm sys}(aE_n) = |aE_{n, {\rm Gauss}} - aE_{n, {\rm Cauchy}}|\,. 
\end{equation}
In Sec.~\ref{sec:numerical_results}, we will discuss numerical evidence that any such discrepancy 
 is smaller than the statistical errors, confirming that,
despite qualitative and quantitative differences in the smeared spectral shapes due to the choice of the kernel, $\Delta_\sigma$, 
the peak positions are statistically consistent~\cite{Bennett:2024cqv}.

The amplitude of the fit functions, $f_{\sigma}^{(k)}(E)$ in
Eq.~\eqref{eq:sum_of_kernels} contains the matrix elements of the interpolating meson or  baryon operator. Hence, 
Eq.~\eqref{eq:sum_of_kernels} can be used to simultaneously  determine such matrix elements, and we devote the rest of this subsection to outline the procedure we apply to extract these quantities from spectral density fits. 

\begin{center}
   \begin{figure}
    \includegraphics[width=0.90\linewidth]{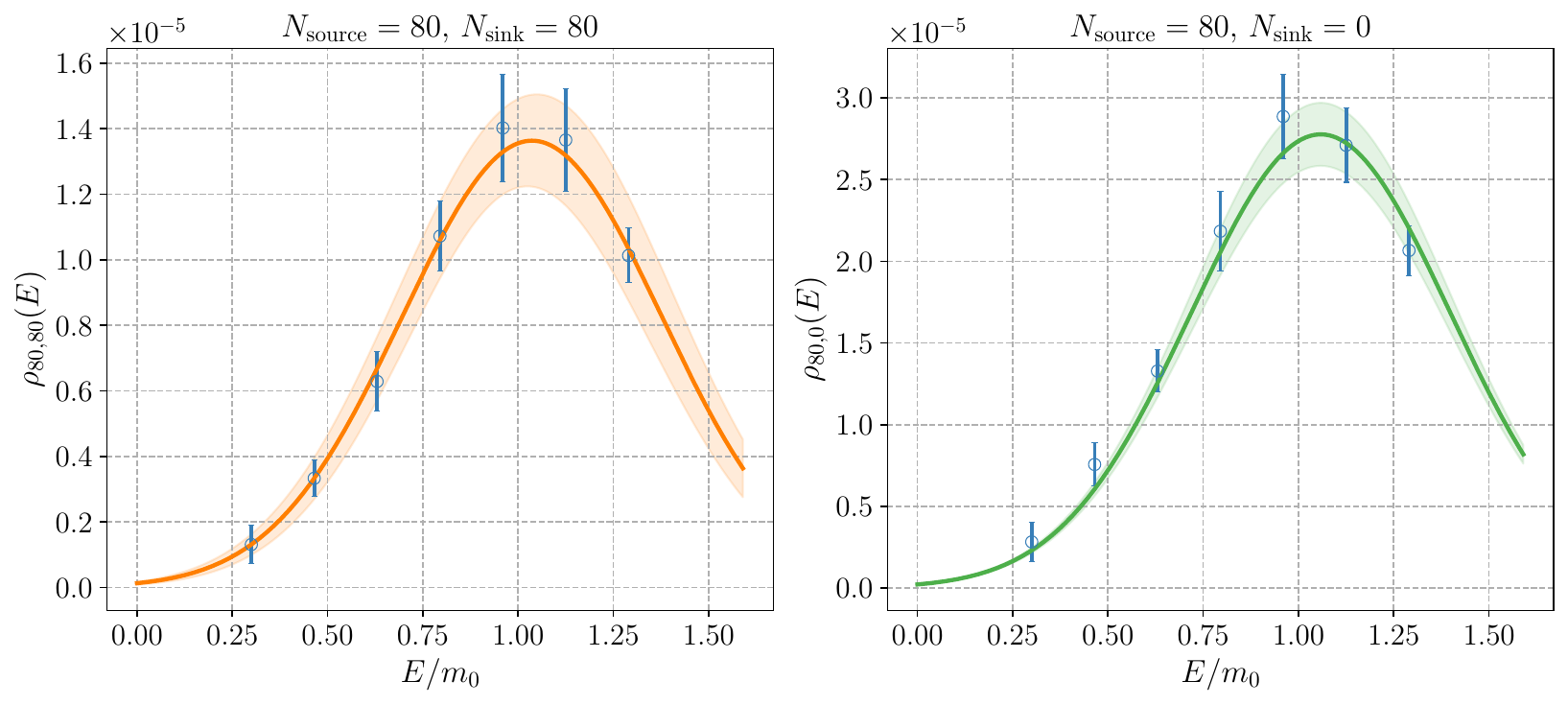}
    \caption{Examples of the result of a simultaneous fit of the spectral densities, involving different smearing levels (Tab.~\ref{table:E1_matrix_mesons}), defined in Eqs.~(\ref{eq:sp_dens_N80_N80}) and~(\ref{eq:sp_dens_N80_N0}). The two spectral densities and the best fit functions are shown as a function of the energy, $E$, normalized to the mass of the ground state, $m_0$, in the relevant channel. The height of the fitting bands is our estimate of the uncertainty, inclusive of both statistical and systematic components, summed in quadrature.  In these examples, we are measuring the correlation functions involving  vector meson, ${\rm V}$, composed of  $({\rm f})$-type fermions. The underlying numerical data is taken from  ensemble M1 in Tab.~\ref{tab:ensembles}. These measurements use a Gaussian kernel with $\sigma / m_0 = 0.33$.  The two analysis differ by the smearing level of the sink, $N_{\rm sink}=80$ (left panel) and $N_{\rm sink}=0$ (right panel). 
    \label{fig:simultaneous_fitting_ME}}
     \end{figure}
\end{center}

For each operator of interest, we measure the correlation functions corresponding to different levels of Wuppertal smearing of the source and sink, $N_{\rm source}= 80$ and  $N_{\rm sink} = (0,\,80)$, which we  denote as  $C_{80, \, 80} (t)$ and $C_{80, \, 0} (t)$---see Eq.~\eqref{eq:decay_const_corr}. We then estimate the two corresponding spectral densities, expressed in the functional forms for mesons:
\begin{equation}
\label{eq:sp_dens_N80_N80}
    \rho_{\sigma, 80, \, 80} (E) = \sum_{n = 0}^{k-1} \dfrac{|\langle 0 | \mathcal{O}_{80}(t=0) | e_{M, n} \rangle|^2 }{2E_n} \, \, \Delta_\sigma (E - E_n)\,,
\end{equation}
 and
 \begin{equation}
\label{eq:sp_dens_N80_N0}
    \rho_{\sigma, 80, \, 0} (E) = \sum_{n = 0}^{k-1} \dfrac{\langle 0 | \mathcal{O}_{80}(0) | e_{M, n} \rangle \langle e_{M, n} |  \bar{\mathcal{O}}_{0}(0) | 0 \rangle }{2E_n} \, \, \Delta_\sigma (E - E_n)\,.
\end{equation}
Similarly,  for chimera baryons we measure 
\begin{equation}
\label{eq:sp_dens_N80_N80_baryons}
    \rho_{\sigma, 80, \, 80} (E) = \sum_{n = 0}^{k-1} |K_{B, n, 80}|^2 \, \, \Delta_\sigma (E - E_n)\,,
\end{equation}
 and
 \begin{equation}
\label{eq:sp_dens_N80_N0_baryons}
    \rho_{\sigma, 80, \, 0} (E) = \sum_{n = 0}^{k-1} K_{B,n, 80} K_{B,n, 0}^{\dagger} \, \, \Delta_\sigma (E - E_n)\,,
\end{equation}
by performing a simultaneous fit of the two spectral densities, accounting for correlations. We hence extrapolate the unsmeared matrix elements and overlap factors corresponding to the transition between the vacuum and the $n^{th}$ state of the interpolating operator.   An example of such fits is reported in Fig.~\ref{fig:simultaneous_fitting_ME}, for the vector  meson operator, ${\rm V}$, made of
 $({\rm f})$-type fermions, obtained using the numerical data in ensemble M1 in Tab.~\ref{tab:ensembles}.

While the physical results should not depend on our choice of  smearing kernel, $\Delta_\sigma$, we use the difference between the results extracted using Gaussian and Cauchy fits as a further cross-check for our results:
\begin{equation} \label{eq:fitting_systematics_ME} 
    \sigma_{\rm sys}(\langle 0 |  \mathcal{O}(0) | e_{M, n} \rangle) \equiv \left|\langle 0 |  \mathcal{O}(0) | e_{M, n} \rangle_{{\rm Gauss}} - \langle 0 |  \mathcal{O}(0) | e_{M, n} \rangle_{{\rm Cauchy}}\right|\,,
\end{equation}
for mesons, while for chimera baryons,
\begin{equation} \label{eq:fitting_systematics_ME_baryons} 
    \sigma_{\rm sys}(K_n) \equiv \left| K_{n, {\rm Gauss}} - K_{n, {\rm Cauchy}}\right|\,. 
\end{equation}
We show in Sec.~\ref{sec:numerical_results} that this is negligibly small, in comparison with the statistical uncertainties of the matrix elements and overlap factors.

     \begin{figure}[t]
    \includegraphics[width=0.55\linewidth]{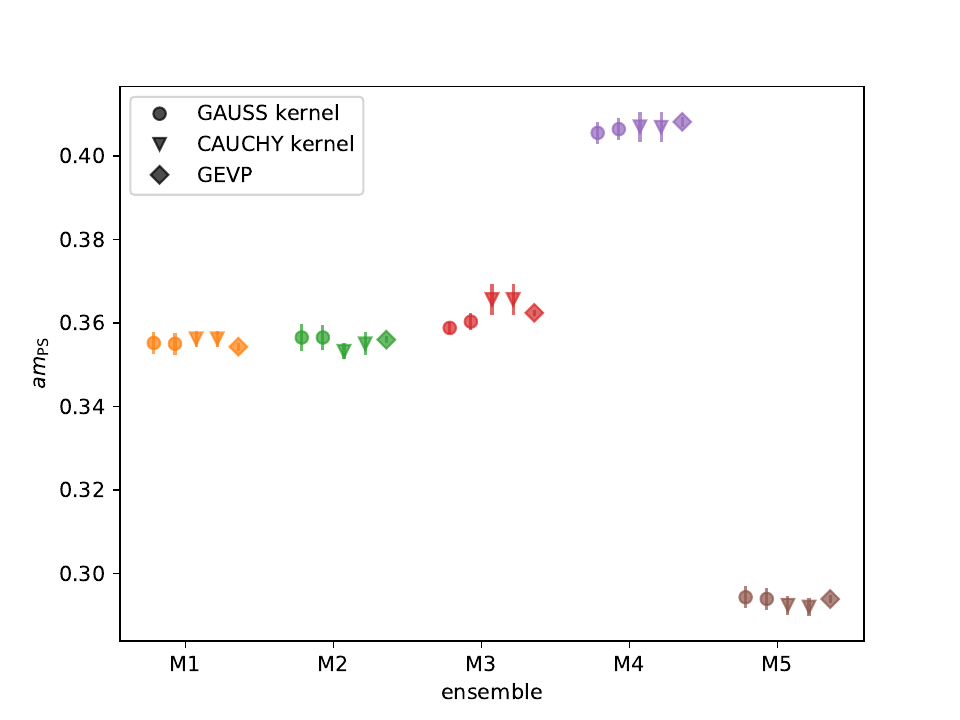}
      \includegraphics[width=0.55\linewidth]{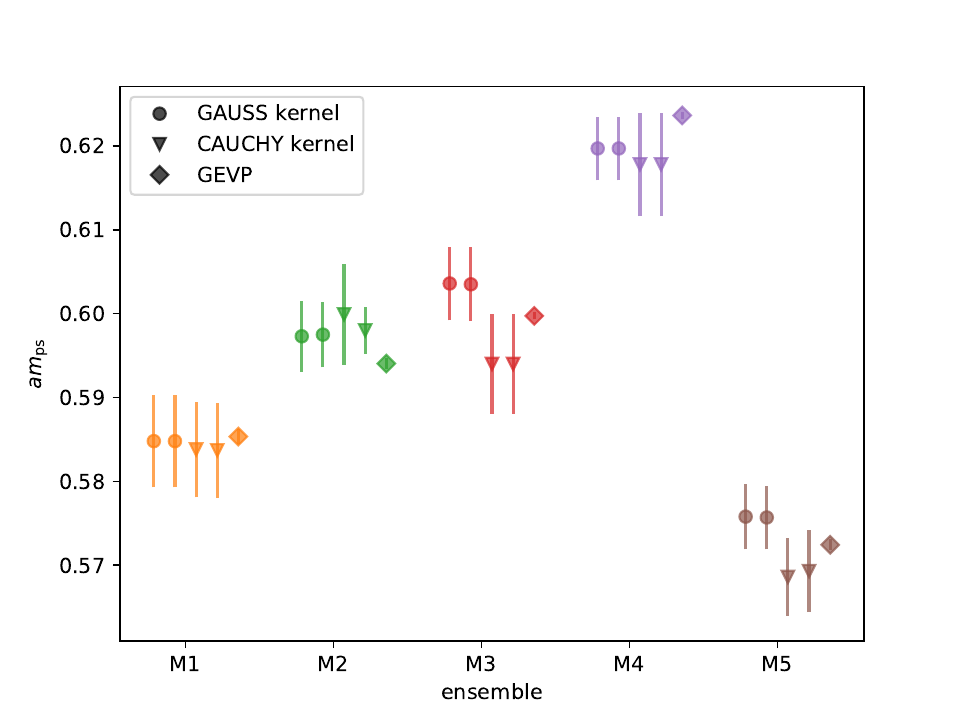}
       \includegraphics[width=0.55\linewidth]{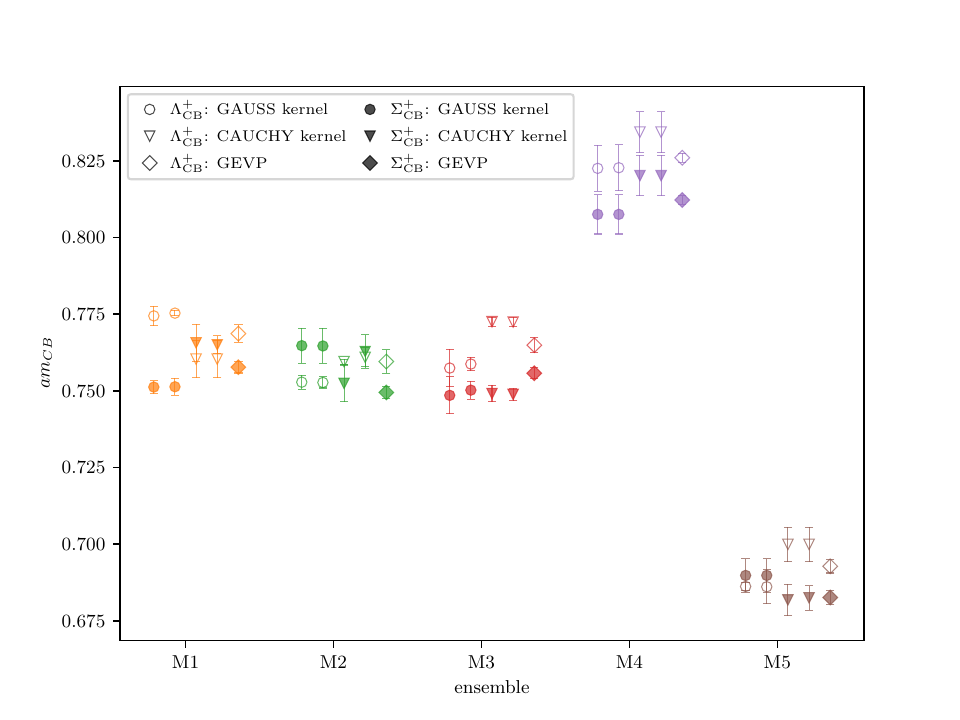}
    \caption{ 
    Representative examples of mass measurements in the $Sp(4)$ theory with $N_{\rm f}=2$ and $N_{\rm as}=3$ hyperquarks, for the lightest flavored mesons composed of $({\rm f})$-type (${\rm PS}$ mesons, top panel) and  $({\rm as})$-type (${\rm ps}$ mesons, 
middle panel) fermions, as well as the lightest chimera 
    baryons ($\Lambda^{+}_{\rm CB}$ and $\Sigma^{+}_{\rm CB}$, bottom panel), in all available ensembles, as summarised in Tab.~\ref{tab:ensembles}. 
    The masses are expressed in lattice units, and the uncertainties displayed in these plots include only the statistical component.
    The five measurements of each bound-state mass (horizontally offset for presentation purposes) are obtained with five different methodologies: the result of the conventional GEVP analysis of (APE and Wuppertal smeared) correlation functions based on the variational method is compared to those obtained with four different choices of smearing kernel, in the HLT reconstruction of the spectral densities---for more details, including all the other mass measurements performed, see Tabs.~\ref{table:E1_results_ground_mesons}--\ref{table:E5_results_second_mesons} for mesons, and 
    Tabs.~\ref{table:E1_results_ground_CB}--\ref{table:E5_results_second_CB} for chimera baryons.
    \label{fig:error comparison}}
     \end{figure}

     \begin{figure}[t]
    \includegraphics[width=0.9\linewidth]{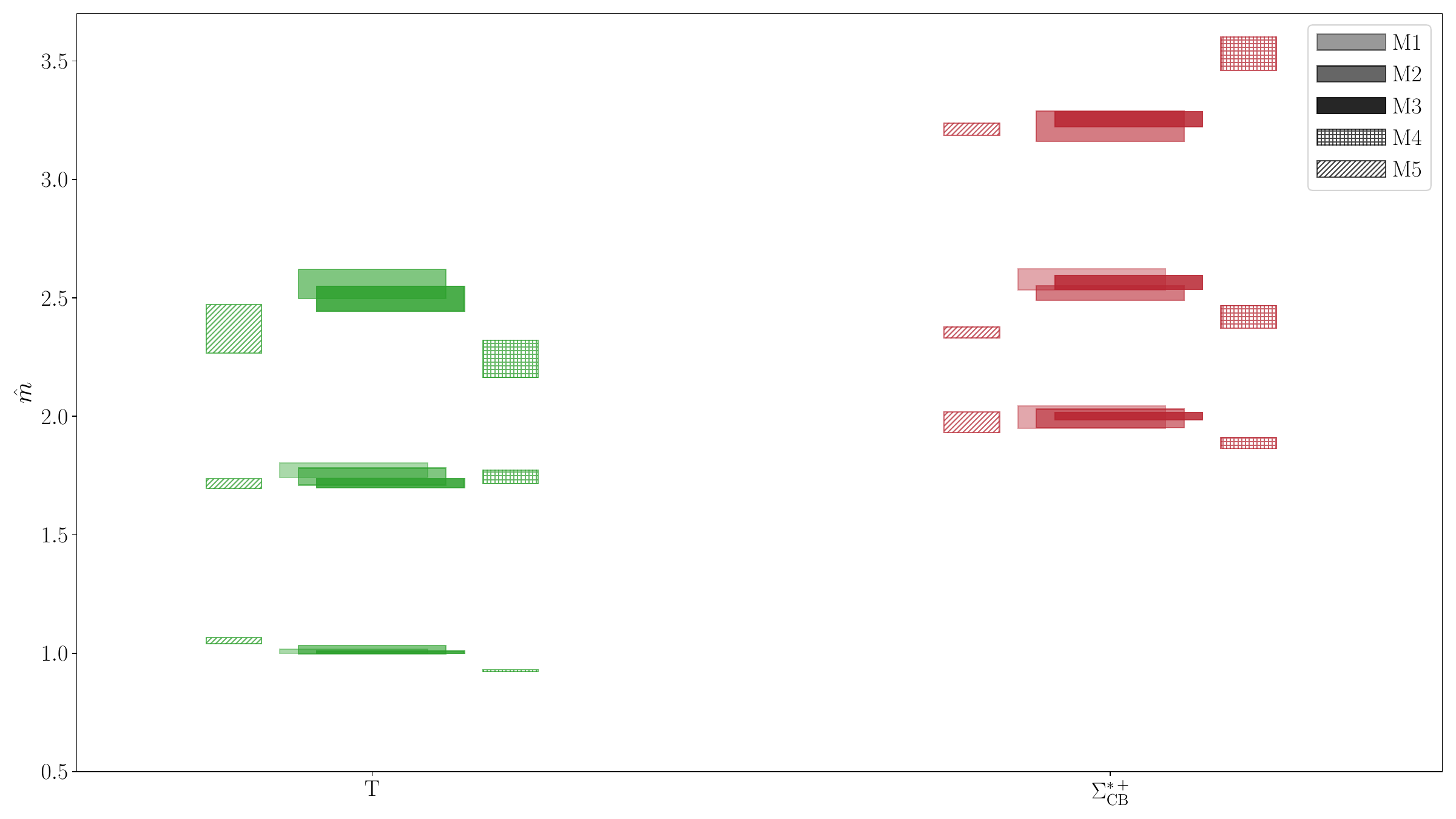}
    \caption{Representative examples of meson ($\text{T}$) and chimera baryon ($\Sigma^{*+}_{\rm CB}$) mass spectra  in the ensembles M1, M2, M3, M4, and M5,  characterized in Tab.~\ref{tab:ensembles}. All measurements have been obtained by fitting the spectral densities. For each channel, a tower of mass eigenvalues, expressed in units of the Wilson flow, $\hat{m} \equiv w_0 \cdot m$, is shown, the elements of which correspond to ground, first and (where available) second excited state. The vertical midpoint of each color block is the numerical result, whereas the height is the uncertainty, comprehensive of statistical and systematic errors, summed in quadrature. Horizontal offsets in the towers have no physical meaning, but are used  to distinguish graphically the different ensembles. 
    The choice of filling color is used to identify ensembles  M1-M3, while different patterns are used to indicate ensembles M4 and M5, as shown in the legend.
    \label{fig:spectrum_ensembles}}
     \end{figure}


\section{Numerical results and discussion}
\label{sec:numerical_results}
In this section, we present our numerical results, along with mass and  matrix element measurements, together with critical comparisons with earlier work. First, we determine the meson spectrum using both a variational analysis and spectral density fits on the expanded statistics ensembles, M1--M5, and we compare our results with the analysis in Ref.~\cite{Bennett:2024cqv}. We then present our new findings on the chimera baryon mass spectrum and on the matrix elements computations for both mesons and chimera baryons.

\subsection{Meson and chimera baryon mass spectra}
\label{sec:updates_results}

Direct comparison of Tab.~\ref{tab:ensembles} in this paper with Tab.~I of Ref.~\cite{Bennett:2024cqv}  shows that we extended the statistics by increasing $N_{\rm conf}$, the number of thermalized and uncorrelated configurations available. This is particularly evident for ensemble M1, which is approximately doubled in size, and M3, approximately tripled.  These larger data sets allow us to improve the statistical analysis compared to the one published in Ref.~\cite{Bennett:2024cqv}. We reconstruct the spectral densities from correlation functions, as detailed in Sec.~\ref{sec:HLT_method}, by using the interpolating operators in Tab.~\ref{tab:operators} and implementing both Wuppertal and APE smearing. The reconstructed spectral densities are then fitted, as explained in Sec.~\ref{sec:fitting_procedure}, to perform spectroscopy measurements.

     \begin{figure}[t]
    \includegraphics[width=0.9\linewidth]{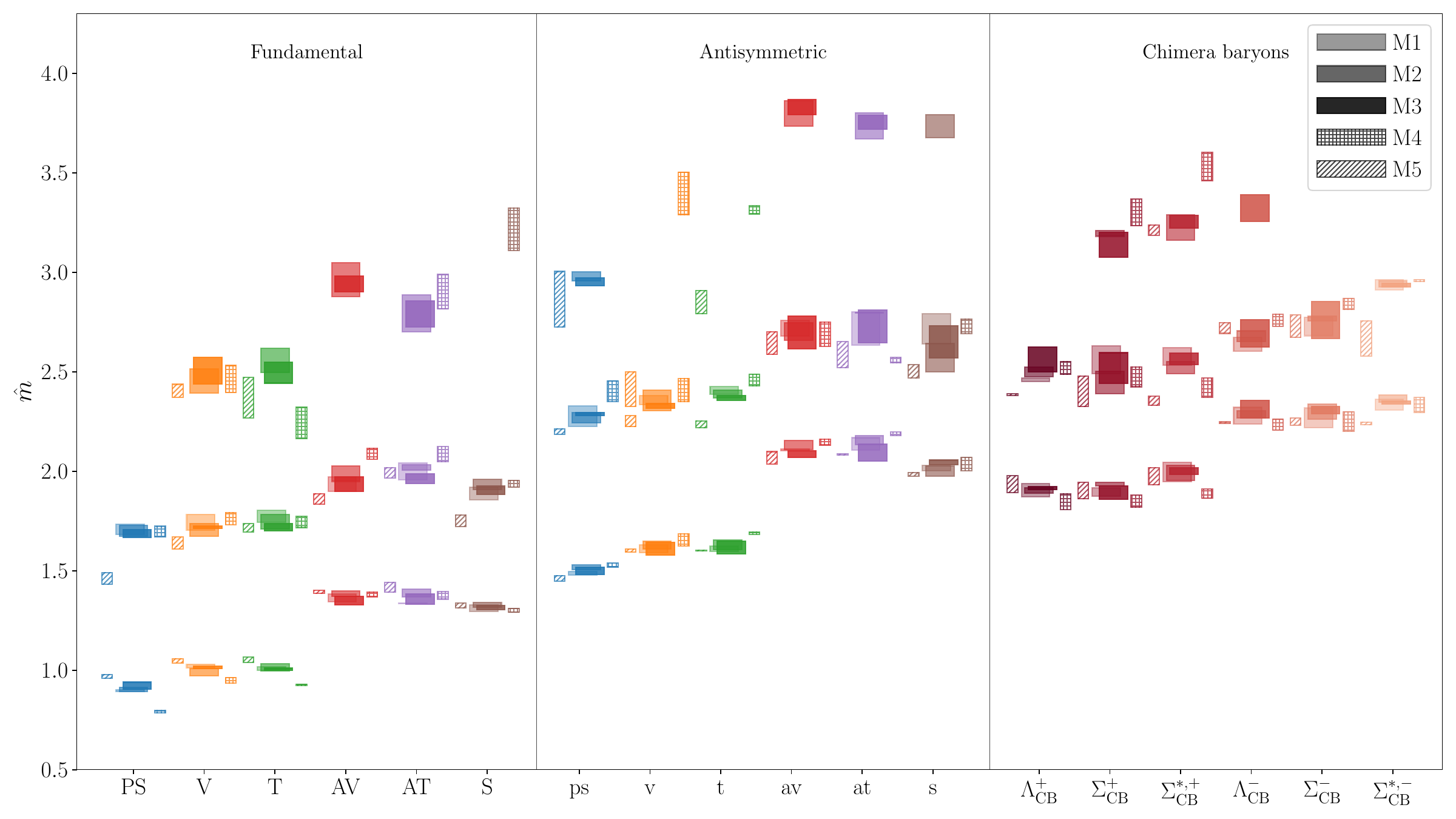}
    \caption{Flavored meson and chimera baryon mass spectra in all available ensembles, as summarized in Tab.~\ref{tab:ensembles}. The meson spectra include both composite states made of  $({\rm f})$-type (Fundamental) and $({\rm as})$-type (Antisymmetric) fermions. The spectrum is found through the fitting analysis of the spectral densities described in the text. For each channel, we show a tower of masses, $\hat{m} \equiv w_0 \cdot m$, expressed in units of the Wilson flow scale, $w_0 $. The particles correspond to ground, first and (where available) second excited states. The vertical midpoint of each color block is the numerical result, while the height is the uncertainty,  inclusive of statistical and systematic errors, summed in quadrature. Horizontal offsets are used to distinguish different ensembles. Different shadings of the same colors differentiate ensembles that differ in time extents ($N_t = 48, \, 64$ and $96$ for ensembles M1, M2, and M3, respectively), while different patterns are used to indicate ensembles that differ also in bare parameters (ensembles M4 and M5).  Six colors distinguish different meson channels. The colors match  meson operators built with the same gamma-matrix structure, but different fermion constituents. We include also the three chimera baryons channels, split according to their two parity projections.
    \label{fig:final_spectrum}}
     \end{figure}

For all ensembles, M1-M5, we list in Appendix~\ref{sec:tables} the numerical results for the masses of mesons in  ground state, first excited state, and (where available) second excited states,  derived both from spectral densities, as well as the variational analysis,  in Tabs.~\ref{table:E1_results_ground_mesons}--\ref{table:E5_results_second_mesons}. These values are expressed in units of the lattice spacing, $a$, and can be converted to Wilson flow units by using Tab.~\ref{tab:ensembles}.  The tables display results across the twelve flavored meson channels listed in Tab.~\ref{tab:operators}, detailing the number of Wuppertal smearing iterations at the source and sink, $N_{\rm source}$ and $N_{\rm sink}$, and the number of fitting functions used in the spectral density reconstruction, $k$, as defined in Eq.~\eqref{eq:sum_of_kernels}. Five separate analyses are included for each correlation function: spectral density reconstructions with Gaussian kernels (using $k$ and $k+1$ functions), or with Cauchy kernels (also using $k$ and $k+1$ functions), and variational analysis. The smearing radius for both Gaussian and Cauchy cases is also provided in the tables. In each case, if the contamination from additional excited states is minimal, one expects consistency between results for the optimal number of peaks $k$ and the $k+1$-peak fits across both kernels, which is confirmed by our measurements, across the whole range analyzed, and within statistical uncertainties.

For chimera baryons, after computing the correlation functions, using spin- and parity-projected operators, we reconstruct spectral densities and fit them.  For each channel listed in Tab.~\ref{tab:operators_baryons} and defined by Eq.~\eqref{eq:parity_projection}, we provide results for all ensembles, M1–M5, including the same five types of analysis performed for mesons: $k$- and $k+1$-peak fits using Gaussian and Cauchy kernels for spectral density reconstruction,
as well as the GEVP results.
 We tabulate our results for chimera baryon ground state, first excited state, and, where available, second excited state, in Tabs.~\ref{table:E1_results_ground_CB}--\ref{table:E5_results_second_CB}.

 In Fig.~\ref{fig:error comparison}, we provide visual examples of the comparison between spectroscopy measurements obtained with different methodologies. We present the lightest states in each of the sectors of bound states of interest in the $Sp(4)$ theory: mesons composed, respectively, of $({\rm f})$-type and $({\rm as})$-type fermions, as well as chimera baryons (including the two candidates relevant to TPC).
The measurements obtained with spectral density techniques yield results that, within the statistical errors, are independent of the type of kernel used. The  direct measurements, obtained with state-of-the-art conventional application of APE and Wuppertal smearing to the interesting correlation functions, are also consistent with the spectra density results. We repeated this exercise for all available ensembles and in all accessible channels, including also excited states, when possible, yielding equivalent outcomes---see Tabs.~\ref{table:E1_results_ground_mesons}--\ref{table:E5_results_second_mesons} for mesons, and 
    Tabs.~\ref{table:E1_results_ground_CB}--\ref{table:E5_results_second_CB} for chimera baryons---even for our measurements of matrix elements. 
    These results demonstrate that going from the correlation function to the smeared spectral function does not degrade the information about the spectrum. While this statement depends in principle on the radius of the smearing kernel, we have shown that, even with moderate radius sizes, the spectrum can be reconstructed with a good degree of precision, preserving the information contained in the correlation functions.  

For both mesons and chimera baryons, the results obtained in ensembles M1-M3  (which differ only by the time extent of the lattice) show good agreement with one another, and display a trend towards error reduction with increased temporal lattice extent, $N_t$. As observed in Section~V of Ref.~\cite{Bennett:2024cqv}, the improvement is expected, as in the HLT method an expansion of $t_{\rm max} < a N_t$ elements is performed, hence the longer the lattice temporal size is, the more accurate the spectral density reconstruction is expected to be. Having more temporal sites results in progressively lower statistical uncertainties, while the four energy estimates from spectral density fits are consistent with one another, indicating that residual systematic effects are negligible. These estimates also agree with the variational analysis results. 

Figure~\ref{fig:spectrum_ensembles} displays representative examples of mass measurements for meson ($\text{T}$) and chimera baryon ($\Sigma^{* +}_{\rm CB}$), demonstrating the overall improvement achieved by extending $N_t$. 
The masses are expressed in Wilson flow units, $\hat{m} \equiv w_0 \cdot m$. We combine statistical errors and systematic effects in quadrature. The systematic effects include artefacts due to additional excited-states contamination from the $k+1$-peak fits and choice of smearing kernels, with differences evaluated as the maximum spread between lattice results from $k$- and $k+1$-peak obtained with Gaussian and Cauchy kernels ($k$-G, $(k+1)$-G, $k$-C, and $(k+1)$-C).

Our results for meson masses are consistent with numerical estimates in Ref.~\cite{Bennett:2024cqv}, yet  show a general improvement, with reduced statistical uncertainties, as expected with increased statistics, and better access to first and second excited states, in line with expectations that a larger value of $N_{\rm conf}$ improves spectral density reconstruction. Moreover, a progressively increasing number of states appears in ensembles while considering larger time extent ($N_t > 96$ for M3 and $N_t > 64$ for M4 and M5), compared to Ref.~\cite{Bennett:2024cqv}.

As described by Eq.~(\ref{eq:fitting_systematics}), the use of multiple smearing kernels serves as a check against potential systematic effects in the spectral density reconstruction. Our results confirm that  such systematic effects are under control, the spectroscopy being consistent  across analysis methodology, inclusive of $k$- and $k+1$-peaks fits across Gaussian, Cauchy kernels, and GEVP analysis.

A summary display of our measurements of the mass spectrum, in all ensembles, showing both flavored mesons  and chimera baryons, is presented in Fig.~\ref{fig:final_spectrum}. This has been obtained by combining the results for mesons in Tabs.~\ref{table:E1_results_ground_mesons}--\ref{table:E5_results_second_mesons} and for chimera baryons in Tabs.~\ref{table:E1_results_ground_CB}--\ref{table:E5_results_second_CB}.  All measurements are expressed in units of the Wilson flow, $\hat{m} \equiv w_0 \cdot m$.

     \begin{figure}[t]
    \includegraphics[width=0.8\linewidth]{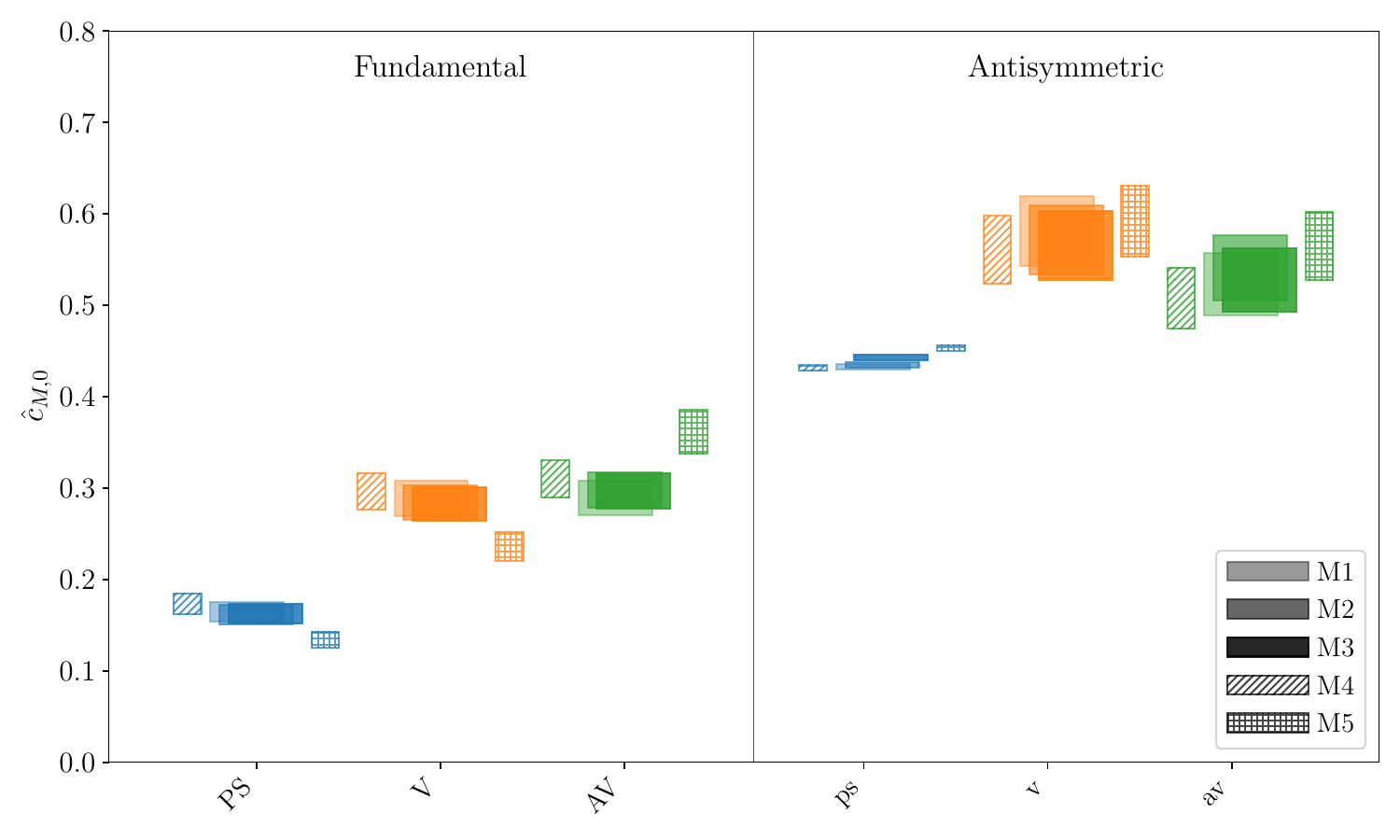}
    \caption{Matrix elements, $\hat{c}_{M,0} \equiv w_0^2\, \langle 0| \mathcal{O}^{M} |e_{M,0} \rangle$, for ground state flavored mesons, measured in all available ensembles, summarized in Tab.~\ref{tab:ensembles}, obtained through fitting spectral densities. Results are expressed in units of the Wilson flow scale, $w_0$. The vertical midpoint of each color block is the numerical result, while the height represents the combined statistical and systematic uncertainties. Horizontal offsets are used to distinguish different ensembles. Different shadings of the same color differentiate ensembles that change only in time extents ($N_t = 48, \, 64$, and $96$ for ensembles M1, M2, and M3, respectively), whereas no filling color and patterns are used to indicate ensembles that differ also in bare parameters (ensembles M4 and M5).
    \label{fig:final_matrixelements}}
     \end{figure}

\subsection{Mesons matrix elements and chimera baryons overlap factors}
\label{sec:matrix_elements}

As discussed in Secs.~\ref{sec:f_p_corr} and \ref{sec:fitting_procedure}, we extract matrix elements and overlap factors both by using correlation function fitting techniques also by fitting spectral density amplitudes. This approach allows us to determine the matrix elements associated with the interpolating operators listed in Tab.~\ref{tab:operators} for flavored meson matrix elements. For the renormalized meson matrix elements we introduce the shorthand notation:
\beqs
c_{M,n} \equiv Z_{M}^{\rm R}\langle 0| \mathcal{O}^{\rm R} |e_{M,n} \rangle\,,
\eeqs
 with $\mathcal{O}^{\rm R}$ any meson interpolating operator entering in Eqs.~\eqref{eq:fpi},~\eqref{eq:fV}, and~\eqref{eq:fAV}, and $Z_{M}^{\mathrm R}$ the renormalization coefficients for fermions transforming according to the representation $\rm{R}$ of the gauge group. We focus our analysis on the ground state, $n=0$,  due to the limited available statistics---a similar analysis can be used also for excited states, but would require using higher statistics, and an enlarged basis of Wuppertal smeared operators.
For the chimera baryon operators listed in Tab.~\ref{tab:operators_baryons}, the overlap factors of interest, $K_{B,0}$,  are defined in Eqs.~\eqref{eq:overlap_factors} and~\eqref{eq:correlation_matrix_baryonic}---notice the different normalization and dimensionality, and the appearance of spinors in the defining relations. The renormalization constants are provided in Sect.~\ref{sec:f_p_corr}. The renormalized matrix elements are reported in Tabs.~\ref{table:E1_matrix_mesons}--\ref{table:E5_matrix_mesons}, for mesons. For chimera baryons, the renormalized overlap factors are tabulated in Tabs.~\ref{table:E1_matrix_CB}--\ref{table:E5_matrix_CB}.

The tables detail our results for matrix elements and overlap factors, obtained through simultaneous spectral density fits (as described in Sec.~\ref{sec:fitting_procedure}). Measurements obtained by using Gaussian ($a^{2}c_{M,0}\text{-G}$ for mesons, $a^{3}K_{B,0}\text{-G}$ for chimera baryons) and Cauchy ($a^{2}c_{M,0}\text{-C}$  for mesons, $a^{3}K_{B,0}\text{-C}$ for chimera baryons) kernels are shown next to one another, along with results derived from the correlation function fitting techniques outlined in Sec.~\ref{sec:f_p_corr}. As with spectral results, physical matrix elements and overlap factors do not depend on the choice of smearing kernel, which is confirmed by the numerical results, showing consistency across methodologies. We use the discrepancies as an estimation of the systematic effects, via Eq.~\eqref{eq:fitting_systematics_ME}, and find them to be smaller than the statistical uncertainties. Our spectral-density analysis yields results that also agree with those obtained through traditional correlation function fitting techniques, supporting the validity of this novel approach. The general agreement across the results for ensembles M1--M3 is expected, given that the three ensembles differ only in the temporal extent, $N_t$, of the lattice. This is observed both for mesonic matrix elements, in Tabs.~\ref{table:E1_matrix_mesons}--\ref{table:E3_matrix_mesons}, and for chimera baryons overlap factors, in Tabs.~\ref{table:E1_matrix_CB}--\ref{table:E3_matrix_CB}.

          \begin{figure}[t]
    \includegraphics[width=0.8\linewidth]{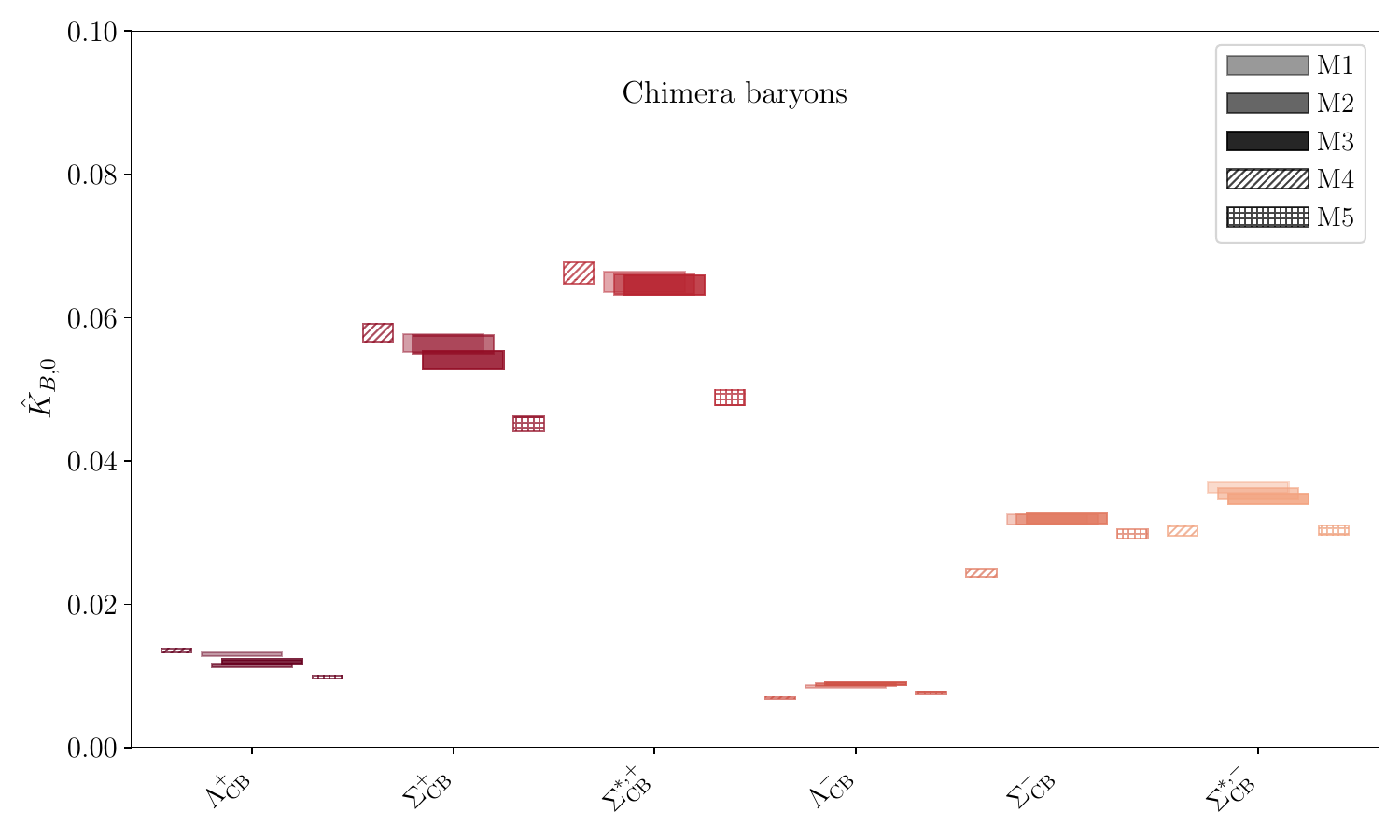}
    \caption{Chimera baryon overlap factors, $\hat{K}_{B,0} \equiv w_0^{3} \cdot K_{B,0}$, expressed in units of the Wilson flow scale, $w_0$, in all available ensembles summarized in Tab.~\ref{tab:ensembles}, extracted through spectral densities fitting analysis. The vertical midpoint of each color block is the numerical result, while the height represents the uncertainty, comprehensive of statistical and systematic errors, summed in quadrature. Horizontal offsets are used to distinguish different ensembles. Different shadings of the same color differentiate ensembles that differ in time extents ($N_t = 48, \, 64$, and $96$ for ensembles M1, M2, and M3, respectively), whereas different  patterns are used to indicate ensembles that differ also in bare parameters (ensembles M4 and M5).     \label{fig:final_overlapfactors}}
     \end{figure}

The results for mesons and baryons (restricted to the ground state) are shown in Figs.~\ref{fig:final_matrixelements} and~\ref{fig:final_overlapfactors}, respectively.  As for  the masses, for both mesons and chimera baryons, the measurements obtained in ensembles  M1-M3, which differ only by the time extent of the lattice, generally agree with one another, and show a trend of improvement with increased temporal lattice extent, $N_t$. 
The matrix elements and overlap factors are expressed in Wilson flow units, $\hat{c}_{M,0} \equiv w_0^2 \cdot c_{M,0}$ and $\hat{K}_{B,0} \equiv w_0^{3} \cdot K_{B,0}$. We combined statistical and systematic errors in quadrature, although the systematic errors are smaller than the statistical ones. The systematic effects include artefacts due to excited-state contamination and choice of smearing kernels, with differences evaluated as the maximum spread between lattice results from $k$- and $(k+1)$-peak in Gaussian and Cauchy fits ($k$-G, $(k+1)$-G, $k$-C, and $(k+1)$-C). \\

We could not perform the continuum extrapolation with the available ensembles, hence we expect our measurements to be affected by (discretization) lattice artefacts. Yet,  some interesting qualitative pattern emerges. In the meson sector, we find that among the matrix elements, $\hat{c}_{M, 0}$, those in the pseudoscalar channel are smaller than those in the vector and axial-vector channels. Furthermore, these values tend to be larger in the antisymmetric representation than in the fundamental one. Turning to chimera baryons, we observe that the ground-state overlap factors, $\hat{K}_{B, 0}$, for the $B=\Lambda_{\rm CB}$ states are smaller than those for the $\Sigma_{\rm CB}$ and $\Sigma^*_{\rm CB}$ states, which exhibit comparable magnitudes. Odd-parity states generally display smaller overlap factors than their even-parity counterparts. \\

It is informative to compare our results for the overlap factors of chimera baryons with those obtained in other theories. In particular, we can draw a comparison with analogous quantities computed in QCD for proton decay~\cite{Aoki:2017puj}, as well as the overlap factors relevant for partial top-compositeness in a $SU(4)$ gauge theory with $N_{\rm f} = 2$ fermions in the fundamental representation and $N_{\rm as} = 2$ in the two-index antisymmetric representation~\cite{Ayyar:2018glg}. In both cases, a key dimensionless quantity of interest is the ratio $ K_{B,0} / f_{\rm PS}^3$. 
The reported values for this ratio in Ref.~\cite{Aoki:2017puj} and~\cite{Ayyar:2018glg} are, respectively:\footnote{In both references, the decay constant is normalized so that as $F_\pi = F_{\rm PS} \simeq 130$ MeV. To match our conventions, we rescaled the measurement, that for QCD would lead to $f_{\pi}\simeq 93$ MeV. For the $SU(4)$ theory, we use the decay constant computed for mesons composed of fermions transforming in the fundamental representation. It is also worth to be mentioned that our values are renormalized at scale $\mu = 1/a$, which differs for our lattice spacings from the scale used in the QCD results, which is chosen to be $\mu = 2$ GeV. This fact is not expected to defeat the purposes of our order-of-magnitude estimation, since the running effect is logarithmic.}
\begin{equation}
   \left( \dfrac{K_{{B}, 0}}{f^3_{\rm PS}}\right)_{\rm{QCD}} \sim 20\,, \quad\quad \text{and} \quad\quad \left( \dfrac{K_{{B}, 0}}{f^3_{\rm PS}}\right)_{SU(4)} \sim 3 \, .
\end{equation}
For comparison, our measurements in the $\Lambda_{\rm CB}$ and $\Sigma_{\rm CB}$ channels  are fairly consistent across all ensembles considered (M1--M5). Averaging over the ensembles, we find:
\begin{equation}
   \left( \frac{K_{{\Lambda_+}, 0}}{f^3_{\rm PS}}\right) \sim 6\,, 
   \quad\quad \text{and} \quad\quad \left( \frac{K_{{\Sigma_+}, 0}}{f^3_{\rm PS}}\right) \sim 25\,.
\end{equation}
 Despite the  fact that we did not perform a continuum limit extrapolation, and that the matching of the renormalization corrections is done naively using the cutoff scale, it is reassuring to notice that our results for the $\Lambda$ and $\Sigma$ chimera baryons are broadly comparable with those in the literature, at the order-of-magnitude level.

We notice that in Ref.~\cite{Ayyar:2018glg} the authors normalize the overlap factor to the decay constant of the mesons emerging from the fermions transforming in the antisymmetric representation, because that case is of relevance to the CHM based on $SU(4)$, as would be 
 appropriate also for the model in Ref.~\cite{Cacciapaglia:2019ixa}.
In our case, for the CHM proposed  in Refs.~\cite{Barnard:2013zea,Ferretti:2013kya}, the Higgs sector arises from the fermions transforming in the fundamental sector. Because we find that $f_{\rm PS}<f_{\rm ps}$, across all our measurements, our result for the overlap factors are enhanced in respect to those in Ref.~\cite{Ayyar:2018glg}. 
Furthermore, we notice that if the chimera baryon of relevance to the TPC mechanism is the $\Sigma$ (rather than $\Lambda$), an additional enhancement factor appears, which is interesting for model-building considerations.  We look forward to seeing how these measurements change when performing the continuum limit extrapolations in future precision studies.

\begin{table}
    \centering
    \caption{The quantities $s_0$, $s_1$, and $s_2$, defined in the body of the paper, measured in all ensembles, and for both  meson species, obtained by combining our best measurements. Errors include both statistical and systematic uncertainties (in quadrature) of the lattice calculations performed for this paper.  Theoretical approximations due to continuum and massless extrapolation, as well as saturation over the ground state mesons, lead to $s_1\neq 0\neq s_2$.
    \label{tab:Weinberg} \\    }
    \begin{tabular}{ |c|c|c|c|c| }
        \hline\hline
       Meson  & Ensemble & $s_0$ & $s_1$ & $s_2$ \\
        \hline
$({\rm f})$-type     & M1 & 0.298(4) & -0.008(19) & 0.005(44) \\
$({\rm f})$-type     & M2 & 0.253(5) & -0.149(32) & -0.310(78) \\
$({\rm f})$-type     & M3 & 0.257(15) & -0.156(47) & -0.270(73) \\
$({\rm f})$-type     & M4 & 0.220(9) & -0.245(31) & -0.384(55) \\
$({\rm f})$-type     & M5 & 0.192(4) & -0.629(25) & -1.561(68) \\
        \hline
$({\rm as})$-type     & M1 & 0.220(4) & -0.116(19) & 0.213(27) \\
$({\rm as})$-type     & M2 & 0.204(3) & -0.193(21) & 0.103(34) \\
$({\rm as})$-type     & M3 & 0.206(5) & -0.194(30) & 0.151(40) \\
$({\rm as})$-type     & M4 & 0.229(6) & -0.137(23) & 0.205(26) \\
$({\rm as})$-type     & M5 & 0.197(5) & -0.206(23) & 0.097(37) \\
\hline  \hline
\end{tabular}
\end{table}

We conclude with another simplified exercise, intended to assess the current precision level of our measurements, but also to illustrate the physics insight that the application of the technology developed for this paper could yield with future high precision lattice studies.
We define the following three quantities,  borrowing, for convenience, the normalization conventions adopted in Ref.~\cite{Bennett:2023qwx},
for mesons made of $({\rm f})$-type fermions:
\beqs
s_0&\equiv&  4\pi   \left( \frac{\hat{f}_{V}^2}{\hat{m}_{{\rm V}}^2}-\frac{ \hat{f}_{{\rm AV}}^2}{ \hat{m}_{{\rm AV}}^2}\right)\,,\\
s_1&\equiv&1-  \frac{\hat{f}_{{\rm AV}}^2+\hat{f}_{\rm PS}^2}{ \hat{f}_{{\rm V}}^2}\,,\\
s_2&\equiv&1-\frac{ \hat{m}_{{\rm AV}}^2 \hat{f}_{{\rm AV}}^2}{\hat{m}_{{\rm V}}^2 \hat{f}_{\rm V}^2}\,,
\eeqs
and analogous definitions for mesons made of $({\rm as})$-type fermions.
The second and third such relations are related to the Weinberg sum rules~\cite{Weinberg:1967kj}, and the first to the 
Peskin-Takeuchi $S$ parameter~\cite{Peskin:1991sw}.
In the continuum limit, for massless hyperquarks, and replacing the right-hand side of these definitions with  summation running over the whole tower of vector and axial-vector states, 
the Weinberg sum rules can be formulated by stating that $s_1=0=s_2$. 
In the same limits, and with the additional requirement that the electroweak $SU(2)_L\times U(1)_Y$ gauge group of the Standard Model be embedded in the global symmetries of this theory such that the vacuum triggers electroweak symmetry breaking as in the Standard Model, then 
$s_0$ yields a measure of (isospin) symmetry breaking effects due to new physics.
Current electroweak precision tests set the bound $s_0=S<0.4$ at $3\sigma$ confidence level.  

We performed the measurements for finite fermion mass and lattice spacing, and measured the decay constants only for the ground-state particles, hence introducing potentially large systematic uncertainties due to these theoretical limitations.
We display these three quantities, computed in all our ensembles, and for both mesons composed of $({\rm f})$-type and $({\rm as})$-type fermions, in Tab.~\ref{tab:Weinberg}.
The results for $s_1$ and $s_2$ are not compatible with zero, consistently with the presence of large systematics. Yet, the central values are about one order of magnitude smaller than those obtained in the quenched approximation, listed in Ref.~\cite{Bennett:2023qwx}, demonstrating how these quantities are sensitive to the dynamics of the underlying theory.
In future measurements, in which we aim at extrapolating our results towards the continuum and massless limits, these quantities will provide a test of the saturation of the Weinberg sum rule on the ground state.

\section{Summary and outlook} 
\label{sec:outlook}

We have developed a new implementation of the HLT algorithm that allows to reconstruct the spectral density 
from two-point correlation functions, to compute 
masses, matrix elements and overlap factors of flavored mesons and 
chimera baryons. 
We have applied these new tools to study the lattice theory with $Sp(4)$ gauge group and matter
 content consisting of $N_{\rm f}=2$ Wilson-Dirac fermions
transforming in the fundamental and $N_{\rm as}=3$ in the 2-index antisymmetric representation,
by generating five lattice ensembles that have the same lattice parameters as
in Ref.~\cite{Bennett:2024cqv,Bennett:2024wda}, but larger statistics.
This theory is the minimal candidate for the completion of a composite Higgs model with top partial compositeness; some of the mesons play the role of the Higgs fields in CHMs, while some of 
the chimera baryons have the right quantum numbers to be identified as top partners in TPC, 
hence this information is important for the phenomenology of  extensions of the Standard Model.
The main elements of novelty of this publication are the application of HLT algorithm to baryon bound states, and to the extraction of matrix elements and overlap factors. The numerical strategy developed for this work has also general validity, 
as it could be used in the study of other gauge theories, including QCD.


The main results we reported in this publication that are relevant to phenomenological studies of new physics models are our measurement of the masses of the lightest bound states in all flavoured channels, together with our estimate
of the overlap factor, $\left( {K_{{\Sigma_+}, 0}}/{f^3_{\rm PS}}\right) \sim 25$. The former completes and complements existing literature and is useful for new physics searches. The latter is one order of magnitude larger than 
that obtained in the $SU(4)$ theory studied in Ref.~\cite{Ayyar:2018glg}. 
This factor enters the estimate of the mass of the top quark in a realistic CHM with TPC,
and hence our preliminary results, if confirmed in the continuum limit, would give this theory prominence as
a  potentially realistic candidate for extensions of the standard model with composite dynamics.
Such finding hence motivates  additional, large scale numerical studies of this theory,
taking advantage of improved lattice action, aimed at 
approaching the continuum limit.


The results presented here are propaedeutic to an ambitious, long-term future research program, part of which is already undergoing.
We focused our analysis on bound states that transform non-trivially under the unbroken global $Sp(4)$ (or $SO(6)$) symmetry that acts on the
$({\rm f})$-type (or $({\rm as})$-type) fermions.  For this study  we used a limited number of ensembles, with fixed values of lattice coupling and comparatively large hyperquark masses. For phenomenological purposes, it may not be necessary to carry out an extrapolation towards the limit of massless hyperquarks, as CHMs do require the presence of explicit sources of symmetry breaking to be viable.
Nevertheless, the continuum limit extrapolation is needed, and it would be useful to perform it in combination with the matching to an 
effective field theory in which the mass dependence of all the lightest bound states (including flavor singlets) can be studied systematically.
In order to pursue such a major endeavour, we envision changing the lattice formulation of the theory, by adopting domain wall fermions. Doing so would result in improving the approach to the continuum limit,
as well as providing numerical access to lower-mass regions in parameter space. 
Encouraging preliminary results have been collected in this direction and will be presented in the near
 future~\cite{Bennett:2025domainwall}.


\begin{acknowledgments}

We would like to thank Giacomo Cacciapaglia, Gabriele Ferretti, Thomas Flacke, Anna Hasenfratz, Chulwoo Jung, and Sarada Rajeev, for very helpful discussions during the “PNU Workshop on Composite Higgs: Lattice study and all”, at Haeundae, Busan, in February 2024. 
E.B. and B.L. are supported by the EPSRC ExCALIBUR programme ExaTEPP project EP/X017168/1. 
E.B. is supported by the STFC Research Software Engineering Fellowship EP/V052489/1. 
E.B., B.L., M.P. and F.Z. are supported by the STFC Consolidated Grant No. ST/X000648/1.
The work of N.F. is supported by the STFC Doctoral Training Grant
No. ST/X508834/1. 
A.L. is funded in part by l’Agence Nationale de la Recherche (ANR), under grant ANR-22-CE31-0011.
D.K.H. is supported by Basic Science Research Program through the National Research Foundation of Korea (NRF) funded by the Ministry of Education (NRF-2017R1D1A1B06033701) and by the NRF grant 2021R1A4A5031460 funded by the Korean government (MSIT). 
L.D.D. and R.C.H. are supported by the STFC grant ST/P000630/1.
L.D.D. is supported by the ExaTEPP project EP/X01696X/1.
J.W.L. is supported by IBS under the project code, IBS-R018-D1. 
H.H. and C.J.D.L. acknowledge support from NSTC Taiwan, through grant number 112-2112-M-A49-021-MY3. C.J.D.L. is also supported by the Taiwanese MoST grant 109-2112-M-009-006-MY3. 
C.J.D.L. is supported by Grants No. 112-2639-M-002-006-ASP and No. 113-2119-M-007-013-.
B.L. and M.P. are supported by the STFC  Consolidated Grant No. ST/T000813/1.
B.L., M.P. and L.D.D. received funding from the European Research Council (ERC) under the European Union’s Horizon 2020 research and innovation program under Grant Agreement No.~813942. 
D.V. is supported by STFC under Consolidated Grant No. ST/X000680/1.

Numerical simulations have been performed on the DiRAC Extreme Scaling service at the University of Edinburgh, and on the DiRAC Data Intensive service at Leicester.
The DiRAC Extreme Scaling service is operated by the Edinburgh Parallel Computing Centre on behalf of the STFC DiRAC HPC Facility (www.dirac.ac.uk). This equipment was funded by BEIS capital funding via STFC capital grant ST/R00238X/1 and STFC DiRAC Operations grant ST/R001006/1. DiRAC is part of the UKRI Digital Research Infrastructure. The DiRAC Data Intensive service (DIaL2 / DIaL [*]) at the University of Leicester, is managed by the University of Leicester Research Computing Service on behalf of the STFC DiRAC HPC Facility (www.dirac.ac.uk). The DiRAC service at Leicester was funded by BEIS, UKRI and STFC capital funding and STFC operations grants. 

\vspace{1.0cm}
{\bf Research Data Access Statement}---The analysis code and data generated for this manuscript can be downloaded from Refs.~\cite{analysis_release} and~\cite{data_release}, respectively. We refer to Ref.~\cite{Bennett:2025neg} for our 
approach to reproducibility and open science.
\vspace{1.0cm}

{\bf Open Access Statement}---For the purpose of open access, the authors have applied a Creative Commons 
Attribution (CC BY) license to any Author Accepted Manuscript version arising.

\end{acknowledgments}

\appendix
\section{Correlation functions smearing techniques} \label{sec:Wuppertal_APE}

Wuppertal smearing~\cite{Gusken:1989qx,Roberts:2012tp,Alexandrou:1990dq} and APE smearing~\cite{APE:1987ehd,Falcioni:1984ei} are well established lattice techniques, typically applied together, the purpose of which is to improve the signal of correlation functions, in particular for the extraction of the mass of the ground state. By applying APE smearing to the gauge links in a given configuration, we smoothen out short-distance  fluctuations in gauge links, which improves the signal of the effective mass plateau. 
The implementation of Wuppertal smearing to the source and sink replaces point-like operators with extended ones, which enhances the overlap with the ground state, and hence the signal-to-noise ratio, and suppresses excited state contamination, so that the plateau in effective mass appears at earlier Euclidean time, which improves our  control over fitting range systematics.

APE smearing is an iterative process involving the staple operator, $S_\mu(x) \equiv \sum_{\pm \nu \neq \mu} U_\nu(x)U_\mu(x+\hat{\nu})U^\dagger_\nu(x+\hat{\mu})$, around each gauge link, $U_{\mu} (x)$. A new,  smeared link, $U^{(m)}_{\mu}(x)$,  is defined by subsequent  modifications of the link
\begin{equation} 
\label{eq} U^{(m)}_{\mu}(x) \equiv \mathcal{P} \left( (1-\alpha_{\textrm{APE}}) U^{(m-1)}_{\mu}(x) + \frac{\alpha_{\textrm{APE}}}{6} S^{(m-1)}_{\mu}(x) \right)\,, 
\end{equation} 
for $m = 1, \dots, N_{\rm APE}$, with initial conditions  $U_{\mu}^0 = U_{\mu}$ and $S^0_{\mu} = S_{\mu}$. One has to specify two parameters: the APE-smearing step size, $\alpha_{\textrm{APE}}$, and the total number of smearing steps, $N_{\rm APE}$.
A projection operator, $\mathcal{P}$, the precise form of which is determined by the group and the representation used for the links, is included in the definition, since when the gauge links are updated at each iteration by summing over neighboring staples, the result may generally fall outside the group manifold.

To illustrate our implementation of Wuppertal smearing, we start by writing the equation for the Green function associate with the Dirac operator for point-like source and sink: 
\begin{equation} 
\label{eq} 
\sum_{y, \beta,  b} D^{R}_{a \alpha , b \beta} (x,y) , S^{b \beta}_{R , c \gamma} (y,0) = \delta_{x, 0} , \delta_{\alpha \gamma} , 
\delta_{ac}\,, 
\end{equation} 
where $D^{R}_{a \alpha , b \beta}$ is the Wilson-Dirac operator in representation $R$, with $\alpha, \beta, \gamma$ denoting spinor indices, and $a, b, c$ generalized color indices. The solution of this equation,  $S^{b \beta}_{R , c \gamma}$, is the hyperquark propagator in representation $R$. For instance, the two-point correlation function of mesons can then be schematically written as: 
\begin{equation}
\label{eq:correlator_propagator}
    C(t) = \langle \mathcal{O}_{R} (t) \, \bar{\mathcal{O}}_{R} (0) \rangle = \Big\langle \sum_{\vec{x}} \Tr\, \left[ \left( \Gamma \,  S_R(x,0) \, \bar{\Gamma} \, S_R (x,0) \right)  \right] \Big\rangle\,,
\end{equation}
where $\Gamma$ and $\bar{\Gamma} \equiv \gamma^0 \Gamma^\dagger \gamma^0$ depend on the spin structure of the interpolating operator, $\mathcal{O}_{R}$, which is bilinear in fermion fields.

Wuppertal smearing consists of replacing  the delta function, $\delta_{x,0}$, on the right-hand side of 
Eq.~(\ref{eq})  with a new function of the coordinates, $q_R^{(n+1)}(x)$, defined through an iterative diffusion process that take the following form:
 \beqs
 q^{(n+1)}(x)&\equiv& \frac{1}{1+6\varepsilon_{\rm R}}\left[q^{(n)}(x)\frac{}{}+\frac{}{}
 \varepsilon_{\rm R}\sum_{\hat\mu} U_{\mu}(x) q^{(n)}(x+\hat{\mu})\right]\,,
 \eeqs
 with $\varepsilon_{\rm R}$ the smearing step size, and $q^{(0)}(x)=\delta_{x,0}$.
One then  performs the inversion of the Dirac operator, producing a smeared propagator $S^{(n)}_R(y, 0)$. For the sink smearing, the same procedure is applied to the source-smeared propagator without requiring additional inversions.
Finally, one modifies  Eq.~\eqref{eq:correlator_propagator}, by rewriting it in terms of the new propagator, $S^R_{(N_{\rm source},N_{\rm sink})} (x,\,0)$ obtained by solving the associated Green function equation in the presence of the extended source and sink:
\begin{equation}
    C_{N_{\rm source}, \, N_{\rm sink}}(t) = \Big\langle \sum_{\vec{x}} \Tr \left[  \Gamma \, \left(S^R_{(N_{\rm source},N_{\rm sink})} (x,\,0) \right)   \, \bar{\Gamma} \, \left(S^R_{(N_{\rm source},N_{\rm sink})} (x,\,0) \right)  \right]\Big\rangle \,.
\end{equation}
The fully smeared propagator, after $N_{\rm source}$ iterations of source smearing and $N_{\rm sink}$ iterations of sink smearing, is denoted as $S^R_{(N_{\rm source}, N_{\rm sink})}(x, 0)$. One then has to specify three parameters, the values of which are chosen to optimize the analysis: the smearing step size, $\varepsilon_{\rm R}$, and the number of source and sink smearing steps, $N_{\rm source}$ and $N_{\rm sink}$.
Measurements of two-point meson correlation functions smeared with APE and Wuppertal smearing are performed using the HiRep code~\cite{DelDebbio:2008zf,HiRepSUN,HiRepSpN}.

\section{Renormalization of chimera baryons overlap factors}
\label{sec:renormalisation_CB}

In this Appendix, we summarize  our main   results for the one-loop renormalization of the local chimera baryon operators relevant to this paper.
We write the chimera baryon operator in the following form:
\begin{equation}
\label{eq:CB_operator}
O^{\mu}_{\rm CB}(x)\equiv\left\{\left[Q_{1 \, \alpha}^{a} ({\cal C} \Gamma_{1})^{\alpha \beta} Q_{2 \, \beta}^{b}\right] \Omega_{a d} \Omega^{b c} \Gamma_{2}^{\mu {\gamma}} \Psi_{\gamma}^{c d}\right\}(x)\,,
\end{equation}
as in Eq.~\eqref{eq:baryon_ops},
where $ \Gamma_1 = \{ \gamma_5,\,  \gamma_i\} $ and $\Gamma_2^{\mu \gamma} = \mathbb{1}^{\mu \gamma}$.\footnote{For the charge-conjugation operator, we make use of the conventions $\mathcal{C}^{2}=\mathbb{1}, \,\mathcal{C}^{\dagger}=-\mathcal{C}, \, \mathcal{C}^{T}=-\mathcal{C}$ and the commutation rule $\mathcal{C} \gamma_{\mu} \mathcal{C}^{-1}=-\gamma_{\mu}^{T} \Rightarrow \gamma_{\mu}^{T} \mathcal{C}=-\mathcal{C} \gamma_{\mu}$ and $\gamma_5^{T}=\gamma_5$.} 
For each such operator, we introduce the 4-point function
\begin{equation}
\label{eq:four_point_corr}
\Gamma^{\mu \alpha \beta \gamma}\left(x, x_{1}, x_{2}, x_{3}\right)\equiv\left\langle O^{\mu}_{\rm CB}(x) \Psi_{c d}^{\alpha}\left(x_{3}\right) Q_{1 \, a}^{\beta}\left(x_{2}\right) Q_{2\, b}^{\gamma}\left(x_{3}\right)\right\rangle \left[ \dfrac{1}{2N_c} \Omega^{ad} \Omega^{bc}\right] \,,
\end{equation}
where we have separated the term $ \dfrac{1}{2N_c} \Omega^{ad} \Omega^{bc}$, serves to project the fermionic lines attached to the operator onto a given color structure.

We will write the result for the renormalization factors at renormalization scale $\mu = 1/a$ in the following form:
\begin{equation}
\label{eq:final_result_CB_ren}
Z_{\rm{CB}, \Gamma} [\mu = 1/a]=1+\frac{g^{2}}{16 \pi^{2}}\left(\left[C^{\rm f}+\frac{1}{2} C^{\rm as}\right] \Delta_{\Sigma_1}+\Delta_{\rm CB} [\Gamma, \mu = 1/a]\right)\,,
\end{equation}
where $\Gamma = \{ \gamma_5, \gamma_i \}$ and
\beqs
\label{eq:matching_factors}
\Delta_{\rm{CB}}[\gamma_5] &=& -26.67\,, \quad\quad
\Delta_{\rm{CB}}[\gamma_i] \,=\, 18.12\,, \quad\quad
\Delta_{\Sigma_1} \,=\, -12.82\,,
\eeqs
and where the eigenvalues for the quadratic Casimir operators, evaluated  in the $Sp(2N=4)$ gauge theory, are $C^{\rm f} =5/4$ and $C^{\rm as} = 2$.

\begin{figure}[t]
  \centering
  \begin{minipage}[t]{0.30\linewidth}
  \centering
  \includegraphics[width=\linewidth]{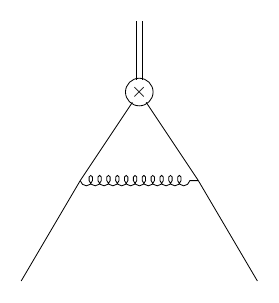}
\end{minipage}
\hspace{1.9em}
\begin{minipage}[t]{0.30\linewidth}
  \centering
  \includegraphics[width=\linewidth]{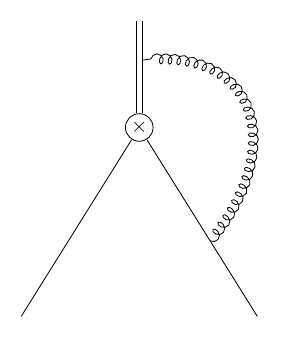}
\end{minipage}
  \caption{One-loop Feynman diagrams contributing vertex corrections to the renormalization of the chimera baryon operators. Left panel: gluon exchange between two $({\rm f})$-type fermions. Right panel: gluon exchange between one $({\rm f})$-type and $({\rm as})$-type fermion. A third diagram is obtained by exchanging the two fundamental fermion legs in the second.}
  \label{fig:combined_diagrams}
\end{figure}

Two types of diagrams contribute to the one-loop correction, both in the continuum and on the lattice:
vertex corrections are shown in Fig~\ref{fig:combined_diagrams}, where (\rm{f})-type and (\rm{as})-type fermions are denoted by single and double solid lines, and in addition we included also fermion self-energy diagrams in our analysis.
For the latter, the self-energy contribution to  Eq.~\eqref{eq:final_result_CB_ren} can be adapted from Ref.~\cite{Martinelli:1982mw}.
For the former, the vertex diagrams are evaluated both in the continuum, in the \(\overline{\text{MS}}\) scheme, and on the lattice using the actions in Eqs.~\eqref{eq:Lagrangian} and~\eqref{eq:lattice_fermion_action}, respectively. Part of the algebra has been handled with FORM~\cite{Kuipers:2012rf}, a symbolic manipulation system designed for high-performance algebraic computations in quantum field theory. We provide details of both calculations in the next subsections.
 Finally, the coupling in Eq.~\eqref{eq:final_result_CB_ren} is defined using a mean-field approximation for the link variable, which effectively incorporates the effects of tadpole diagrams. In this formulation, the improved coupling is expressed as \(\tilde{g}^2 = \frac{g^2}{\langle P \rangle}\), where \(\langle P \rangle\) represents the average value of the plaquette, and \(g\) is the bare gauge coupling. 
 
\subsection*{Continuum renormalization (\(\overline{\text{MS}}\) scheme)}
We report here our  result for the perturbative calculation performed in the continuum theory. We adopt dimensional regularization, and  the $\overline{\text{MS}}$ renormalization scheme.\footnote{The finite parts of the one-loop integrals for the local operator are scheme dependent. Here, we chose the dimensional regularization with $\overline{\text{MS}}$ scheme. Other choices are possible. One is the Breitenlohner–Maison–'t Hooft–Veltman (BMHV)~\cite{Breitenlohner:1975hg,Breitenlohner:1977hr,tHooft:1972tcz} renormalization scheme, which prescribes a non-standard definition of $\gamma_5$ while operating in generic $d$ dimensions.} 
Assuming the external momenta of the two $({\rm f})$-type fermions to approach the null value $p_1, p_2 \to 0$, the sum of the Feynman integrals for the vertex functions corresponding to the Feynman diagrams in Fig.~\ref{fig:combined_diagrams} is given by the following expressions, for $\Gamma_1 = \gamma_5$:
\begin{equation}
\begin{aligned}
\mathcal{I}[ \gamma_5 ] = \frac{g^2}{16\pi^2}   \gamma_5 \Bigg\{ &C^{\rm f} \left[ 6 + 4 \log\frac{\mu^2}{\Lambda^2}  + 4 \delta_{\overline{\text{MS}}} \right] \\
+ &C^{\rm f} (2-N_c) \left[ 12 + 8 \log\frac{\mu^2}{\Lambda^2} + 8 \delta_{\overline{\text{MS}}}  \right] \Bigg\} \,,
\end{aligned}
\end{equation}

where
\begin{equation}
    \delta_{\overline{\text{MS}}} = \dfrac{1}{\epsilon} +\log 4\pi - \gamma_E \,.
\end{equation}
For the chimera baryon operator with  $\Gamma_1 = \gamma_i$, we find
\begin{equation}
\begin{aligned}
\mathcal{I}[ \gamma_i ] = \frac{g^2}{16\pi^2}   \gamma_i \Bigg\{ &C^{\rm f} \left[ -\frac{1}{2} - \log\frac{\mu^2}{\Lambda^2}  - \delta_{\overline{\text{MS}}}\right] \\
+ &C^{\rm f}(2-N_c) \left[ 1 + 2\log\frac{\mu^2}{\Lambda^2} + 2 \delta_{\overline{\text{MS}}} \right] \Bigg\} \, .
\end{aligned}
\end{equation}



\subsection*{Lattice results}
In order to compute the finite part of the integrals in lattice perturbation theory, we make use of the multidimensional integrals Monte-Carlo evaluation Vegas package~\cite{peter_lepage_2025_14834979}. For $\Gamma_1 =  \gamma_5$, we arrive at the following expression:
\begin{equation}
L[ \gamma_5] = \frac{g^2}{16 \pi^2}  \gamma_5 \left[ -4 C^{\rm f} \log(a^2 \Lambda^2) +11.748\, C^{\rm f}  - 8 C^{\rm f} (2-N_c) \log(a^2 \Lambda^2) + 4.206\,C^{\rm f} (2-N_c)  \right]\,,
\end{equation}
where $\Lambda$ is an infrared regulator (a gluon mass) inserted in the lattice gluon propagators. \\
For $\Gamma_1 =  \gamma_\alpha$, our result is the following:
\begin{equation}
L[\gamma_i] = \frac{g^2}{16 \pi^2}   \gamma_i \, \left[ C^{\rm f} \left( \log(a^2 \Lambda^2) - 7.2644(11) \right) + C^{\rm f} (2-N_c) \left( -2 \log(a^2 \Lambda^2) + 4.8870(16) \right) \right]\,.
\end{equation}
As for the results in the continuum theory,  the finite parts depend on the regularization scheme chosen.

By matching the results of the continuum and of the lattice perturbative calculations~\cite{Martinelli:1982mw,Martinelli:1983be,Capitani:2002mp}, the matching coefficients at renormalization scale $\mu$ for the vertex function read:
\begin{eqnarray}
    \Delta_{\rm CB} [\gamma_5, \mu] = 5.74 C^{\rm f}  + 7.79 C^{\rm f} (2 - N_c)  + 8 C^{\rm f} \log(a \mu) + 16 C^{\rm f}(2-N_c) \log(a \mu) , \\
    \Delta_{\rm CB} [\gamma_i, \mu] = 6.72 C^{\rm f}  - 3.89 C^{\rm f} (2 - N_c)  - 2 C^{\rm f} \log(a \mu) + 4 C^{\rm f}(2-N_c) \log(a \mu) ,
\end{eqnarray}
and it is customary to choose the renormalization scale $\mu = 1/a$, where the logarithms vanish.  
Therefore, summing this contribution with the fermion self-energy contributions~\cite{Martinelli:1982mw}, the renormalization condition in $\overline{\text{MS}}$ scheme for the 4-point correlation function in Eq.~\eqref{eq:four_point_corr}, amputated, that reads as follows:
\begin{equation}
\Gamma_{\overline{\text{MS}}} = \left[ 1 + \frac{g^2}{16\pi^2} \left( \left[C^{\rm f} + \frac{1}{2} C^{\rm as} \right] \Delta_{\Sigma_1} + \Delta_{\rm CB}[\Gamma, \mu = 1/a] \right) \right] \Gamma_{\text{lattice}}\,,
\end{equation}
with 
 matching coefficients summarized in Eq.~\eqref{eq:matching_factors}.


\section{Tables}
\label{sec:tables}

We list here all our main numerical results, from the meson mass spectra in Tabs.~\ref{table:E1_results_ground_mesons}--\ref{table:E5_results_second_mesons}, to the baryon masses in Tabs.~\ref{table:E1_results_ground_CB}--\ref{table:E5_results_second_CB},
from the meson matrix elements in Tabs.~\ref{table:E1_matrix_mesons}--\ref{table:E5_matrix_mesons} to the chimera baryon matrix elements in Tabs.~\ref{table:E1_matrix_CB}--\ref{table:E5_matrix_CB}. The content of the tables is described in details in the captions. Missing entries denote cases in which the analysis did not yield a signal strong enough to perform the measurements.

\begin{table}
    \caption{\label{table:E1_results_ground_mesons} Numerical results for the ground state mass of the mesonic spectrum in ensemble M1. $k$-G stands for $k$-Gauss fit, ($k+1$)-G is $k+1$-Gauss fit, $k$-C stands for $k$-Cauchy function fit, ($k+1$)-G $k+1$-Cauchy function fit, $C$ indicates the mesonic channel considered, $am_C$ is the result of the GEVP analysis, $\sigma_G$ is the energy smearing radius used for the Gaussian fits, $\sigma_C$ for the Cauchy fit.}
    \centering
    \begin{tabular}{ |c|c|c|c|c|c|c|c|c|c|c| }
    \hline  \hline
    $C$ & $k$ & $N_{\rm source}$ & $N_{\rm sink}$ & $aE_0$ $k$-G & $aE_0$ ($k+1$)-G & $aE_0$ $k$-C & $aE_0$ ($k+1$)-C & $am_C$ & $\sigma_{G} / m_C$ & $\sigma_C / m_C$ \\
    \hline
PS & 3 & 80 & 80 & 0.3562(24) & 0.3562(36) & 0.3559(21) & 0.3559(21) & 0.3543(60) & 0.18 & 0.2 \\
V & 3 & 80 & 80 & 0.4019(34) & 0.40200(120) & 0.4051(19) & 0.4052(18) & 0.4018(90) & 0.2 & 0.26 \\
T & 3 & 80 & 80 & 0.4030(36) & 0.40300(120) & 0.4002(28) & 0.3999(22) & 0.4016(33) & 0.2 & 0.26 \\
AV & 2 & 80 & 80 & 0.5337(45) & 0.5335(53) & 0.5411(13) & 0.5374(18) & 0.5353(73) & 0.2 & 0.18 \\
AT & 3 & 80 & 80 & 0.5293(26) & 0.5292(53) & 0.5294(23) & 0.5300(24) & 0.525(15) & 0.18 & 0.18 \\
S & 3 & 80 & 80 & 0.5137(44) & 0.5136(51) & 0.5201(52) & 0.52000(90) & 0.5145(87) & 0.18 & 0.2 \\
\hline
ps & 3 & 80 & 80 & 0.5877(15) & 0.5877(59) & 0.5898(33) & 0.5897(33) & 0.5853(18) & 0.18 & 0.2 \\
v & 3 & 80 & 80 & 0.6467(54) & 0.6463(65) & 0.6386(64) & 0.6386(64) & 0.6447(18) & 0.2 & 0.26 \\
t & 3 & 80 & 80 & 0.6437(42) & 0.6443(64) & 0.6388(17) & 0.6421(42) & 0.6447(60) & 0.2 & 0.26 \\
av & 2 & 80 & 80 & 0.8342(46) & 0.8343(83) & 0.8362(21) & 0.8361(30) & 0.828(30) & 0.2 & 0.18 \\
at & 3 & 80 & 80 & 0.8476(61) & 0.8476(85) & 0.8365(64) & 0.8383(75) & 0.8420(97) & 0.18 & 0.18 \\
s & 3 & 80 & 80 & 0.7946(49) & 0.7951(24) & 0.7998(21) & 0.7992(23) & 0.7923(51) & 0.18 & 0.2 \\
\hline \hline
\end{tabular}
\end{table}

\begin{table}    
    \centering
    \caption{\label{table:E1_results_first_mesons} Numerical results for the first excited  state mass of the mesonic spectrum in ensemble M1. $k$-G stands for $k$-Gauss fit, ($k+1$)-G is $k+1$-Gauss fit, $k$-C stands for $k$-Cauchy function fit, ($k+1$)-G $k+1$-Cauchy function fit, $C$ indicates the mesonic channel considered, $am_C$ is the result of the GEVP analysis, $\sigma_G$ is the energy smearing radius used for the Gaussian fits, $\sigma_C$ for the Cauchy fit.}
    \begin{tabular}{ |c|c|c|c|c|c|c|c|c|c|c| }
    \hline  \hline
    $C$ & $k$ & $N_{\rm source}$ & $N_{\rm sink}$ & $aE_1$ $k$-G & $aE_1$ ($k+1$)-G & $aE_1$ $k$-C & $aE_1$ ($k+1$)-C & $am_C$ & $\sigma_{G} / m_C$ & $\sigma_C / m_C$ \\
    \hline
PS & 3 & 80 & 80 & 0.6775(98) & 0.6781(68) & 0.672(13) & 0.672(13) & 0.675(60) & 0.18 & 0.2 \\
V & 3 & 80 & 80 & 0.7114(14) & 0.7114(71) & 0.6957(15) & 0.6957(24) & 0.703(18) & 0.2 & 0.26 \\
T & 3 & 80 & 80 & 0.7034(16) & 0.7033(21) & 0.693(14) & 0.691(12) & 0.697(22) & 0.2 & 0.26 \\
AV & 2 & 80 & 80 & 0.709(14) & 0.7088(71) & 0.6977(98) & 0.706(12) & 0.703(75) & 0.2 & 0.18 \\
AT & 3 & 80 & 80 & 0.710(17) & 0.7097(71) & 0.69300(30) & 0.6936(34) & 0.707(44) & 0.18 & 0.18 \\
S & 3 & 80 & 80 & 0.719(17) & 0.7190(72) & 0.721(60) & 0.7288(78) & 0.722(63) & 0.18 & 0.2 \\
\hline
ps & 3 & 80 & 80 & 0.903(13) & 0.903(17) & 0.918(60) & 0.918(19) & 0.914(24) & 0.18 & 0.2 \\
v & 3 & 80 & 80 & 0.942(18) & 0.9427(94) & 0.9353(94) & 0.9353(94) & 0.949(14) & 0.2 & 0.26 \\
t & 3 & 80 & 80 & 0.95450(30) & 0.9545(95) & 0.950(18) & 0.947(21) & 0.942(17) & 0.2 & 0.26 \\
av & 2 & 80 & 80 & 1.079(16) & 1.079(11) & 1.083(26) & 1.080(21) & 1.092(28) & 0.2 & 0.18 \\
at & 3 & 80 & 80 & 1.0881(90) & 1.088(11) & 1.111(11) & 1.111(11) & 1.097(59) & 0.18 & 0.18 \\
s & 3 & 80 & 80 & 1.047(36) & 1.047(30) & 1.077(11) & 1.077(15) & 1.062(27) & 0.18 & 0.2 \\
\hline \hline
\end{tabular}
\end{table}

\begin{table}    
    \centering
    \caption{\label{table:E1_results_second_mesons} Numerical results for the second excited state mass of the mesonic spectrum in ensemble M1.  $k$-G stands for $k$-Gauss fit, ($k+1$)-G is $k+1$-Gauss fit, $k$-C stands for $k$-Cauchy function fit, ($k+1$)-G $k+1$-Cauchy function fit, $C$ indicates the mesonic channel considered, $am_C$ is the result of the GEVP analysis, $\sigma_G$ is the energy smearing radius used for the Gaussian fits, $\sigma_C$ for the Cauchy fit.}
    \begin{tabular}{ |c|c|c|c|c|c|c|c|c|c|c| }
    \hline  \hline
    $C$ & $k$ & $N_{\rm source}$ & $N_{\rm sink}$ & $aE_2$ $k$-G & $aE_2$ ($k+1$)-G & $aE_2$ $k$-C & $aE_2$ ($k+1$)-C & $am_C$ & $\sigma_{G} / m_C$ & $\sigma_C / m_C$ \\
    \hline
PS & 3 & 80 & 80 & - & - & - & - & 1.09(17) & 0.18 & 0.2 \\
V & 3 & 80 & 80 & 0.948(14) & 0.9480(95) & 1.040(31) & 1.046(25) & 1.02(60) & 0.2 & 0.26 \\
T & 3 & 80 & 80 & 0.935(17) & 0.9343(93) & 0.975(13) & 0.9723(73) & 0.96(19) & 0.2 & 0.26 \\
AV & 2 & 80 & 80 & - & - & - & - & 1.03(16) & 0.2 & 0.18 \\
AT & 3 & 80 & 80 & - & - & - & - & 1.02(19) & 0.18 & 0.18 \\
S & 3 & 80 & 80 & - & - & - & - & - & 0.18 & 0.2 \\
\hline
ps & 3 & 80 & 80 & 1.178(42) & 1.178(12) & 1.232(63) & 1.245(59) & 1.21(12) & 0.18 & 0.2 \\
v & 2 & 80 & 80 & - & - & - & - & - & 0.2 & 0.26 \\
t & 2 & 80 & 80 & - & - & - & - & - & 0.2 & 0.26 \\
av & 2 & 80 & 80 & - & - & - & - & - & 0.2 & 0.18 \\
at & 3 & 80 & 80 & 1.567(54) & 1.571(16) & 1.591(16) & 1.591(16) & 1.56(11) & 0.18 & 0.18 \\
s & 3 & 80 & 80 & 1.464(14) & 1.464(15) & 1.504(36) & 1.497(15) & 1.479(75) & 0.18 & 0.2 \\
\hline \hline
\end{tabular}
\end{table}

\begin{table}    
    \centering
    \caption{\label{table:E2_results_ground_mesons} Numerical results for the ground state mass of the mesonic spectrum in ensemble M2.  $k$-G stands for $k$-Gauss fit, ($k+1$)-G is $k+1$-Gauss fit, $k$-C stands for $k$-Cauchy function fit, ($k+1$)-G $k+1$-Cauchy function fit, $C$ indicates the mesonic channel considered, $am_C$ is the result of the GEVP analysis, $\sigma_G$ is the energy smearing radius used for the Gaussian fits, $\sigma_C$ for the Cauchy fit.}
    \begin{tabular}{ |c|c|c|c|c|c|c|c|c|c|c| }
    \hline  \hline
    $C$ & $k$ & $N_{\rm source}$ & $N_{\rm sink}$ & $aE_0$ $k$-G & $aE_0$ ($k+1$)-G & $aE_0$ $k$-C & $aE_0$ ($k+1$)-C & $am_C$ & $\sigma_{G} / m_C$ & $\sigma_C / m_C$ \\
    \hline
PS & 3 & 80 & 80 & 0.3569(29) & 0.3570(36) & 0.3543(90) & 0.3542(27) & 0.3560(19) & 0.18 & 0.18 \\
V & 3 & 80 & 80 & 0.3992(16) & 0.3992(12) & 0.3920(39) & 0.3920(39) & 0.3965(27) & 0.2 & 0.2 \\
T & 3 & 80 & 80 & 0.3939(60) & 0.3938(39) & 0.4010(12) & 0.4010(12) & 0.3965(29) & 0.2 & 0.2 \\
AV & 3 & 80 & 80 & 0.5428(36) & 0.5431(54) & 0.5475(25) & 0.5476(24) & 0.5430(21) & 0.18 & 0.18 \\
AT & 3 & 80 & 80 & 0.5405(39) & 0.5409(54) & 0.5484(55) & 0.5484(55) & 0.5424(66) & 0.18 & 0.23 \\
S & 3 & 80 & 80 & 0.5262(42) & 0.5255(53) & 0.5295(24) & 0.5295(24) & 0.5248(66) & 0.18 & 0.18 \\
\hline
ps & 3 & 80 & 80 & 0.5952(58) & 0.5954(18) & 0.59980(30) & 0.5948(21) & 0.5940(13) & 0.18 & 0.18 \\
v & 3 & 80 & 80 & 0.6367(62) & 0.6377(64) & 0.6444(64) & 0.6381(18) & 0.6379(22) & 0.2 & 0.2 \\
t & 3 & 80 & 80 & 0.6345(51) & 0.6349(63) & 0.6444(64) & 0.6444(64) & 0.6374(24) & 0.2 & 0.2 \\
av & 3 & 80 & 80 & 0.8341(53) & 0.8346(83) & 0.8433(84) & 0.8433(84) & 0.8351(74) & 0.18 & 0.18 \\
at & 3 & 80 & 80 & 0.8437(61) & 0.8433(84) & 0.8524(85) & 0.8524(85) & 0.8442(83) & 0.18 & 0.23 \\
s & 3 & 80 & 80 & 0.7914(39) & 0.7927(79) & 0.8006(42) & 0.8012(47) & 0.7985(21) & 0.18 & 0.18 \\
\hline \hline
\end{tabular}
\end{table}

\begin{table}    
    \centering
    \caption{\label{table:E2_results_first_mesons} Numerical results for the first excited  state mass  of the mesonic spectrum in ensemble M2.  $k$-G stands for $k$-Gauss fit, ($k+1$)-G is $k+1$-Gauss fit, $k$-C stands for $k$-Cauchy function fit, ($k+1$)-G $k+1$-Cauchy function fit, $C$ indicates the mesonic channel considered, $am_C$ is the result of the GEVP analysis, $\sigma_G$ is the energy smearing radius used for the Gaussian fits, $\sigma_C$ for the Cauchy fit.}
    \begin{tabular}{ |c|c|c|c|c|c|c|c|c|c|c| }
    \hline  \hline
    $C$ & $k$ & $N_{\rm source}$ & $N_{\rm sink}$ & $aE_1$ $k$-G & $aE_1$ ($k+1$)-G & $aE_1$ $k$-C & $aE_1$ ($k+1$)-C & $am_C$ & $\sigma_{G} / m_C$ & $\sigma_C / m_C$ \\
    \hline
PS & 3 & 80 & 80 & 0.6725(91) & 0.6731(67) & 0.679(11) & 0.679(12) & 0.677(15) & 0.18 & 0.18 \\
V & 3 & 80 & 80 & 0.6637(81) & 0.6635(66) & 0.6727(81) & 0.6737(75) & 0.669(36) & 0.2 & 0.2 \\
T & 3 & 80 & 80 & 0.6769(96) & 0.6775(68) & 0.6905(41) & 0.6904(12) & 0.683(23) & 0.2 & 0.2 \\
AV & 3 & 80 & 80 & 0.791(18) & 0.7911(79) & 0.786(15) & 0.791(18) & 0.792(21) & 0.18 & 0.18 \\
AT & 3 & 80 & 80 & 0.802(13) & 0.8022(24) & 0.7984(35) & 0.7992(85) & 0.808(26) & 0.18 & 0.23 \\
S & 3 & 80 & 80 & 0.761(14) & 0.7648(76) & 0.7590(96) & 0.7593(99) & 0.765(26) & 0.18 & 0.18 \\
\hline
ps & 3 & 80 & 80 & 0.8971(33) & 0.8970(27) & 0.8969(27) & 0.9071(84) & 0.908(22) & 0.18 & 0.18 \\
v & 3 & 80 & 80 & 0.942(12) & 0.9415(94) & 0.9216(77) & 0.9314(21) & 0.930(19) & 0.2 & 0.2 \\
t & 3 & 80 & 80 & 0.938(19) & 0.9378(94) & 0.9442(94) & 0.9442(94) & 0.930(24) & 0.2 & 0.2 \\
av & 3 & 80 & 80 & 1.069(11) & 0.9378(94) & 1.055(33) & 1.052(33) & 1.055(43) & 0.18 & 0.18 \\
at & 3 & 80 & 80 & 1.1058(29) & 1.106(11) & 1.106(11) & 1.106(11) & 1.12(12) & 0.18 & 0.23 \\
s & 3 & 80 & 80 & 1.044(13) & 1.040(30) & 1.017(30) & 1.018(14) & 1.034(77) & 0.18 & 0.18 \\
\hline \hline
\end{tabular}
\end{table}

\begin{table}    
    \centering
    \caption{\label{table:E2_results_second_mesons} Numerical results for the second excited state mass of the mesonic spectrum in ensemble M2.  $k$-G stands for $k$-Gauss fit, ($k+1$)-G is $k+1$-Gauss fit, $k$-C stands for $k$-Cauchy function fit, ($k+1$)-G $k+1$-Cauchy function fit, $C$ indicates the mesonic channel considered, $am_C$ is the result of the GEVP analysis, $\sigma_G$ is the energy smearing radius used for the Gaussian fits, $\sigma_C$ for the Cauchy fit.}
    \begin{tabular}{ |c|c|c|c|c|c|c|c|c|c|c| }
    \hline  \hline
    $C$ & $k$ & $N_{\rm source}$ & $N_{\rm sink}$ & $aE_2$ $k$-G & $aE_2$ ($k+1$)-G & $aE_2$ $k$-C & $aE_2$ ($k+1$)-C & $am_C$ & $\sigma_{G} / m_C$ & $\sigma_C / m_C$ \\
    \hline
PS & 3 & 80 & 80 & - & - & - & - & - & 0.18 & 0.18 \\
V & 3 & 80 & 80 & 0.969(24) & 0.9706(97) & 0.924(22) & 0.9136(99) & - & 0.2 & 0.2 \\
T & 3 & 80 & 80 & 1.012(22) & 1.012(30) & 1.021(12) & 1.017(13) & - & 0.2 & 0.2 \\
AV & 3 & 80 & 80 & 1.169(33) & 1.171(12) & 1.187(26) & 1.196(22) & - & 0.18 & 0.18 \\
AT & 3 & 80 & 80 & 1.105(37) & 1.105(11) & 1.075(29) & 1.068(23) & - & 0.18 & 0.23 \\
S & 3 & 80 & 80 & - & - & - & - & - & 0.18 & 0.18 \\
\hline
ps & 3 & 80 & 80 & 1.176(24) & 1.178(12) & 1.348(13) & 1.272(24) & 1.267(74) & 0.18 & 0.18 \\
v & 2 & 80 & 80 & - & - & - & - & - & 0.2 & 0.2 \\
t & 2 & 80 & 80 & - & - & - & - & - & 0.2 & 0.2 \\
av & 3 & 80 & 80 & 1.533(14) & 1.533(15) & 1.5528(60) & 1.533(11) & 1.529(24) & 0.18 & 0.18 \\
at & 3 & 80 & 80 & 1.513(89) & 1.514(15) & 1.511(15) & 1.486(21) & 1.58(12) & 0.18 & 0.23 \\
s & 3 & 80 & 80 & 1.466(12) & 1.466(15) & 1.435(11) & 1.438(12) & 1.463(24) & 0.18 & 0.18 \\ 
\hline \hline
\end{tabular}
\end{table}

\begin{table}    
    \centering
    \caption{\label{table:E3_results_ground_mesons} Numerical results for the ground state mass of the mesonic spectrum in ensemble M3.  $k$-G stands for $k$-Gauss fit, ($k+1$)-G is $k+1$-Gauss fit, $k$-C stands for $k$-Cauchy function fit, ($k+1$)-G $k+1$-Cauchy function fit, $C$ indicates the mesonic channel considered, $am_C$ is the result of the GEVP analysis, $\sigma_G$ is the energy smearing radius used for the Gaussian fits, $\sigma_C$ for the Cauchy fit.}
    \begin{tabular}{ |c|c|c|c|c|c|c|c|c|c|c| }
    \hline  \hline
    $C$ & $k$ & $N_{\rm source}$ & $N_{\rm sink}$ & $aE_0$ $k$-G & $aE_0$ ($k+1$)-G & $aE_0$ $k$-C & $aE_0$ ($k+1$)-C & $am_C$ & $\sigma_{G} / m_C$ & $\sigma_C / m_C$ \\
    \hline
PS & 3 & 80 & 80 & 0.3608(25) & 0.3602(36) & 0.3626(37) & 0.3626(60) & 0.3624(15) & 0.23 & 0.22 \\
V & 3 & 80 & 80 & 0.4008(27) & 0.4009(12) & 0.4025(24) & 0.4021(16) & 0.4027(18) & 0.24 & 0.25 \\
T & 3 & 80 & 80 & 0.4002(34) & 0.4002(12) & 0.3981(18) & 0.39810(30) & 0.4025(18) & 0.28 & 0.28 \\
AV & 3 & 80 & 80 & 0.5346(53) & 0.4002(12) & 0.5429(28) & 0.5431(28) & 0.5393(66) & 0.18 & 0.25 \\
AT & 3 & 80 & 80 & 0.5376(15) & 0.5377(54) & 0.5476(60) & 0.5477(21) & 0.5431(62) & 0.25 & 0.25 \\
S & 3 & 80 & 80 & 0.5181(39) & 0.5170(52) & 0.5210(27) & 0.5210(21) & 0.5159(68) & 0.23 & 0.23 \\
\hline
ps & 3 & 80 & 80 & 0.6016(53) & 0.6009(18) & 0.5940(59) & 0.5940(59) & 0.59973(87) & 0.23 & 0.22 \\
v & 3 & 80 & 80 & 0.6386(64) & 0.6009(18) & 0.6514(65) & 0.6514(65) & 0.6454(17) & 0.24 & 0.25 \\
t & 3 & 80 & 80 & 0.6405(64) & 0.6009(18) & 0.6532(13) & 0.6522(29) & 0.6469(22) & 0.28 & 0.28 \\
av & 3 & 80 & 80 & 0.8268(12) & 0.8268(83) & 0.8335(59) & 0.8329(58) & 0.836(30) & 0.18 & 0.25 \\
at & 3 & 80 & 80 & 0.84620(90) & 0.8460(85) & 0.8296(83) & 0.8296(30) & 0.838(11) & 0.25 & 0.25 \\
s & 3 & 80 & 80 & 0.8057(41) & 0.8054(81) & 0.8100(81) & 0.8100(81) & 0.8024(87) & 0.23 & 0.23 \\
\hline \hline
\end{tabular}
\end{table}
    
\begin{table}    
    \centering
    \caption{\label{table:E3_results_first_mesons} Numerical results for the first excited  state mass of the mesonic spectrum in ensemble M3.  $k$-G stands for $k$-Gauss fit, ($k+1$)-G is $k+1$-Gauss fit, $k$-C stands for $k$-Cauchy function fit, ($k+1$)-G $k+1$-Cauchy function fit, $C$ indicates the mesonic channel considered, $am_C$ is the result of the GEVP analysis, $\sigma_G$ is the energy smearing radius used for the Gaussian fits, $\sigma_C$ for the Cauchy fit.}
    \begin{tabular}{ |c|c|c|c|c|c|c|c|c|c|c| }
    \hline  \hline
    $C$ & $k$ & $N_{\rm source}$ & $N_{\rm sink}$ & $aE_1$ $k$-G & $aE_1$ ($k+1$)-G & $aE_1$ $k$-C & $aE_1$ ($k+1$)-C & $am_C$ & $\sigma_{G} / m_C$ & $\sigma_C / m_C$ \\
    \hline
PS & 3 & 80 & 80 & 0.6485(48) & 0.6521(65) & 0.655(14) & 0.6499(76) & 0.656(16) & 0.23 & 0.22 \\
V & 3 & 80 & 80 & 0.6822(72) & 0.6822(68) & 0.6815(35) & 0.6812(60) & 0.690(12) & 0.24 & 0.25 \\
T & 3 & 80 & 80 & 0.6809(69) & 0.6808(68) & 0.684(11) & 0.684(30) & 0.690(14) & 0.28 & 0.28 \\
AV & 3 & 80 & 80 & 0.7668(77) & 0.6808(68) & 0.7801(68) & 0.7796(63) & 0.778(17) & 0.18 & 0.25 \\
AT & 3 & 80 & 80 & 0.7775(65) & 0.7774(78) & 0.7850(94) & 0.7849(89) & 0.789(18) & 0.25 & 0.25 \\
S & 3 & 80 & 80 & 0.7571(96) & 0.7613(76) & 0.7543(45) & 0.7543(51) & 0.766(60) & 0.23 & 0.23 \\
\hline
ps & 3 & 80 & 80 & 0.9064(34) & 0.9064(91) & 0.9060(91) & 0.9060(91) & 0.9172(74) & 0.23 & 0.22 \\
v & 3 & 80 & 80 & 0.9224(92) & 0.9064(91) & 0.9224(92) & 0.9224(92) & 0.937(14) & 0.24 & 0.25 \\
t & 3 & 80 & 80 & 0.9187(92) & 0.9064(91) & 0.9194(74) & 0.9198(61) & 0.934(19) & 0.28 & 0.28 \\
av & 3 & 80 & 80 & 1.1019(49) & 1.102(11) & 1.0692(37) & 1.0690(60) & 1.088(14) & 0.18 & 0.25 \\
at & 3 & 80 & 80 & 1.113(30) & 1.113(11) & 1.081(11) & 1.0810(60) & 1.096(60) & 0.25 & 0.25 \\
s & 3 & 80 & 80 & 1.0506(15) & 1.051(11) & 1.083(11) & 1.083(11) & 1.062(16) & 0.23 & 0.23 \\
\hline \hline
\end{tabular}
\end{table}

\begin{table}    
    \centering
    \caption{\label{table:E3_results_second_mesons} Numerical results for the second excited state mass of the mesonic spectrum in ensemble M3.  $k$-G stands for $k$-Gauss fit, ($k+1$)-G is $k+1$-Gauss fit, $k$-C stands for $k$-Cauchy function fit, ($k+1$)-G $k+1$-Cauchy function fit, $C$ indicates the mesonic channel considered, $am_C$ is the result of the GEVP analysis, $\sigma_G$ is the energy smearing radius used for the Gaussian fits, $\sigma_C$ for the Cauchy fit.}
    \begin{tabular}{ |c|c|c|c|c|c|c|c|c|c|c| }
    \hline  \hline
    $C$ & $k$ & $N_{\rm source}$ & $N_{\rm sink}$ & $aE_2$ $k$-G & $aE_2$ ($k+1$)-G & $aE_2$ $k$-C & $aE_2$ ($k+1$)-C & $am_C$ & $\sigma_{G} / m_C$ & $\sigma_C / m_C$ \\
    \hline
PS & 3 & 80 & 80 & - & - & - & - & - & 0.23 & 0.22 \\
V & 3 & 80 & 80 & 0.992(27) & 0.9927(99) & 1.006(13) & 1.003(14) & 1.049(32) & 0.24 & 0.25 \\
T & 3 & 80 & 80 & 0.989(21) & 0.9888(99) & 0.934(30) & 0.935(11) & 1.057(48) & 0.28 & 0.28 \\
AV & 3 & 80 & 80 & 1.165(15) & 1.166(12) & 1.162(17) & 1.158(14) & - & 0.18 & 0.25 \\
AT & 3 & 80 & 80 & 1.107(25) & 1.105(11) & 1.027(21) & 1.029(22) & 1.085(73) & 0.25 & 0.25 \\
S & 3 & 80 & 80 & - & - & - & - & - & 0.23 & 0.23 \\
\hline
ps & 3 & 80 & 80 & 1.170(12) & - & 1.170(12) & 1.170(12) & 1.264(49) & 0.23 & 0.22 \\
v & 3 & 80 & 80 & - & - & - & - & - & 0.24 & 0.25 \\
t & 3 & 80 & 80 & - & - & - & - & - & 0.28 & 0.28 \\
av & 3 & 80 & 80 & 1.517(14) & 1.518(15) & 1.426(66) & 1.468(61) & - & 0.18 & 0.25 \\
at & 3 & 80 & 80 & 1.484(13) & 1.488(15) & 1.534(15) & 1.531(29) & - & 0.25 & 0.25 \\
s & 3 & 80 & 80 & 1.490(11) & 1.491(15) & 1.492(15) & 1.490(33) & - & 0.23 & 0.23 \\
\hline \hline
\end{tabular}
\end{table}

\begin{table}    
    \centering
    \caption{\label{table:E4_results_ground_mesons} Numerical results for the ground state mass in ensemble M4.  $k$-G stands for $k$-Gauss fit, ($k+1$)-G is $k+1$-Gauss fit, $k$-C stands for $k$-Cauchy function fit, ($k+1$)-G $k+1$-Cauchy function fit, $C$ indicates the mesonic channel considered, $am_C$ is the result of the GEVP analysis, $\sigma_G$ is the energy smearing radius used for the Gaussian fits, $\sigma_C$ for the Cauchy fit.}
    \begin{tabular}{ |c|c|c|c|c|c|c|c|c|c|c| }
    \hline  \hline
    $C$ & $k$ & $N_{\rm source}$ & $N_{\rm sink}$ & $aE_0$ $k$-G & $aE_0$ ($k+1$)-G & $aE_0$ $k$-C & $aE_0$ ($k+1$)-C & $am_C$ & $\sigma_{G} / m_C$ & $\sigma_C / m_C$ \\
    \hline
PS & 3 & 80 & 80 & 0.4097(32) & 0.4094(41) & 0.4071(37) & 0.4071(37) & 0.4081(25) & 0.24 & 0.24 \\
V & 3 & 80 & 80 & 0.4425(44) & 0.4094(41) & 0.4472(38) & 0.4469(37) & 0.4474(35) & 0.23 & 0.2 \\
T & 3 & 80 & 80 & 0.4396(33) & 0.4399(44) & 0.4448(17) & 0.4449(16) & 0.4413(24) & 0.23 & 0.25 \\
AV & 2 & 80 & 80 & 0.5885(54) & 0.5889(59) & 0.5913(12) & 0.5906(39) & 0.5863(27) & 0.2 & 0.2 \\
AT & 2 & 80 & 80 & 0.5882(27) & 0.5925(59) & 0.5988(24) & 0.5988(12) & 0.5930(77) & 0.25 & 0.25 \\
S & 2 & 80 & 80 & 0.5598(28) & 0.5620(56) & 0.5642(43) & 0.5640(34) & 0.56388(27) & 0.24 & 0.2 \\
\hline
ps & 3 & 80 & 80 & 0.6234(15) & 0.6234(62) & 0.6178(62) & 0.6178(62) & 0.62363(95) & 0.24 & 0.24 \\
v & 3 & 80 & 80 & 0.6766(30) & 0.6767(68) & 0.6742(47) & 0.6739(41) & 0.6698(13) & 0.23 & 0.2 \\
t & 3 & 80 & 80 & 0.67670(30) & 0.6767(68) & 0.6757(34) & 0.6767(68) & 0.6697(14) & 0.23 & 0.25 \\
av & 2 & 80 & 80 & 0.87360(90) & 0.8736(87) & 0.8604(18) & 0.8673(43) & 0.8651(72) & 0.2 & 0.2 \\
at & 2 & 80 & 80 & 0.8794(39) & 0.8803(88) & 0.8807(88) & 0.8807(88) & 0.872(13) & 0.25 & 0.25 \\
s & 2 & 80 & 80 & 0.8385(24) & 0.8387(84) & 0.8354(57) & 0.8353(56) & 0.8307(66) & 0.24 & 0.2 \\
\hline \hline
\end{tabular}
\end{table}

\begin{table}    
    \centering
    \caption{\label{table:E4_results_first_mesons} Numerical results for the first excited state mass of the mesonic spectrum in ensemble M4. $k$-G stands for $k$-Gauss fit, ($k+1$)-G is $k+1$-Gauss fit, $k$-C stands for $k$-Cauchy function fit, ($k+1$)-G $k+1$-Cauchy function fit, $C$ indicates the mesonic channel considered, $am_C$ is the result of the GEVP analysis, $\sigma_G$ is the energy smearing radius used for the Gaussian fits, $\sigma_C$ for the Cauchy fit.}
    \begin{tabular}{ |c|c|c|c|c|c|c|c|c|c|c| }
    \hline  \hline
    $C$ & $k$ & $N_{\rm source}$ & $N_{\rm sink}$ & $aE_1$ $k$-G & $aE_1$ ($k+1$)-G & $aE_1$ $k$-C & $aE_1$ ($k+1$)-C & $am_C$ & $\sigma_{G} / m_C$ & $\sigma_C / m_C$ \\
    \hline
PS & 3 & 80 & 80 & 0.618(11) & 0.6159(62) & 0.613(15) & 0.613(15) & 0.615(47) & 0.24 & 0.24 \\
V & 3 & 80 & 80 & 0.6929(69) & 0.6159(62) & 0.705(12) & 0.705(12) & 0.700(32) & 0.23 & 0.2 \\
T & 3 & 80 & 80 & 0.726(11) & 0.7257(73) & 0.7251(88) & 0.7250(85) & 0.731(60) & 0.23 & 0.25 \\
AV & 2 & 80 & 80 & 0.787(12) & 0.7780(78) & 0.785(11) & 0.7861(62) & 0.782(55) & 0.2 & 0.2 \\
AT & 2 & 80 & 80 & 0.838(18) & 0.8418(84) & 0.847(16) & 0.845(12) & 0.843(29) & 0.25 & 0.25 \\
S & 2 & 80 & 80 & 0.743(15) & 0.7403(74) & 0.739(16) & 0.739(12) & - & 0.24 & 0.2 \\
\hline
ps & 3 & 80 & 80 & 0.932(11) & 0.9326(93) & 0.9298(93) & 0.9298(93) & 0.941(24) & 0.24 & 0.24 \\
v & 3 & 80 & 80 & 0.962(26) & 0.9613(96) & 0.9514(95) & 0.9514(95) & 0.963(27) & 0.23 & 0.2 \\
t & 3 & 80 & 80 & 0.9447(21) & 0.9447(94) & 0.9453(18) & 0.952(21) & 0.955(35) & 0.23 & 0.25 \\
av & 2 & 80 & 80 & 1.104(24) & 1.104(11) & 1.112(28) & 1.117(60) & 1.113(15) & 0.2 & 0.2 \\
at & 2 & 80 & 80 & 1.115(90) & 1.120(11) & 1.0925(62) & 1.111(13) & 1.101(91) & 0.25 & 0.25 \\
s & 2 & 80 & 80 & 1.060(19) & 1.060(11) & 1.057(14) & 1.057(14) & 1.068(59) & 0.24 & 0.2 \\
\hline \hline
\end{tabular}
\end{table}

\begin{table}    
    \centering
    \caption{\label{table:E4_results_second_mesons} Numerical results for the second excited state mass of the mesonic spectrum in ensemble M4.  $k$-G stands for $k$-Gauss fit, ($k+1$)-G is $k+1$-Gauss fit, $k$-C stands for $k$-Cauchy function fit, ($k+1$)-G $k+1$-Cauchy function fit, $C$ indicates the mesonic channel considered, $am_C$ is the result of the GEVP analysis, $\sigma_G$ is the energy smearing radius used for the Gaussian fits, $\sigma_C$ for the Cauchy fit.}
    \begin{tabular}{ |c|c|c|c|c|c|c|c|c|c|c| }
    \hline  \hline
    $C$ & $k$ & $N_{\rm source}$ & $N_{\rm sink}$ & $aE_2$ $k$-G & $aE_2$ ($k+1$)-G & $aE_2$ $k$-C & $aE_2$ ($k+1$)-C & $am_C$ & $\sigma_{G} / m_C$ & $\sigma_C / m_C$ \\
    \hline
PS & 3 & 80 & 80 & - & - & - & - & - & 0.24 & 0.24 \\
V & 3 & 80 & 80 & 1.016(14) & 1.016(30) & 1.000(29) & 1.003(28) & 0.978(96) & 0.23 & 0.2 \\
T & 3 & 80 & 80 & 1.008(43) & 1.001(30) & 0.998(31) & 0.999(27) & 1.00(30) & 0.23 & 0.25 \\
AV & 2 & 80 & 80 & - & - & - & - & 1.16(60) & 0.2 & 0.2 \\
AT & 2 & 80 & 80 & - & - & - & - & - & 0.25 & 0.25 \\
S & 2 & 80 & 80 & - & - & - & - & - & 0.24 & 0.2 \\
\hline
ps & 3 & 80 & 80 & 1.207(57) & 1.210(12) & 1.224(44) & 1.157(36) & 1.18(12) & 0.24 & 0.24 \\
v & 3 & 80 & 80 & 1.200(36) & 1.220(30) & 1.218(57) & 1.218(61) & 1.21(34) & 0.23 & 0.2 \\
t & 3 & 80 & 80 & 1.204(25) & 1.204(12) & 1.260(11) & 1.284(21) & 1.21(39) & 0.23 & 0.25 \\
av & 2 & 80 & 80 & - & - & - & - & - & 0.2 & 0.2 \\
at & 2 & 80 & 80 & - & - & - & - & - & 0.25 & 0.25 \\
s & 2 & 80 & 80 & - & - & - & - & - & 0.24 & 0.2 \\
\hline \hline
\end{tabular}
\end{table}

\begin{table}    
    \centering
    \caption{\label{table:E5_results_ground_mesons} Numerical results for the ground state mass of the mesonic spectrum in ensemble M5.  $k$-G stands for $k$-Gauss fit, ($k+1$)-G is $k+1$-Gauss fit, $k$-C stands for $k$-Cauchy function fit, ($k+1$)-G $k+1$-Cauchy function fit, $C$ indicates the mesonic channel considered, $am_C$ is the result of the GEVP analysis, $\sigma_G$ is the energy smearing radius used for the Gaussian fits, $\sigma_C$ for the Cauchy fit.}
    \begin{tabular}{ |c|c|c|c|c|c|c|c|c|c|c| }
    \hline  \hline
    $C$ & $k$ & $N_{\rm source}$ & $N_{\rm sink}$ & $aE_0$ $k$-G & $aE_0$ ($k+1$)-G & $aE_0$ $k$-C & $aE_0$ ($k+1$)-C & $am_C$ & $\sigma_{G} / m_C$ & $\sigma_C / m_C$ \\
    \hline
PS & 2 & 80 & 80 & 0.2941(23) & 0.2937(29) & 0.2933(27) & 0.2933(27) & 0.2939(18) & 0.2 & 0.2 \\
V & 3 & 80 & 80 & 0.3577(25) & 0.3576(36) & 0.3528(14) & 0.3525(15) & 0.3361(28) & 0.2 & 0.25 \\
T & 3 & 80 & 80 & 0.3432(24) & 0.3433(34) & 0.3438(15) & 0.3438(14) & 0.3418(24) & 0.2 & 0.25 \\
AV & 2 & 80 & 80 & 0.5121(26) & 0.5087(51) & 0.5131(51) & 0.5130(30) & 0.5077(81) & 0.2 & 0.2 \\
AT & 2 & 80 & 80 & 0.5109(26) & 0.5110(51) & 0.5054(38) & 0.5046(35) & 0.506(11) & 0.2 & 0.2 \\
S & 2 & 80 & 80 & 0.4832(16) & 0.4831(48) & 0.4810(39) & 0.4799(44) & 0.478(30) & 0.2 & 0.2 \\
\hline
ps & 2 & 80 & 80 & 0.5677(35) & 0.5676(57) & 0.5676(36) & 0.5679(39) & 0.5724(13) & 0.2 & 0.2 \\
v & 3 & 80 & 80 & 0.6259(32) & 0.6259(63) & 0.6148(61) & 0.6181(14) & 0.6208(17) & 0.2 & 0.25 \\
t & 3 & 80 & 80 & 0.6257(32) & 0.6250(63) & 0.6272(63) & 0.62720(30) & 0.6208(19) & 0.2 & 0.25 \\
av & 2 & 80 & 80 & 0.7954(62) & 0.7954(24) & 0.7973(55) & 0.7972(51) & 0.7987(27) & 0.2 & 0.2 \\
at & 2 & 80 & 80 & 0.8128(17) & 0.8127(81) & 0.8125(15) & 0.8097(13) & 0.806(14) & 0.2 & 0.2 \\
s & 2 & 80 & 80 & 0.7687(53) & 0.7686(77) & 0.7564(76) & 0.7566(24) & 0.7646(68) & 0.2 & 0.2 \\
\hline \hline
\end{tabular}
\end{table}

\begin{table}    
    \centering
    \caption{\label{table:E5_results_first_mesons} Numerical results for the first excited state mass of the mesonic spectrum in ensemble M5.  $k$-G stands for $k$-Gauss fit, ($k+1$)-G is $k+1$-Gauss fit, $k$-C stands for $k$-Cauchy function fit, ($k+1$)-G $k+1$-Cauchy function fit, $C$ indicates the mesonic channel considered, $am_C$ is the result of the GEVP analysis, $\sigma_G$ is the energy smearing radius used for the Gaussian fits, $\sigma_C$ for the Cauchy fit.}
    \begin{tabular}{ |c|c|c|c|c|c|c|c|c|c|c| }
    \hline  \hline
    $C$ & $k$ & $N_{\rm source}$ & $N_{\rm sink}$ & $aE_1$ $k$-G & $aE_1$ ($k+1$)-G & $aE_1$ $k$-C & $aE_1$ ($k+1$)-C & $am_C$ & $\sigma_{G} / m_C$ & $\sigma_C / m_C$ \\
    \hline
PS & 2 & 80 & 80 & 0.6303(96) & 0.6296(63) & 0.629(11) & 0.629(11) & 0.631(60) & 0.2 & 0.2 \\
V & 3 & 80 & 80 & 0.644(30) & 0.6437(64) & 0.6546(90) & 0.6544(28) & 0.650(33) & 0.2 & 0.25 \\
T & 3 & 80 & 80 & 0.648(30) & 0.6485(65) & 0.650(11) & 0.650(11) & 0.649(21) & 0.2 & 0.25 \\
AV & 2 & 80 & 80 & 0.769(11) & 0.7756(78) & 0.771(15) & 0.773(14) & 0.774(15) & 0.2 & 0.2 \\
at & 2 & 80 & 80 & 0.775(14) & 0.7746(77) & 0.777(14) & 0.779(15) & 0.777(18) & 0.2 & 0.2 \\
S & 2 & 80 & 80 & 0.720(30) & 0.7201(72) & 0.7196(75) & 0.7195(62) & 0.729(27) & 0.2 & 0.2 \\
\hline
ps & 2 & 80 & 80 & 0.8923(15) & 0.8920(89) & 0.874(19) & 0.873(16) & 0.884(14) & 0.2 & 0.2 \\
v & 3 & 80 & 80 & 0.915(18) & 0.9153(92) & 0.8942(89) & 0.8960(74) & 0.905(60) & 0.2 & 0.25 \\
t & 3 & 80 & 80 & 0.903(60) & 0.9069(91) & 0.9129(91) & 0.9129(91) & 0.899(24) & 0.2 & 0.25 \\
av & 2 & 80 & 80 & 0.999(30) & 0.9069(91) & 0.976(17) & 0.979(15) & 0.984(56) & 0.2 & 0.2 \\
at & 2 & 80 & 80 & 0.9492(88) & 0.9496(95) & 0.9499(95) & 0.9499(95) & 0.94(11) & 0.2 & 0.2 \\
s & 2 & 80 & 80 & 1.003(14) & 1.004(30) & 1.01350(270) & 1.013(12) & 1.016(33) & 0.2 & 0.2 \\
\hline \hline
\end{tabular}
\end{table}

\begin{table}    
    \centering
    \caption{\label{table:E5_results_second_mesons} Numerical results for the second excited  state mass of the mesonic spectrum in ensemble M5.  $k$-G stands for $k$-Gauss fit, ($k+1$)-G is $k+1$-Gauss fit, $k$-C stands for $k$-Cauchy function fit, ($k+1$)-G $k+1$-Cauchy function fit, $C$ indicates the mesonic channel considered, $am_C$ is the result of the GEVP analysis, $\sigma_G$ is the energy smearing radius used for the Gaussian fits, $\sigma_C$ for the Cauchy fit.}
    \begin{tabular}{ |c|c|c|c|c|c|c|c|c|c|c| }
    \hline  \hline
    $C$ & $k$ & $N_{\rm source}$ & $N_{\rm sink}$ & $aE_2$ $k$-G & $aE_2$ ($k+1$)-G & $aE_2$ $k$-C & $aE_2$ ($k+1$)-C & $am_C$ & $\sigma_{G} / m_C$ & $\sigma_C / m_C$ \\
    \hline
PS & 2 & 80 & 80 & - & - & - & - & - & 0.2 & 0.2 \\
V & 3 & 80 & 80 & 0.916(25) & 0.9151(92) & 0.900(13) & 0.898(13) & 0.90(30) & 0.2 & 0.25 \\
T & 3 & 80 & 80 & 0.830(29) & 0.8331(83) & 0.867(19) & 0.865(18) & 0.85(17) & 0.2 & 0.25 \\
AV & 2 & 80 & 80 & - & - & - & - & - & 0.2 & 0.2 \\
AT & 2 & 80 & 80 & 1.079(32) & 1.078(11) & 1.083(22) & 1.092(25) & 1.10(90) & 0.2 & 0.2 \\
S & 2 & 80 & 80 & 1.199(38) & 1.195(12) & 1.204(38) & 1.206(35) & 1.193(77) & 0.2 & 0.2 \\
\hline
ps & 2 & 80 & 80 & - & - & - & - & - & 0.2 & 0.2 \\
v & 3 & 80 & 80 & 1.259(42) & 1.261(13) & 1.279(13) & 1.265(60) & 1.293(89) & 0.2 & 0.25 \\
t & 3 & 80 & 80 & 1.2317(74) & 1.231(12) & 1.234(13) & 1.224(13) & 1.21(11) & 0.2 & 0.25 \\
av & 2 & 80 & 80 & - & - & - & - & - & 0.2 & 0.2 \\
at & 2 & 80 & 80 & - & - & - & - & - & 0.2 & 0.2 \\
s & 2 & 80 & 80 & - & - & - & - & - & 0.2 & 0.2 \\
\hline \hline
\end{tabular}
\end{table}

\newpage
\begin{table}
    \centering
    \caption{\label{table:E1_results_ground_CB} Numerical results for the ground state mass of the chimera baryon spectrum in ensemble M1. $k$-G stands for $k$-Gauss fit, ($k+1$)-G is $k+1$-Gauss fit, $k$-C stands for $k$-Cauchy function fit, ($k+1$)-G $k+1$-Cauchy function fit, $C$ indicates the chimera baryons channel considered, $am_C$ is the result of the GEVP analysis, $\sigma_G$ is the energy smearing radius used for the Gaussian fits, $\sigma_C$ for the Cauchy fit.}
    \begin{tabular}{ |c|c|c|c|c|c|c|c|c|c|c| }
    \hline  \hline
    $C$ & $k$ & $N_{\rm source}$ & $N_{\rm sink}$ & $aE_0$ $k$-G & $aE_0$ ($k+1$)-G & $aE_0$ $k$-C & $aE_0$ ($k+1$)-C & $am_C$ & $\sigma_{G} / m_C$ & $\sigma_C / m_C$ \\
    \hline
$\Lambda^{+}_{\rm CB}$ & 2 & 80 & 80 & 0.7757(62) & 0.77560(90) & 0.7757(62) & 0.7757(62) & 0.7686(59) & 0.32 & 0.32 \\
$\Sigma^{+}_{\rm CB}$ & 2 & 80 & 80 & 0.7582(73) & 0.7580(74) & 0.7592(66) & 0.7517(37) & 0.7577(41) & 0.3 & 0.3 \\
$\Sigma^{* \, +}_{\rm CB}$ & 2 & 80 & 80 & 0.7918(63) & 0.7918(63) & 0.7918(63) & 0.7918(63) & 0.7842(59) & 0.2 & 0.2 \\
$\Lambda^{-}_{\rm CB}$ & 2 & 80 & 80 & 0.8880(71) & 0.8880(71) & 0.9039(55) & 0.9033(62) & 0.897(13) & 0.22 & 0.22 \\
$\Sigma^{-}_{\rm CB}$ & 2 & 80 & 80 & 0.8999(72) & 0.8999(72) & 0.9181(73) & 0.9181(73) & 0.909(15) & 0.18 & 0.18 \\
$\Sigma^{* \, -}_{\rm CB}$ & 2 & 80 & 80 & 0.9061(44) & 0.9061(45) & 0.8989(72) & 0.8989(72) & 0.907(15) & 0.2 & 0.2 \\
\hline \hline
\end{tabular}
\end{table}

\begin{table}    
    \centering
    \caption{\label{table:E1_results_first_CB} Numerical results for the first excited  state mass of the chimera baryon spectrum in ensemble M1. $k$-G stands for $k$-Gauss fit, ($k+1$)-G is $k+1$-Gauss fit, $k$-C stands for $k$-Cauchy function fit, ($k+1$)-G $k+1$-Cauchy function fit, $C$ indicates the chimera baryons channel considered, $am_C$ is the result of the GEVP analysis, $\sigma_G$ is the energy smearing radius used for the Gaussian fits, $\sigma_C$ for the Cauchy fit.}
    \begin{tabular}{ |c|c|c|c|c|c|c|c|c|c|c| }
    \hline  \hline
    $C$ & $k$ & $N_{\rm source}$ & $N_{\rm sink}$ & $aE_1$ $k$-G & $aE_1$ ($k+1$)-G & $aE_1$ $k$-C & $aE_1$ ($k+1$)-C & $am_C$ & $\sigma_{G} / m_C$ & $\sigma_C / m_C$ \\
    \hline
$\Lambda^{+}_{\rm CB}$ & 2 & 80 & 80 & 0.9754(98) & 0.9756(37) & 0.9754(98) & 0.9760(35) & 0.995(26) & 0.32 & 0.32 \\
$\Sigma^{+}_{\rm CB}$ & 2 & 80 & 80 & 0.990(20) & 0.990(19) & 1.016(10) & 1.016(10) & 0.998(21) & 0.3 & 0.3 \\
$\Sigma^{* \, +}_{\rm CB}$ & 2 & 80 & 80 & 1.043(10) & 1.043(10) & 1.036(26) & 1.0424(41) & 1.026(20) & 0.2 & 0.2 \\
$\Lambda^{-}_{\rm CB}$ & 2 & 80 & 80 & 1.005(10) & 1.005(10) & 1.0051(37) & 1.0062(35) & 1.026(76) & 0.22 & 0.22 \\
$\Sigma^{-}_{\rm CB}$ & 2 & 80 & 80 & 1.095(11) & 1.096(17) & 1.082(11) & 1.082(11) & 1.103(60) & 0.18 & 0.18 \\
$\Sigma^{* \, -}_{\rm CB}$ & 2 & 80 & 80 & 1.0451(60) & 1.0451(60) & 1.044(10) & 1.044(10) & 1.063(66) & 0.2 & 0.2 \\
\hline \hline
\end{tabular}
\end{table}

\begin{table}    
    \centering
    \caption{\label{table:E1_results_second_CB} Numerical results for the second excited state mass of the chimera baryon spectrum in ensemble M1.  $k$-G stands for $k$-Gauss fit, ($k+1$)-G is $k+1$-Gauss fit, $k$-C stands for $k$-Cauchy function fit, ($k+1$)-G $k+1$-Cauchy function fit, $C$ indicates the chimera baryons channel considered, $am_C$ is the result of the GEVP analysis, $\sigma_G$ is the energy smearing radius used for the Gaussian fits, $\sigma_C$ for the Cauchy fit.}
    \begin{tabular}{ |c|c|c|c|c|c|c|c|c|c|c| }
    \hline  \hline
    $C$ & $k$ & $N_{\rm source}$ & $N_{\rm sink}$ & $aE_2$ $k$-G & $aE_2$ ($k+1$)-G & $aE_2$ $k$-C & $aE_2$ ($k+1$)-C & $am_C$ & $\sigma_{G} / m_C$ & $\sigma_C / m_C$ \\
    \hline
$\Lambda^{+}_{\rm CB}$ & 2 & 80 & 80 & - & - & - & - & 1.33(12) & 0.32 & 0.32 \\
$\Sigma^{+}_{\rm CB}$ & 2 & 80 & 80 & - & - & - & - & 1.345(79) & 0.3 & 0.3 \\
$\Sigma^{* \, +}_{\rm CB}$ & 2 & 80 & 80 & - & - & - & - & 1.377(59) & 0.2 & 0.2 \\
$\Lambda^{-}_{\rm CB}$ & 2 & 80 & 80 & - & - & - & - & - & 0.22 & 0.22 \\
$\Sigma^{-}_{\rm CB}$ & 2 & 80 & 80 & - & - & - & - & - & 0.18 & 0.18 \\
$\Sigma^{* \, -}_{\rm CB}$ & 2 & 80 & 80 & - & - & - & - & - & 0.2 & 0.2 \\
\hline \hline
\end{tabular}
\end{table}
    
\begin{table}
    \centering
    \caption{\label{table:E2_results_ground_CB} Numerical results for the ground state mass of the chimera baryon spectrum in ensemble M2.  $k$-G stands for $k$-Gauss fit, ($k+1$)-G is $k+1$-Gauss fit, $k$-C stands for $k$-Cauchy function fit, ($k+1$)-G $k+1$-Cauchy function fit, $C$ indicates the chimera baryons channel considered, $am_C$ is the result of the GEVP analysis, $\sigma_G$ is the energy smearing radius used for the Gaussian fits, $\sigma_C$ for the Cauchy fit.}
    \begin{tabular}{ |c|c|c|c|c|c|c|c|c|c|c| }
    \hline  \hline
    $C$ & $k$ & $N_{\rm source}$ & $N_{\rm sink}$ & $aE_0$ $k$-G & $aE_0$ ($k+1$)-G & $aE_0$ $k$-C & $aE_0$ ($k+1$)-C & $am_C$ & $\sigma_{G} / m_C$ & $\sigma_C / m_C$ \\
    \hline
$\Lambda^{+}_{\rm CB}$ & 2 & 80 & 80 & 0.7525(12) & 0.7526(15) & 0.7575(61) & 0.7575(15) & 0.7595(78) & 0.3 & 0.3 \\
$\Sigma^{+}_{\rm CB}$ & 3 & 80 & 80 & 0.7676(61) & 0.7676(61) & 0.7652(33) & 0.7652(31) & 0.7494(43) & 0.23 & 0.23 \\
$\Sigma^{* \, +}_{\rm CB}$ & 2 & 80 & 80 & 0.7675(18) & 0.7676(21) & 0.7827(63) & 0.7827(63) & 0.7750(57) & 0.2 & 0.2 \\
$\Lambda^{-}_{\rm CB}$ & 3 & 80 & 80 & 0.9039(72) & 0.9039(72) & 0.9039(72) & 0.9039(72) & 0.913(18) & 0.33 & 0.33 \\
$\Sigma^{-}_{\rm CB}$ & 2 & 80 & 80 & 0.9292(74) & 0.9292(74) & 0.9289(21) & 0.9290(15) & 0.920(29) & 0.18 & 0.18 \\
$\Sigma^{* \, -}_{\rm CB}$ & 2 & 80 & 80 & 0.9353(75) & 0.9353(75) & 0.9353(75) & 0.9353(75) & 0.926(22) & 0.2 & 0.2 \\
\hline \hline
\end{tabular}
\end{table}

\begin{table}    
    \centering
    \caption{\label{table:E2_results_first_CB} Numerical results for the first excited  state mass of the chimera baryon spectrum in ensemble M2.  $k$-G stands for $k$-Gauss fit, ($k+1$)-G is $k+1$-Gauss fit, $k$-C stands for $k$-Cauchy function fit, ($k+1$)-G $k+1$-Cauchy function fit, $C$ indicates the chimera baryons channel considered, $am_C$ is the result of the GEVP analysis, $\sigma_G$ is the energy smearing radius used for the Gaussian fits, $\sigma_C$ for the Cauchy fit.}
    \begin{tabular}{ |c|c|c|c|c|c|c|c|c|c|c| }
    \hline  \hline
    $C$ & $k$ & $N_{\rm source}$ & $N_{\rm sink}$ & $aE_1$ $k$-G & $aE_1$ ($k+1$)-G & $aE_1$ $k$-C & $aE_1$ ($k+1$)-C & $am_C$ & $\sigma_{G} / m_C$ & $\sigma_C / m_C$ \\
    \hline
$\Lambda^{+}_{\rm CB}$ & 2 & 80 & 80 & 0.9881(99) & 0.9880(99) & 0.9880(99) & 0.9880(99) & 1.008(55) & 0.3 & 0.3 \\
$\Sigma^{+}_{\rm CB}$ & 3 & 80 & 80 & 0.988(23) & 0.988(22) & 0.9671(82) & 0.9673(86) & 0.979(41) & 0.23 & 0.23 \\
$\Sigma^{* \, +}_{\rm CB}$ & 2 & 80 & 80 & 0.9765(98) & 0.9766(98) & 0.978(13) & 0.9768(62) & 0.992(59) & 0.2 & 0.2 \\
$\Lambda^{-}_{\rm CB}$ & 3 & 80 & 80 & 1.059(11) & 1.059(11) & 1.059(11) & 1.059(11) & - & 0.33 & 0.33 \\
$\Sigma^{-}_{\rm CB}$ & 2 & 80 & 80 & 1.095(11) & 1.095(11) & 1.0945(45) & 1.095(11) & 1.075(59) & 0.18 & 0.18 \\
$\Sigma^{* \, -}_{\rm CB}$ & 2 & 80 & 80 & 1.157(12) & 1.157(12) & 1.164(12) & 1.1635(13) & 1.18(12) & 0.2 & 0.2 \\
\hline \hline
\end{tabular}
\end{table}

\begin{table}    
    \centering
    \caption{\label{table:E2_results_second_CB} Numerical results for the second excited state mass of the chimera baryon spectrum in ensemble M2.  $k$-G stands for $k$-Gauss fit, ($k+1$)-G is $k+1$-Gauss fit, $k$-C stands for $k$-Cauchy function fit, ($k+1$)-G $k+1$-Cauchy function fit, $C$ indicates the chimera baryons channel considered, $am_C$ is the result of the GEVP analysis, $\sigma_G$ is the energy smearing radius used for the Gaussian fits, $\sigma_C$ for the Cauchy fit.}
    \begin{tabular}{ |c|c|c|c|c|c|c|c|c|c|c| }
    \hline  \hline
    $C$ & $k$ & $N_{\rm source}$ & $N_{\rm sink}$ & $aE_2$ $k$-G & $aE_2$ ($k+1$)-G & $aE_2$ $k$-C & $aE_2$ ($k+1$)-C & $am_C$ & $\sigma_{G} / m_C$ & $\sigma_C / m_C$ \\
    \hline
$\Lambda^{+}_{\rm CB}$ & 2 & 80 & 80 & - & - & - & - & - & 0.3 & 0.3 \\
$\Sigma^{+}_{\rm CB}$ & 3 & 80 & 80 & 1.283(27) & 1.283(28) & 1.271(23) & 1.270(21) & 1.31(22) & 0.23 & 0.23 \\
$\Sigma^{* \, +}_{\rm CB}$ & 2 & 80 & 80 & 1.275(25) & 1.275(25) & 1.2724(84) & 1.2743(40) & - & 0.2 & 0.2 \\
$\Lambda^{-}_{\rm CB}$ & 3 & 80 & 80 & - & - & - & - & - & 0.33 & 0.33 \\
$\Sigma^{-}_{\rm CB}$ & 2 & 80 & 80 & - & - & - & - & - & 0.18 & 0.18 \\
$\Sigma^{* \, -}_{\rm CB}$ & 2 & 80 & 80 & - & - & - & - & - & 0.2 & 0.2 \\
\hline \hline
\end{tabular}
\end{table}

\begin{table}    
    \centering
    \caption{\label{table:E3_results_ground_CB} Numerical results for the ground state mass of the chimera baryon spectrum in ensemble M3.  $k$-G stands for $k$-Gauss fit, ($k+1$)-G is $k+1$-Gauss fit, $k$-C stands for $k$-Cauchy function fit, ($k+1$)-G $k+1$-Cauchy function fit, $C$ indicates the chimera baryons channel considered, $am_C$ is the result of the GEVP analysis, $\sigma_G$ is the energy smearing radius used for the Gaussian fits, $\sigma_C$ for the Cauchy fit.}
    \begin{tabular}{ |c|c|c|c|c|c|c|c|c|c|c| }
    \hline  \hline
    $C$ & $k$ & $N_{\rm source}$ & $N_{\rm sink}$ & $aE_0$ $k$-G & $aE_0$ ($k+1$)-G & $aE_0$ $k$-C & $aE_0$ ($k+1$)-C & $am_C$ & $\sigma_{G} / m_C$ & $\sigma_C / m_C$ \\
    \hline
$\Lambda^{+}_{\rm CB}$ & 2 & 80 & 80 & 0.7579(24) & 0.7594(22) & 0.7574(61) & 0.7574(61) & 0.7649(52) & 0.27 & 0.27 \\
$\Sigma^{+}_{\rm CB}$ & 3 & 80 & 80 & 0.7636(61) & 0.7636(61) & 0.7502(40) & 0.7503(44) & 0.7557(36) & 0.28 & 0.28 \\
$\Sigma^{* \, +}_{\rm CB}$ & 3 & 80 & 80 & 0.7928(63) & 0.7928(63) & 0.7928(63) & 0.7928(63) & 0.7865(45) & 0.18 & 0.18 \\
$\Lambda^{-}_{\rm CB}$ & 3 & 80 & 80 & 0.9332(75) & 0.9332(75) & 0.9167(56) & 0.9165(52) & 0.925(11) & 0.25 & 0.25 \\
$\Sigma^{-}_{\rm CB}$ & 3 & 80 & 80 & 0.9138(73) & 0.9138(73) & 0.9138(73) & 0.9138(73) & 0.923(20) & 0.32 & 0.32 \\
$\Sigma^{* \, -}_{\rm CB}$ & 2 & 80 & 80 & 0.9295(36) & 0.9292(30) & 0.9286(74) & 0.9286(74) & 0.938(15) & 0.24 & 0.24 \\
\hline \hline
\end{tabular}
\end{table}

\begin{table}    
    \centering
    \caption{\label{table:E3_results_first_CB} Numerical results for the first excited  state mass of the chimera baryon spectrum in ensemble M3.  $k$-G stands for $k$-Gauss fit, ($k+1$)-G is $k+1$-Gauss fit, $k$-C stands for $k$-Cauchy function fit, ($k+1$)-G $k+1$-Cauchy function fit, $C$ indicates the chimera baryons channel considered, $am_C$ is the result of the GEVP analysis, $\sigma_G$ is the energy smearing radius used for the Gaussian fits, $\sigma_C$ for the Cauchy fit.}
    \begin{tabular}{ |c|c|c|c|c|c|c|c|c|c|c| }
    \hline  \hline
    $C$ & $k$ & $N_{\rm source}$ & $N_{\rm sink}$ & $aE_1$ $k$-G & $aE_1$ ($k+1$)-G & $aE_1$ $k$-C & $aE_1$ ($k+1$)-C & $am_C$ & $\sigma_{G} / m_C$ & $\sigma_C / m_C$ \\
    \hline
$\Lambda^{+}_{\rm CB}$ & 2 & 80 & 80 & 1.0157(33) & 1.0155(25) & 1.040(10) & 1.040(10) & 1.022(12) & 0.27 & 0.27 \\
$\Sigma^{+}_{\rm CB}$ & 3 & 80 & 80 & 1.028(10) & 1.028(10) & 0.9983(59) & 0.9981(41) & 1.0150(89) & 0.28 & 0.28 \\
$\Sigma^{* \, +}_{\rm CB}$ & 3 & 80 & 80 & 1.015(23) & 1.013(24) & 1.019(20) & 1.016(14) & 1.011(19) & 0.18 & 0.18 \\
$\Lambda^{-}_{\rm CB}$ & 3 & 80 & 80 & 1.127(11) & 1.127(11) & 1.100(31) & 1.115(30) & 1.102(27) & 0.25 & 0.25 \\
$\Sigma^{-}_{\rm CB}$ & 3 & 80 & 80 & 1.154(11) & 1.154(11) & 1.117(11) & 1.117(11) & 1.134(31) & 0.32 & 0.32 \\
$\Sigma^{* \, -}_{\rm CB}$ & 2 & 80 & 80 & 1.164(12) & 1.164(12) & 1.163(12) & 1.1633(37) & 1.184(14) & 0.24 & 0.24 \\
\hline \hline
\end{tabular}
\end{table}

\begin{table}    
    \centering
    \caption{\label{table:E3_results_second_CB} Numerical results for the second excited state mass of the chimera baryon spectrum in ensemble M3.  $k$-G stands for $k$-Gauss fit, ($k+1$)-G is $k+1$-Gauss fit, $k$-C stands for $k$-Cauchy function fit, ($k+1$)-G $k+1$-Cauchy function fit, $C$ indicates the chimera baryons channel considered, $am_C$ is the result of the GEVP analysis, $\sigma_G$ is the energy smearing radius used for the Gaussian fits, $\sigma_C$ for the Cauchy fit.}
    \begin{tabular}{ |c|c|c|c|c|c|c|c|c|c|c| }
    \hline  \hline
    $C$ & $k$ & $N_{\rm source}$ & $N_{\rm sink}$ & $aE_2$ $k$-G & $aE_2$ ($k+1$)-G & $aE_2$ $k$-C & $aE_2$ ($k+1$)-C & $am_C$ & $\sigma_{G} / m_C$ & $\sigma_C / m_C$ \\
    \hline
$\Lambda^{+}_{\rm CB}$ & 2 & 80 & 80 & - & - & - & - & 1.377(46) & 0.27 & 0.27 \\
$\Sigma^{+}_{\rm CB}$ & 3 & 80 & 80 & 1.294(25) & 1.294(25) & 1.293(30) & 1.293(20) & 1.352(28) & 0.28 & 0.28 \\
$\Sigma^{* \, +}_{\rm CB}$ & 3 & 80 & 80 & 1.286(14) & 1.289(13) & 1.299(26) & 1.299(26) & 1.325(24) & 0.18 & 0.18 \\
$\Lambda^{-}_{\rm CB}$ & 3 & 80 & 80 & 1.317(26) & 1.317(26) & 1.313(36) & 1.313(37) & - & 0.25 & 0.25 \\
$\Sigma^{-}_{\rm CB}$ & 3 & 80 & 80 & - & - & - & - & - & 0.32 & 0.32 \\
$\Sigma^{* \, -}_{\rm CB}$ & 2 & 80 & 80 & - & - & - & - & - & 0.24 & 0.24 \\
\hline \hline
\end{tabular}
\end{table}

\begin{table}    
    \centering
    \caption{\label{table:E4_results_ground_CB} Numerical results for the ground state mass of the chimera baryon spectrum in ensemble M4.  $k$-G stands for $k$-Gauss fit, ($k+1$)-G is $k+1$-Gauss fit, $k$-C stands for $k$-Cauchy function fit, ($k+1$)-G $k+1$-Cauchy function fit, $C$ indicates the chimera baryons channel considered, $am_C$ is the result of the GEVP analysis, $\sigma_G$ is the energy smearing radius used for the Gaussian fits, $\sigma_C$ for the Cauchy fit.}
    \begin{tabular}{ |c|c|c|c|c|c|c|c|c|c|c| }
    \hline  \hline
    $C$ & $k$ & $N_{\rm source}$ & $N_{\rm sink}$ & $aE_0$ $k$-G & $aE_0$ ($k+1$)-G & $aE_0$ $k$-C & $aE_0$ ($k+1$)-C & $am_C$ & $\sigma_{G} / m_C$ & $\sigma_C / m_C$ \\
    \hline
$\Lambda^{+}_{\rm CB}$ & 2 & 80 & 80 & 0.8177(65) & 0.8177(65) & 0.8343(67) & 0.8343(67) & 0.8260(32) & 0.3 & 0.3 \\
$\Sigma^{+}_{\rm CB}$ & 3 & 80 & 80 & 0.8039(64) & 0.8039(64) & 0.8201(66) & 0.8201(66) & 0.8122(28) & 0.25 & 0.25 \\
$\Sigma^{* \, +}_{\rm CB}$ & 2 & 80 & 80 & 0.8346(67) & 0.8346(67) & 0.8514(68) & 0.8514(68) & 0.8434(42) & 0.25 & 0.25 \\
$\Lambda^{-}_{\rm CB}$ & 3 & 80 & 80 & 0.9484(76) & 0.9484(76) & 0.9484(76) & 0.9485(14) & 0.959(30) & 0.27 & 0.27 \\
$\Sigma^{-}_{\rm CB}$ & 2 & 80 & 80 & 0.9505(76) & 0.9504(76) & 0.9504(76) & 0.9504(76) & 0.961(19) & 0.25 & 0.25 \\
$\Sigma^{* \, -}_{\rm CB}$ & 2 & 80 & 80 & 0.9445(76) & 0.9445(76) & 0.9445(76) & 0.9466(22) & 0.955(26) & 0.3 & 0.3 \\
\hline \hline
\end{tabular}
\end{table}

\begin{table}    
    \centering
    \caption{\label{table:E4_results_first_CB} Numerical results for the first excited state mass of the chimera baryon spectrum in ensemble M4. $k$-G stands for $k$-Gauss fit, ($k+1$)-G is $k+1$-Gauss fit, $k$-C stands for $k$-Cauchy function fit, ($k+1$)-G $k+1$-Cauchy function fit, $C$ indicates the chimera baryons channel considered, $am_C$ is the result of the GEVP analysis, $\sigma_G$ is the energy smearing radius used for the Gaussian fits, $\sigma_C$ for the Cauchy fit.}
    \begin{tabular}{ |c|c|c|c|c|c|c|c|c|c|c| }
    \hline  \hline
    $C$ & $k$ & $N_{\rm source}$ & $N_{\rm sink}$ & $aE_1$ $k$-G & $aE_1$ ($k+1$)-G & $aE_1$ $k$-C & $aE_1$ ($k+1$)-C & $am_C$ & $\sigma_{G} / m_C$ & $\sigma_C / m_C$ \\
    \hline
$\Lambda^{+}_{\rm CB}$ & 2 & 80 & 80 & 1.008(10) & 1.008(10) & 1.0079(16) & 1.008(10) & 1.023(38) & 0.3 & 0.3 \\
$\Sigma^{+}_{\rm CB}$ & 3 & 80 & 80 & 1.015(10) & 1.015(10) & 0.990(18) & 0.984(12) & 1.001(38) & 0.25 & 0.25 \\
$\Sigma^{* \, +}_{\rm CB}$ & 2 & 80 & 80 & 0.9947(99) & 0.9947(99) & 0.9947(99) & 0.9947(99) & 1.012(36) & 0.25 & 0.25 \\
$\Lambda^{-}_{\rm CB}$ & 3 & 80 & 80 & 1.150(11) & 1.150(11) & 1.150(11) & 1.151(13) & 1.171(28) & 0.27 & 0.27 \\
$\Sigma^{-}_{\rm CB}$ & 2 & 80 & 80 & 1.164(23) & 1.168(31) & 1.173(38) & 1.153(13) & 1.173(43) & 0.25 & 0.25 \\
$\Sigma^{* \, -}_{\rm CB}$ & 2 & 80 & 80 & 1.1274(76) & 1.1285(82) & 1.164(12) & 1.157(12) & 1.143(88) & 0.3 & 0.3 \\
\hline \hline
\end{tabular}
\end{table}

\begin{table}    
    \centering
    \caption{\label{table:E4_results_second_CB} Numerical results for the second excited state mass of the chimera baryon spectrum in ensemble M4.  $k$-G stands for $k$-Gauss fit, ($k+1$)-G is $k+1$-Gauss fit, $k$-C stands for $k$-Cauchy function fit, ($k+1$)-G $k+1$-Cauchy function fit, $C$ indicates the chimera baryons channel considered, $am_C$ is the result of the GEVP analysis, $\sigma_G$ is the energy smearing radius used for the Gaussian fits, $\sigma_C$ for the Cauchy fit.}
    \begin{tabular}{ |c|c|c|c|c|c|c|c|c|c|c| }
    \hline  \hline
    $C$ & $k$ & $N_{\rm source}$ & $N_{\rm sink}$ & $aE_2$ $k$-G & $aE_2$ ($k+1$)-G & $aE_2$ $k$-C & $aE_2$ ($k+1$)-C & $am_C$ & $\sigma_{G} / m_C$ & $\sigma_C / m_C$ \\
    \hline
$\Lambda^{+}_{\rm CB}$ & 2 & 80 & 80 & - & - & - & - & 1.28(15) & 0.3 & 0.3 \\
$\Sigma^{+}_{\rm CB}$ & 3 & 80 & 80 & - & - & - & - & - & 0.25 & 0.25 \\
$\Sigma^{* \, +}_{\rm CB}$ & 2 & 80 & 80 & 1.361(27) & 1.358(11) & 1.3613(10) & 1.3608(26) & 1.329(71) & 0.25 & 0.25 \\
$\Lambda^{-}_{\rm CB}$ & 3 & 80 & 80 & - & - & - & - & - & 0.27 & 0.27 \\
$\Sigma^{-}_{\rm CB}$ & 2 & 80 & 80 & - & - & - & - & - & 0.25 & 0.25 \\
$\Sigma^{* \, -}_{\rm CB}$ & 2 & 80 & 80 & - & - & - & - & - & 0.3 & 0.3 \\
\hline \hline
\end{tabular}
\end{table}

\begin{table}    
    \centering
    \caption{\label{table:E5_results_ground_CB} Numerical results for the ground state mass of the chimera baryon spectrum in ensemble M5.  $k$-G stands for $k$-Gauss fit, ($k+1$)-G is $k+1$-Gauss fit, $k$-C stands for $k$-Cauchy function fit, ($k+1$)-G $k+1$-Cauchy function fit, $C$ indicates the chimera baryons channel considered, $am_C$ is the result of the GEVP analysis, $\sigma_G$ is the energy smearing radius used for the Gaussian fits, $\sigma_C$ for the Cauchy fit.}
    \begin{tabular}{ |c|c|c|c|c|c|c|c|c|c|c| }
    \hline  \hline
    $C$ & $k$ & $N_{\rm source}$ & $N_{\rm sink}$ & $aE_0$ $k$-G & $aE_0$ ($k+1$)-G & $aE_0$ $k$-C & $aE_0$ ($k+1$)-C & $am_C$ & $\sigma_{G} / m_C$ & $\sigma_C / m_C$ \\
    \hline
$\Lambda^{+}_{\rm CB}$ & 2 & 80 & 80 & 0.6861(55) & 0.6861(55) & 0.6999(56) & 0.6999(56) & 0.6928(46) & 0.25 & 0.25 \\
$\Sigma^{+}_{\rm CB}$ & 3 & 80 & 80 & 0.6762(54) & 0.6762(54) & 0.6859(40) & 0.6868(37) & 0.6826(47) & 0.3 & 0.3 \\
$\Sigma^{* \, +}_{\rm CB}$ & 2 & 80 & 80 & 0.6999(56) & 0.6999(56) & 0.7081(50) & 0.7010(19) & 0.7062(62) & 0.2 & 0.2 \\
$\Lambda^{-}_{\rm CB}$ & 3 & 80 & 80 & 0.8398(81) & 0.8398(81) & 0.8299(14) & 0.8297(66) & 0.839(15) & 0.3 & 0.3 \\
$\Sigma^{-}_{\rm CB}$ & 2 & 80 & 80 & 0.8524(68) & 0.8524(68) & 0.8356(67) & 0.8356(67) & 0.845(19) & 0.2 & 0.2 \\
$\Sigma^{* \, -}_{\rm CB}$ & 2 & 80 & 80 & 0.8546(75) & 0.8574(83) & 0.8666(69) & 0.8666(69) & 0.858(18) & 0.25 & 0.25 \\
\hline \hline
\end{tabular}
\end{table}

\begin{table}    
    \centering
    \caption{\label{table:E5_results_first_CB} Numerical results for the first excited state mass of the chimera baryon spectrum in ensemble M5.  $k$-G stands for $k$-Gauss fit, ($k+1$)-G is $k+1$-Gauss fit, $k$-C stands for $k$-Cauchy function fit, ($k+1$)-G $k+1$-Cauchy function fit, $C$ indicates the chimera baryons channel considered, $am_C$ is the result of the GEVP analysis, $\sigma_G$ is the energy smearing radius used for the Gaussian fits, $\sigma_C$ for the Cauchy fit.}
    \begin{tabular}{ |c|c|c|c|c|c|c|c|c|c|c| }
    \hline  \hline
    $C$ & $k$ & $N_{\rm source}$ & $N_{\rm sink}$ & $aE_1$ $k$-G & $aE_1$ ($k+1$)-G & $aE_1$ $k$-C & $aE_1$ ($k+1$)-C & $am_C$ & $\sigma_{G} / m_C$ & $\sigma_C / m_C$ \\
    \hline
$\Lambda^{+}_{\rm CB}$ & 2 & 80 & 80 & 0.942(12) & 0.943(13) & 0.9356(94) & 0.9356(94) & 0.951(38) & 0.25 & 0.25 \\
$\Sigma^{+}_{\rm CB}$ & 3 & 80 & 80 & 0.907(24) & 0.904(22) & 0.919(10) & 0.913(18) & 0.906(45) & 0.3 & 0.3 \\
$\Sigma^{* \, +}_{\rm CB}$ & 2 & 80 & 80 & 0.8982(46) & 0.8982(46) & 0.916(20) & 0.8988(21) & 0.912(45) & 0.2 & 0.2 \\
$\Lambda^{-}_{\rm CB}$ & 3 & 80 & 80 & 1.025(11) & 1.025(11) & 1.025(16) & 1.024(14) & 1.043(41) & 0.3 & 0.3 \\
$\Sigma^{-}_{\rm CB}$ & 2 & 80 & 80 & 1.055(11) & 1.055(11) & 1.055(11) & 1.055(11) & 1.038(43) & 0.2 & 0.2 \\
$\Sigma^{* \, -}_{\rm CB}$ & 2 & 80 & 80 & 1.0983(13) & 1.0992(95) & 1.098(11) & 1.098(11) & 1.115(24) & 0.25 & 0.25 \\
\hline \hline
\end{tabular}
\end{table}

\begin{table}    
\centering
\caption{\label{table:E5_results_second_CB} Numerical results for the second excited  state mass of the chimera baryon spectrum in ensemble M5.  $k$-G stands for $k$-Gauss fit, ($k+1$)-G is $k+1$-Gauss fit, $k$-C stands for $k$-Cauchy function fit, ($k+1$)-G $k+1$-Cauchy function fit, $C$ indicates the chimera baryons channel considered, $am_C$ is the result of the GEVP analysis, $\sigma_G$ is the energy smearing radius used for the Gaussian fits, $\sigma_C$ for the Cauchy fit.}
\begin{tabular}{ |c|c|c|c|c|c|c|c|c|c|c| }
    \hline  \hline
    $C$ & $k$ & $N_{\rm source}$ & $N_{\rm sink}$ & $aE_2$ $k$-G & $aE_2$ ($k+1$)-G & $aE_2$ $k$-C & $aE_2$ ($k+1$)-C & $am_C$ & $\sigma_{G} / m_C$ & $\sigma_C / m_C$ \\
    \hline
$\Lambda^{+}_{\rm CB}$ & 2 & 80 & 80 & - & - & - & - & 1.287(73) & 0.25 & 0.25 \\
$\Sigma^{+}_{\rm CB}$ & 3 & 80 & 80 & 1.226(25) & 1.226(25) & 1.178(24) & 1.183(23) & 1.204(96) & 0.3 & 0.3 \\
$\Sigma^{* \, +}_{\rm CB}$ & 2 & 80 & 80 & 1.312(26) & 1.312(26) & 1.288(14) & 1.2681(74) & 1.287(58) & 0.2 & 0.2 \\
$\Lambda^{-}_{\rm CB}$ & 3 & 80 & 80 & - & - & - & - & 1.11(12) & 0.3 & 0.3 \\
$\Sigma^{-}_{\rm CB}$ & 2 & 80 & 80 & - & - & - & - & 0.963(95) & 0.2 & 0.2 \\
$\Sigma^{* \, -}_{\rm CB}$ & 2 & 80 & 80 & - & - & - & - & - & 0.25 & 0.25 \\
\hline \hline
\end{tabular}
\end{table}

\begin{table}    
\centering
\caption{\label{table:E1_matrix_mesons} Numerical results for the mesons matrix elements in ensemble M1.  G stands for Gaussian fit, C stands for Cauchy function fit, $C$ indicates the mesonic channel considered, $c_{M, 0}$ is the result of the correlator fitting analysis, $\sigma_G$ is the energy smearing radius used for the Gaussian fits, $\sigma_C$ for the Cauchy fit.}
\begin{tabular}{ |c |c|c|c|c|c|c|c|c|c|c| }
    \hline  \hline
    $C$ & $a^{2}c_{M, 0}$-G & $a^{2}c_{M, 0}$-C & $a^{2}c_{M,0} (\hbox{corr.})$ & $\sigma_{G} / m_C$ & $\sigma_{C} / m_C$ \\
    \hline
   PS & 0.0259(4) & 0.0259(4) & 0.0262(4) & 0.33 & 0.32 \\ 
V & 0.0454(8) & 0.0454(8) & 0.0446(6) & 0.3 & 0.22 \\ 
AV & 0.0454(7) & 0.0454(7) & 0.0446(19) & 0.2 & 0.18 \\ 
\hline
ps & 0.0680(13) & 0.0680(14) & 0.0689(16) & 0.33 & 0.32 \\ 
v & 0.0914(22) & 0.0914(24) & 0.0930(26) & 0.3 & 0.22 \\ 
av & 0.0822(16) & 0.0822(17) & 0.0825(33) & 0.2 & 0.18 \\ 
    \hline \hline
\end{tabular}
\end{table}

\begin{table}    
\centering
\caption{\label{table:E2_matrix_mesons} Numerical results for the mesons matrix elements in ensemble M2.  G stands for Gaussian fit, C stands for Cauchy function fit, $C$ indicates the mesonic channel considered, $c_{M, 0}$ is the result of the correlator fitting analysis, $\sigma_G$ is the energy smearing radius used for the Gaussian fits, $\sigma_C$ for the Cauchy fit.}
\begin{tabular}{ |c |c|c|c|c|c|c|c|c|c|c| }
    \hline  \hline
    $C$ & $a^{2}c_{M, 0}$-G & $a^{2}c_{M, 0}$-C & $a^{2}c_{M,0} (\hbox{corr.})$ & $\sigma_{G} / m_C$ & $\sigma_{C} / m_C$ \\
    \hline
   PS & 0.0252(4) & 0.0252(4) & 0.0249(4) & 0.35 & 0.3 \\ 
V & 0.0444(7) & 0.0444(8) & 0.0435(6) & 0.28 & 0.33 \\ 
AV & 0.0466(7) & 0.0466(8) & 0.0479(28) & 0.3 & 0.18 \\ 
\hline
ps & 0.0679(13) & 0.0679(14) & 0.0682(18) & 0.35 & 0.3 \\ 
v & 0.0894(22) & 0.0894(24) & 0.0862(34) & 0.28 & 0.33 \\ 
av & 0.0845(16) & 0.0845(18) & 0.0848(38) & 0.3 & 0.18 \\ 
    \hline \hline
\end{tabular}
\end{table}

\begin{table}    
\centering
\caption{\label{table:E3_matrix_mesons} Numerical results for the mesons matrix elements in ensemble M3.  G stands for Gaussian fit, C stands for Cauchy function fit, $C$ indicates the mesonic channel considered, $c_{M, 0}$ is the result of the correlator fitting analysis, $\sigma_G$ is the energy smearing radius used for the Gaussian fits, $\sigma_C$ for the Cauchy fit.}
\begin{tabular}{ |c |c|c|c|c|c|c|c|c|c|c| }
    \hline  \hline
    $C$ & $a^{2}c_{M, 0}$-G & $a^{2}c_{M, 0}$-C & $a^{2}c_{M,0} (\hbox{corr.})$ & $\sigma_{G} / m_C$ & $\sigma_{C} / m_C$ \\
    \hline
   PS & 0.0256(4) & 0.0256(4) & 0.0256(5) & 0.3 & 0.27 \\ 
V & 0.0443(7) & 0.0443(8) & 0.0417(21) & 0.28 & 0.25 \\ 
AV & 0.0466(7) & 0.0466(8) & 0.0469(16) & 0.28 & 0.32 \\ 
\hline
ps & 0.0695(13) & 0.0695(14) & 0.0697(20) & 0.3 & 0.27 \\ 
v & 0.0887(22) & 0.0886(24) & 0.0900(33) & 0.28 & 0.25 \\ 
av & 0.0827(16) & 0.0827(17) & 0.0830(46) & 0.28 & 0.32 \\ 
    \hline \hline
\end{tabular}
\end{table}

\begin{table}    
\centering
\caption{\label{table:E4_matrix_mesons} Numerical results for the mesons matrix elements in ensemble M4.  G stands for Gaussian fit, C stands for Cauchy function fit, $C$ indicates the mesonic channel considered, $c_{M, 0}$ is the result of the correlator fitting analysis, $\sigma_G$ is the energy smearing radius used for the Gaussian fits, $\sigma_C$ for the Cauchy fit.}
\begin{tabular}{ |c |c|c|c|c|c|c|c|c|c|c| }
    \hline  \hline
    $C$ & $a^{2}c_{M, 0}$-G & $a^{2}c_{M, 0}$-C & $a^{2}c_{M,0} (\hbox{corr.})$ & $\sigma_{G} / m_C$ & $\sigma_{C} / m_C$ \\
    \hline
   PS & 0.0310(5) & 0.0310(5) & 0.0309(7) & 0.3 & 0.3 \\ 
V & 0.0529(9) & 0.0529(10) & 0.0477(16) & 0.3 & 0.27 \\ 
AV & 0.0553(8) & 0.0553(9) & 0.0561(12) & 0.25 & 0.25 \\
\hline
ps & 0.0770(15) & 0.0770(16) & 0.0770(22) & 0.3 & 0.3 \\ 
v & 0.1003(24) & 0.1003(27) & 0.1019(36) & 0.3 & 0.27 \\ 
av & 0.0906(17) & 0.0906(19) & 0.0908(26) & 0.25 & 0.25 \\ 
    \hline \hline
\end{tabular}
\end{table}

\begin{table}    
\centering
\caption{\label{table:E5_matrix_mesons} Numerical results for the mesons matrix elements in ensemble M5.  G stands for Gaussian fit, C stands for Cauchy function fit, $C$ indicates the mesonic channel considered, $c_{M, 0}$ is the result of the correlator fitting analysis, $\sigma_G$ is the energy smearing radius used for the Gaussian fits, $\sigma_C$ for the Cauchy fit.}
\begin{tabular}{ |c |c|c|c|c|c|c|c|c|c|c| }
    \hline  \hline
    $C$ & $a^{2}c_{M, 0}$-G & $a^{2}c_{M, 0}$-C & $a^{2}c_{M,0} (\hbox{corr.})$ & $\sigma_{G} / m_C$ & $\sigma_{C} / m_C$ \\
    \hline
   PS & 0.0185(3) & 0.0185(3) & 0.0190(4) & 0.25 & 0.25 \\ 
V & 0.0325(5) & 0.0325(6) & 0.0301(8) & 0.3 & 0.3 \\ 
AV & 0.0498(7) & 0.0498(8) & 0.0506(11) & 0.25 & 0.2 \\ 
\hline
ps & 0.0625(12) & 0.0625(13) & 0.0628(18) & 0.25 & 0.25 \\ 
v & 0.0814(20) & 0.0814(22) & 0.0825(20) & 0.3 & 0.3 \\ 
av & 0.0778(15) & 0.0778(16) & 0.0784(36) & 0.25 & 0.2 \\ 
    \hline \hline
\end{tabular}
\end{table}

\begin{table}    
\centering
\caption{\label{table:E1_matrix_CB} Numerical results for the chimera baryons overlap factors in ensemble M1.  G stands for Gaussian fit, C stands for Cauchy function fit, $C$ indicates the mesonic channel considered, $K_{B,0}$ is the result of the correlator fitting analysis, $\sigma_G$ is the energy smearing radius used for the Gaussian fits, $\sigma_C$ for the Cauchy fit.}
\begin{tabular}{ |c |c|c|c|c|c|c|c|c|c|c| }
    \hline  \hline
    $C$ & $a^{3}K_{B, 0}$-G & $a^{3}K_{B,0}$-C & $a^{3}K_{B,0} (\hbox{corr.})$ & $\sigma_{G} / m_C$ & $\sigma_{C} / m_C$ \\
    \hline
   $\Lambda^{+}_{\rm CB}$ & 0.000813(11) & 0.000812(12) & 0.000813(11) & 0.32 & 0.32 \\
$\Sigma^{+}_{\rm CB}$ & 0.003523(15) & 0.003527(17) & 0.003527(14) & 0.3 & 0.3 \\
$\Sigma^{* \, +}_{\rm CB}$ & 0.004054(18) & 0.004062(19) & 0.004058(18) & 0.2 & 0.2 \\
$\Lambda^{-}_{\rm CB}$ & 0.000533(8) & 0.000533(8) & 0.000534(10) & 0.22 & 0.22 \\
$\Sigma^{-}_{\rm CB}$ & 0.001984(9) & 0.001984(9) & 0.001986(23) & 0.18 & 0.18 \\
$\Sigma^{* \, -}_{\rm CB}$ & 0.002264(10) & 0.002267(10) & 0.002265(16) & 0.2 & 0.2 \\

    \hline \hline
\end{tabular}
\end{table}

\begin{table}    
\centering
\caption{\label{table:E2_matrix_CB} Numerical results for the chimera baryons overlap factors in ensemble M2.  G stands for Gaussian fit, C stands for Cauchy function fit, $C$ indicates the mesonic channel considered, $K_{B,0}$ is the result of the correlator fitting analysis, $\sigma_G$ is the energy smearing radius used for the Gaussian fits, $\sigma_C$ for the Cauchy fit.}
\begin{tabular}{ |c |c|c|c|c|c|c|c|c|c|c| }
    \hline  \hline
    $C$ & $a^{3}K_{B, 0}$-G & $a^{3}K_{B,0}$-C & $a^{3}K_{B,0} (\hbox{corr.})$ & $\sigma_{G} / m_C$ & $\sigma_{C} / m_C$ \\
    \hline
   $\Lambda^{+}_{\rm CB}$ & 0.000711(10) & 0.000711(11) & 0.000709(20) & 0.3 & 0.3 \\
$\Sigma^{+}_{\rm CB}$ & 0.003476(15) & 0.003473(16) & 0.003476(12) & 0.23 & 0.23 \\
$\Sigma^{* \, +}_{\rm CB}$ & 0.003994(17) & 0.003990(19) & 0.003994(17) & 0.2 & 0.2 \\
$\Lambda^{-}_{\rm CB}$ & 0.000540(8) & 0.000540(8) & 0.000540(14) & 0.33 & 0.33 \\
$\Sigma^{-}_{\rm CB}$ & 0.001968(9) & 0.001968(9) & 0.001971(33) & 0.18 & 0.18 \\
$\Sigma^{* \, -}_{\rm CB}$ & 0.002192(9) & 0.002192(10) & 0.002193(25) & 0.2 & 0.2 \\
    \hline \hline
\end{tabular}
\end{table}

\begin{table}    
\centering
\caption{\label{table:E3_matrix_CB} Numerical results for the chimera baryons overlap factors in ensemble M3.  G stands for Gaussian fit, C stands for Cauchy function fit, $C$ indicates the mesonic channel considered, $K_{B,0}$ is the result of the correlator fitting analysis, $\sigma_G$ is the energy smearing radius used for the Gaussian fits, $\sigma_C$ for the Cauchy fit.}
\begin{tabular}{ |c |c|c|c|c|c|c|c|c|c|c| }
    \hline  \hline
    $C$ & $a^{3}K_{B, 0}$-G & $a^{3}K_{B,0}$-C & $a^{3}K_{B,0} (\hbox{corr.})$ & $\sigma_{G} / m_C$ & $\sigma_{C} / m_C$ \\
    \hline
   $\Lambda^{+}_{\rm CB}$ & 0.000748(10) & 0.000748(11) & 0.000749(13) & 0.27 & 0.27 \\
$\Sigma^{+}_{\rm CB}$ & 0.003362(15) & 0.003359(16) & 0.003362(12) & 0.28 & 0.28 \\
$\Sigma^{* \, +}_{\rm CB}$ & 0.004012(17) & 0.004012(19) & 0.004017(13) & 0.18 & 0.18 \\
$\Lambda^{-}_{\rm CB}$ & 0.000550(8) & 0.000550(9) & 0.000550(9) & 0.25 & 0.25 \\
$\Sigma^{-}_{\rm CB}$ & 0.001989(9) & 0.001986(9) & 0.001989(15) & 0.32 & 0.32 \\
$\Sigma^{* \, -}_{\rm CB}$ & 0.002158(9) & 0.002158(10) & 0.002159(25) & 0.24 & 0.24 \\

    \hline \hline
\end{tabular}
\end{table}

\begin{table}    
\centering
\caption{\label{table:E4_matrix_CB} Numerical results for the chimera baryons overlap factors in ensemble M4.  G stands for Gaussian fit, C stands for Cauchy function fit, $C$ indicates the mesonic channel considered, $K_{ B,0}$ is the result of the correlator fitting analysis, $\sigma_G$ is the energy smearing radius used for the Gaussian fits, $\sigma_C$ for the Cauchy fit.}
\begin{tabular}{ |c |c|c|c|c|c|c|c|c|c|c| }
    \hline  \hline
    $C$ & $a^{3}K_{B, 0}$-G & $a^{3}K_{B,0}$-C & $a^{3}K_{B,0} (\hbox{corr.})$ & $\sigma_{G} / m_C$ & $\sigma_{C} / m_C$ \\
    \hline
   $\Lambda^{+}_{\rm CB}$ & 0.001022(15) & 0.001022(16) & 0.001022(8) & 0.3 & 0.3 \\
$\Sigma^{+}_{\rm CB}$ & 0.004365(19) & 0.004362(20) & 0.004362(12) & 0.25 & 0.25 \\
$\Sigma^{* \, +}_{\rm CB}$ & 0.004993(21) & 0.004993(23) & 0.004999(19) & 0.25 & 0.25 \\
$\Lambda^{-}_{\rm CB}$ & 0.000522(8) & 0.000522(8) & 0.000522(22) & 0.27 & 0.27 \\
$\Sigma^{-}_{\rm CB}$ & 0.001833(8) & 0.001833(9) & 0.001835(34) & 0.25 & 0.25 \\
$\Sigma^{* \, -}_{\rm CB}$ & 0.002281(10) & 0.002280(11) & 0.002281(32) & 0.3 & 0.3 \\
    \hline \hline
\end{tabular}
\end{table}

\begin{table}    
\centering
\caption{\label{table:E5_matrix_CB} Numerical results for the chimera baryons overlap factors in ensemble M5.  G stands for Gaussian fit, C stands for Cauchy function fit, $C$ indicates the mesonic channel considered, $K_{B,0}$ is the result of the correlator fitting analysis, $\sigma_G$ is the energy smearing radius used for the Gaussian fits, $\sigma_C$ for the Cauchy fit.}
\begin{tabular}{ |c |c|c|c|c|c|c|c|c|c|c| }
    \hline  \hline
    $C$ & $a^{3}K_{B, 0}$-G & $a^{3}K_{B,0}$-C & $a^{3}K_{B,0} (\hbox{corr.})$ & $\sigma_{G} / m_C$ & $\sigma_{C} / m_C$ \\
    \hline
   $\Lambda^{+}_{\rm CB}$ & 0.000503(7) & 0.000503(8) & 0.000503(11) & 0.25 & 0.25 \\
$\Sigma^{+}_{\rm CB}$ & 0.002312(10) & 0.002312(11) & 0.002312(13) & 0.3 & 0.3 \\
$\Sigma^{* \, +}_{\rm CB}$ & 0.002498(11) & 0.002498(12) & 0.002498(18) & 0.2 & 0.2 \\
$\Lambda^{-}_{\rm CB}$ & 0.000392(6) & 0.000392(6) & 0.000392(16) & 0.3 & 0.3 \\
$\Sigma^{-}_{\rm CB}$ & 0.001525(7) & 0.001525(7) & 0.001525(22) & 0.2 & 0.2 \\
$\Sigma^{* \, -}_{\rm CB}$ & 0.001551(7) & 0.001553(7) & 0.001553(31) & 0.25 & 0.25 \\
    \hline \hline
\end{tabular}
\end{table}

\clearpage

\section{Matrix elements obtained from ultralocal operators using stochastic wall sources}
\label{sec:Wall}

For the analysis summarized in Sect.~\ref{sec:matrix_elements}, we extracted the matrix elements of interest from correlation functions involving ultralocal operators and Wuppertal-smeared operators---see Eq.~\eqref{eq:decay_const_corr}---following the process  discussed in Appendix~\ref{sec:Wuppertal_APE}. In this appendix we compare these results to a determination of the hadron-vacuum matrix elements (i.e. the decay constants) based on  operators that have not been smeared, but rather built by inverting  the Dirac operator with the use of stochastic wall sources with $Z_2 \otimes Z_2$ noise~\cite{Boyle:2008rh}, and without  APE smearing.

To measure the pseudoscalar decay constants, we perform a simultaneous fit to the pseudoscalar correlation function, $C_{\rm{PS}} (t) = \langle \mathcal{O}_{\rm{PS}}(t) \, \bar{\mathcal{O}}_{\rm {PS}}(0)\rangle$, and the correlation function between pseudoscalar and axial-vector operator, restricted to the $\mu=0$ component, $C_{\rm{AV},\, \rm{PS}} (t) = \langle \mathcal{O}^{\mu=0}_{\rm{AV}}(t) \, \bar{\mathcal{O}}_{\rm {PS}}(0)  \rangle$, that,
 at large Euclidean times, takes the form
\begin{equation}
    C_{\rm{AV},\rm{PS}}(t) \to \dfrac{f_{\rm PS} \langle 0 | \mathcal{O}_{\rm {PS}}|\rm{PS} \rangle^*}{\sqrt{2}} \left( e^{-m_{\rm PS}t} + e^{-m_{\rm PS}(T-t)} \right) \, .
\end{equation}
The decay constant, $f_{\rm PS}$, is then extracted from this behavior, combined with the matrix element of the pseudoscalar correlation function, $C_{\rm{PS}} (t)$, and its mass. We proceed in the same manner for $({\rm f})$-type as well as    $({\rm as})$-type fermions.

\begin{table}
    \centering
    \caption{Numerical results for decay constants obtained by using stochastic wall sources (denoted by the label ``$loc$") from ensembles M1-M5 and pseudoscalar/vector meson channels, for both types of fermions, and comparison with the results obtained by using different Wuppertal smearing levels (labeled by ``$smear$").}
    \begin{tabular}{ |c|c|c||c|c||c|c||c|c||c|c| }
\hline
Ensemble &  $af_{\rm PS}^{loc}$ &  $af_{\rm PS}^{smear}$ &  $af_{\rm ps}^{loc}$ &  $af_{\rm ps}^{smear}$ &  $af_{\rm V}^{loc}$ &  $af_{\rm V}^{smear}$ &  $af_{\rm v}^{loc}$ &  $af_{\rm v}^{smear}$ \\
\hline \hline
M1 & 0.0500(2) & 0.0503(3) & 0.0832(2) & 0.0813(7) & 0.0770(3) & 0.0771(3) & 0.1028(4) & 0.1015(8) \\
M2 & 0.0494(3) & 0.0487(3) & 0.0822(3) & 0.0805(8) & 0.0729(4) & 0.0729(4) & 0.099(1) & 0.0984(6) \\
M3 & 0.0497(2) & 0.0499(4) & 0.0828(2) & 0.0820(9) & 0.0717(4) & 0.073(2) & 0.1002(8) & 0.0988(6) \\
M4 & 0.0540(2) & 0.0538(2) & 0.0868(2) & 0.0872(2) & 0.0745(5) & 0.076(1) & 0.1068(5) & 0.107(1) \\
M5 & 0.0445(5) & 0.0437(4) & 0.0787(9) & 0.0770(6) & 0.0639(8) & 0.0634(4) & 0.0952(4) & 0.0942(7) \\
\hline \hline
\end{tabular}
    \label{tab:wall_comparison}
\end{table}

\begin{figure}[t]
    \includegraphics[width=0.7\linewidth]{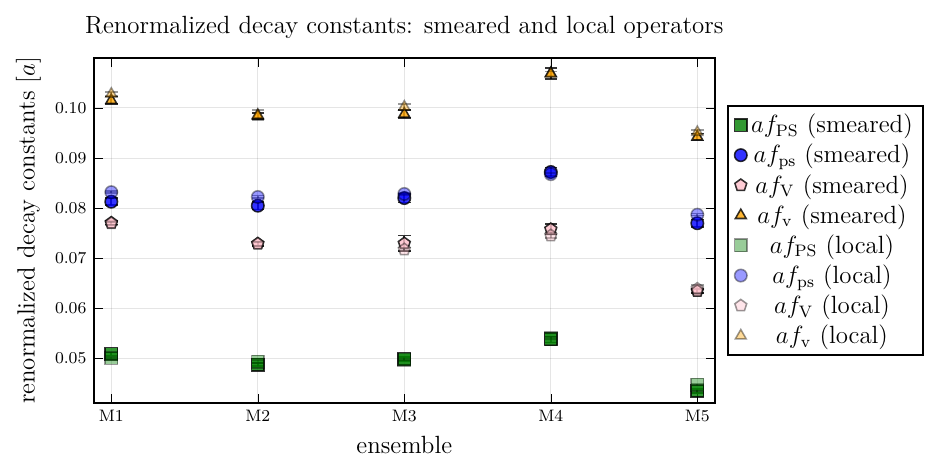}
    \caption{Comparison between matrix elements found through simultaneous fits of smeared-smeared and smeared-unsmeared correlation functions against the adoption of  stochastic wall sources, for all available ensembles, M1-M5, and for  both type of fermions.
    We restrict our attention to the pseudoscalar and vector channel, for which we obtained  the most precise measurements.
    \label{fig:wall_comparison}}
\end{figure}

The fitting results for the $Z_2 \otimes Z_2$ stochastic wall case for ensembles M1-M5 are renormalized according to Eq.~\eqref{eq:ren_const_meson}, and then reported in Tab.~\ref{tab:wall_comparison}. A comparison with simultaneous fits of smeared-smeared and smeared-unsmeared two-point correlation functions is shown in Fig.~\ref{fig:wall_comparison}. From this study, it appears that the two methods present compatible results, with no particular gain with either choice.

\bibliographystyle{JHEP} 
\bibliography{ref}
\end{document}